\documentclass[journal,10pt,twocolumn,final]{IEEEtran}
\usepackage{amsmath,amsfonts}
\usepackage{algorithmic}
\usepackage{algorithm}
\usepackage{array}
\usepackage[caption=false,font=normalsize,labelfont=sf,textfont=sf]{subfig}
\usepackage{textcomp}
\usepackage{stfloats}
\usepackage{url}
\usepackage{verbatim}
\usepackage{graphicx}
\usepackage{cite}
\usepackage{multirow}
\usepackage{makecell}
\usepackage{pifont}
\usepackage{color}
\usepackage{xcolor}
\usepackage{hyperref}   
\usepackage{orcidlink}  
\newcolumntype{C}[1]{>{\centering\let\newline\\\arraybackslash\hspace{0pt}}m{#1}}
\begin{document}

\title{AI Empowered Communication and Radar Modulation Recognition: A Survey}

\author{Wei Huang\orcidlink{0000-0002-8284-2310},~\IEEEmembership{Member,~IEEE,} Pengfei Zhang\orcidlink{0009-0009-6943-728X}, Huai Qin\orcidlink{0009-0009-6128-774X}, Jixuan Zhou\orcidlink{0000-0003-3496-6287}, Hao Zhang\orcidlink{0000-0002-6454-0239},~\IEEEmembership{Senior Member,~IEEE,} Kaitao Meng*\orcidlink{0000-0001-7479-2280},~\IEEEmembership{Member,~IEEE,} Christos Masouros\orcidlink{0000-0002-8259-6615},~\IEEEmembership{Fellow,~IEEE} 
	\thanks{Wei Huang, Pengfei Zhang, Huai Qin and Hao Zhang are with the Faculty of Information Science and Engineering, Ocean University of China (email: hw@ouc.edu.cn; zpf9280@stu.ouc.edu.cn; qinhuai@stu.ouc.edu.cn; qq258357656@163.com).}
	\thanks{Jixuan Zhou is with the Hanjiang National Laboratory (email: zhoujixuan@whu.edu.cn).}
	\thanks{Kaitao Meng is with the Department of Electrical and Electronic Engineering, The University of Manchester, Manchester, UK. Corresponding author: Kaitao Meng (kaitao.meng@manchester.ac.uk).}
	\thanks{Christos Masouros is with the Department of Electrical and Electronic Engineering, University College London, London, UK (email: c.masouros@ucl.ac.uk).}
	\thanks{Manuscript received XX XX, 2026; revised XX XX, 2026.}}

\markboth{IEEE,~Vol.~XX, No.~XX, XXXX~2026}%
{Huang \MakeLowercase{\textit{et al.}}: AI Empowered Communication and Radar Modulation Recognition: A Survey}


\maketitle
\vspace{-20mm}
\begin{abstract}
Automatic modulation recognition (AMR) is of vital importance for ensuring communication and radar reliability, efficient spectrum utilization and resistance to electronic interference. The development of artificial intelligence (AI) technology is reshaping the technological paradigm of AMR, promoting its transition from traditional modes relying on manual features to data-driven intelligent recognition. This change is not only reflected in the significant improvement of recognition accuracy, but also injects strong momentum into the intelligent evolution of both communication and radar systems through algorithm innovation, architecture optimization, and scenario expansion. In order to clarify the current development status and bottlenecks of AMR, and to find breakthrough directions, we make a comprehensive survey of recent AI-based technologies for AMR in this paper, including model-based machine learning (ML) methods and data-driven deep learning (DL) methods. We first investigate the modulation types used in current communication and radar systems. Next, we summarize the typically used features in the field of AMR, and discuss their inherent advantages and disadvantages. Then, we introduce the basic AI models for AMR and conduct a hierarchical investigation of AMR methods for communication and radar. Finally, based on existing research works, we highlight open issues and propose future research directions for AMR.  
\end{abstract}


\section{Introduction}
\IEEEPARstart{W}{ith} rapid development of communication theory and sensing technology, the dynamic channel environment has become increasingly complex. A single modulation configuration is no longer sufficient to meet the communication and radar requirements of modern wireless devices. As more types of communication and radar modulation signals are appearing on the same frequency band and dynamically changing according to various application scenarios, accurately identifying the signal transmission mode of the target sender is of great significance to enhance communication and radar reliability, improve system efficiency, and promote the intelligent utilization of spectrum resources \cite{Van2024AICom,Zhao2025AICom}.

Traditionally, modulation recognition methods mainly rely on manual observation of signal characteristics, such as spectrum and waveform, combined with empirical judgment \cite{Xiao2022Review,Zhang2022DL}. They are suitable for simple signal environments, laboratory testing, or low dynamic scenarios. However, the recognition process is slow and cannot meet the requirements of dynamic scenarios. Automatic modulation recognition (AMR) of communication and radar signals refers to the technology that automatically identifies the modulation type used in signals by analyzing and processing their waveform, time-frequency characteristics, etc. of the received signal without prior knowledge of the signal modulation modes \cite{Xiao2022Review}. AMR aims to enable the communication or radar receiver to have the ability to ``intelligently understand unknown signals", providing basic support for subsequent demodulation, decoding, information acquisition, receive filtering, parameter estimation, or related actions such as interference and spectrum management. Therefore, AMR has become one of the main technology in software defined radio (SDR) and cognitive radio (CR) systems that promotes the intelligence of wireless networks, flexibility of modulation methods, and efficient utilization of spectrum \cite{Haykin2005CR}.

AMR has demonstrated extraordinary value in fields of communication efficiency and reliability \cite{Gupta2022EC}. Spectrum interference is an important reason for the low efficiency of spectrum utilization \cite{Zhu2015AMR,Meng2018AMR,Deepa2024Effcient}, that is, when signals from different services are superimposed on the same frequency band (such as legitimate IoT devices transmitting in the same frequency band), it will cause demodulation failure of both signals due to incompatible modulation methods, resulting in the system making a misdetection of the spectrum being occupied, forming ``invalid occupation". AMR can quickly locate interference sources through modulation feature comparison, thereby eliminating interference and restoring effective spectrum utilization \cite{Hong2019OFDM,Fang2024SrcLoc}, especially in civil aviation ground to air communication. In addition, AMR can be integrated into network management systems to monitor the modulation status of signals across the entire network in real-time. Once modulation abnormalities are detected, the fault source can be quickly located to achieve fault repair and ensure user experience. 

Given the importance of AMR, researchers have conducted extensive and in-depth research in this field. AMR is mainly divided into three categories, including decision theory methods based on maximum likelihood hypothesis testing \cite{Hameed2009HL,Ramezani2013HL,Zhang2017EM}, model-based statistical machine learning (ML) methods \cite{Wang2009SVM,Zhang2017RandomForest}, and data-driven modulation recognition with deep learning (DL) \cite{Hong2017RNN,Wang2017CNN}. The key to the first type of method is constructing a signal likelihood function and setting a threshold, minimizing the misclassification probability via the maximum likelihood criterion. For theoretically optimal, likelihood-based methods require intractable likelihood derivation, incur high computational cost, and often lack robustness across different scenarios. While the model-based statistical ML methods do not require strict derivation of the likelihood function, but extracts instantaneous features, high-order cumulants, cyclic spectrum graph, etc., combined with classifiers such as K-nearest neighbor classification (KNN) \cite{Aslam2012HOCKNN}, support vector machine (SVM) \cite{Wu2005SVM}, or random forest (RF) \cite{Zhang2017RF}. For example, Wu et al. \cite{Wu2005SVM} combined wavelet analysis and SVM, which can effectively identify ten types of unstable signals over a wide range of signal-to-noise ratios (SNRs). These methods have the advantages of low computational complexity and strong engineering applicability, but their limitations lie in excessive reliance on expert experience, poor accuracy at low SNR, and complex classifier design. 

Over the past decades, data-driven DL has made tremendous progress with the continuous development of hardware and algorithms, especially in fields such as speech recognition \cite{Rabiner1989SpeechRec,Yang2025SpeechRec} and computer vision \cite{He2016DRL,Qadri2025CV}. The emergence of DL technology provides new approaches for research and applications of communication and radar modulation recognition. Deep learning alleviates the dependence on hand-crafted feature extraction and improves representation learning capability compared with traditional AMR methods. Early DL models mainly used multi-layer perceptrons (MLPs) to extract the nonlinear mapping relationship from the spectral features of single carrier signals to modulation types \cite{Lu1996CycSpe,Like2008CycSpe}. However, as modulation types become more complex, it is difficult for MLP to fully explore the local correlation relationships of features, resulting in a bottleneck in accuracy performance. Besides, adversarial robustness has recently attracted increasing attention and several attack/defense strategies have been proposed \cite{Sadeghi2019AAC,Zhao2020AAMR,Flowers2020AAWC,Lin2021AAMR,Bao2022AADI,Kim2022AASC,Bahramali2021AAC,Kokalj2019ADL}. \cite{Adesina2023CST,Xu2025AtacSurvey} summarized existing research on adversarial attack and defense technologies in the field of AMR, and proposed a framework for adversarial attack threat modeling.

With the development of DL, convolutional neural networks (CNNs) \cite{Wang2017CNN,Tian2019CNN}, recurrent neural networks (RNNs) \cite{Hong2017RNN,Ghasemzadeh2022RNN}, Transformer \cite{Vaswani2017Att,Zheng2021GPT,Wang2023ChatGPT,Guo2025DSR1} and hybrid models \cite{Chen2020CRNN,Liu2022CGNU} have significantly improved AMR performance. Recent studies further focus on few-shot learning, robustness and lightweight deployment. However, as the parameter scale increases, hybrid models require more labeled data for training, which is often difficult to be met in practical applications with strong environmental dynamic changes. Therefore, the few-shot learning AMR problem has become the focus of recent researchers \cite{Tang2018GAN,Bu2020Transfer}. On this issue, one approach is to conduct data augmentation through generative adversarial networks (GANs) \cite{Tang2018GAN}, and the other is to perform model pre-training via transfer learning (TL) \cite{Bu2020Transfer}. Both of the above approaches have achieved good results and continue to develop towards higher precision. Besides few-shot learning problem in data-driven DL-based AMR, it is also necessary to address emerging issues such as algorithm robustness optimization, model lightweighting, and real-time performance improvement in complex environments. 

\begin{table*}[!h]
	\caption{Comparison with other surveys. \label{tab01}}
	\renewcommand{\arraystretch}{0.5}
	\centering
	\resizebox{1.0\linewidth}{!}{
	\begin{tabular}{|c|c|c|c|c|c|c|c|c|}
		\hline
		Work / Item & Model-based AMR & Data-driven AMR & Communication & Sensing & \makecell{Comparison on \\communication and radar} & SISO & MIMO & Simulation\\
		\hline
		\cite{Van2024AICom} & $\checkmark$ & $\checkmark$ & $\checkmark$ & $\times$ & $\times$ & $\times$ & $\times$ & $\times$\\
		\hline
		\cite{Jdid2021ML} & $\checkmark$ & $\checkmark$ &  $\checkmark$ & $\times$ & $\times$ & $\checkmark$& $\checkmark$ & $\times$ \\
		\hline
		\cite{Xiao2022Review} & $\times$ & $\checkmark$ & $\checkmark$ & $\times$ & $\times$ & $\checkmark$& $\times$ & $\times$\\
		\hline
		\cite{Zhang2022DL} & $\times$ & $\checkmark$ & $\checkmark$ & $\times$ & $\times$ & $\checkmark$& $\times$ & $\checkmark$ \\
		\hline
		\cite{Peng2022Survey} & $\times$ & $\checkmark$ & $\checkmark$ & $\times$ & $\times$ & $\checkmark$& $\times$ & $\checkmark$ \\
		\hline
		\cite{Zhang2024MIMO} & $\checkmark$ &  $\checkmark$ & $\checkmark$ & $\times$ & $\times$ & $\checkmark$ & $\checkmark$ & $\times$ \\		
		\hline
		Our & $\checkmark$ & $\checkmark$ & $\checkmark$ & $\checkmark$ & $\checkmark$ & $\checkmark$ & $\checkmark$ & $\checkmark$ \\		
		\hline
	\end{tabular}
	}
\end{table*}

In recent years, several review works has summarized the development of this field \cite{Van2024AICom,Jdid2021ML,Xiao2022Review,Zhang2022DL,Zhang2024MIMO} including studies on massive MIMO systems \cite{Arshad2020Spectral}, but mainly focused on communication AMR or DL-based AMR, while few discuss radar modulation recognition and unified AI frameworks.

The application of AMR is far beyond communication and radar, and it also plays a key role and has a rich research accumulation in fields such as electronic warfare, signal intelligence, spectrum regulation, underwater acoustic communication, and cognitive radio. However, due to limitations in length and topic coherence, we will focus on comparing the differences between AMR in two typical tasks within current AMR research: communication and radar, which benefits from relatively mature public datasets, the most comprehensive algorithm validation, and the deepest penetration of artificial intelligence technology. In contrast to existing surveys, this work offers a comprehensive and up-to-date overview of AI-based AMR techniques up to 2026\footnote{To ensure a comprehensive and systematic review, the literature considered in this survey was collected from major scientific databases, including IEEE Xplore, Elsevier ScienceDirect, SpringerLink, Web of Science, and Google Scholar. The search covered publications up to January 2026.}, and systematically examines their use in both communication and radar sensing systems. 

In this paper, we first compare the differences between communication and perception tasks in detail from three aspects: modulation method, recognition features, and recognition methods. Next, we divide AI AMR methods into two categories: model-based statistical ML methods and data-driven DL methods. Then, in each category, we summarize related works based on the different input features of the model. Compared to \cite{Van2024AICom}, our research on modulation recognition is more focused on AMR. Compared to \cite{Jdid2021ML,Xiao2022Review,Peng2022Survey,Zhang2022DL,Zhang2024MIMO}, we provide a broader summary of AI-based AMR methods that additionally summarizes the application of model-based methods for AMR. We have improved the experiment in \cite{Zhang2022DL} and conducted simulation tests on classical methods using a different public dataset. Table~\ref{tab01} has shown the difference between our survey and other related works, where our greatest contribution lies in summarizing the modulation signal recognition work used for sensing. 

Based on the current development status of AMR, we have discovered some open issues and provided suggestions for future research directions, mainly aiming at SNR scenarios and few-shot learning scenarios to enhance accuracy, which can be achieved through multimodal and cross domain fusion. Besides, lightweight design of models for practical application deployment is also an important research direction, and AMR for integrated sensing and communication (ISAC) signals in the future is also one of the most important directions in the 6G era. The main contributions are summarized as follows.
\begin{itemize}
	\item We present the conventional modulation modes and features used for AMR, and fundamentally compare the applicability of various features on different modulation modes, which provides guidance for multi-source feature fusion studies.
	\item Given the lack of comparison between communication and radar of AMR in existing research, We discuss the differences of AMR in communication and radar tasks from waveform, features to recognition models.
	\item We give an in-depth analysis and summary of the AI applications in AMR for both communication and radar, including model-based ML methods and data-driven DL methods. We discuss the advantages and disadvantages of various models and test their accuracy and efficiency performance through experiments on public datasets, which serves as vital guidance for further advancing the application of AI in AMR. 
	\item We present significant open issues when applying AI technology in AMR from several perspective, such as accuracy issues in complex channel environments, few-shot learning problem, model complexity problem, recognition of different modulation methods and waveforms, and recognition of future ISAC signals.
\end{itemize}

\begin{figure}[!h]
	\centering
	\includegraphics[width=0.7\linewidth]{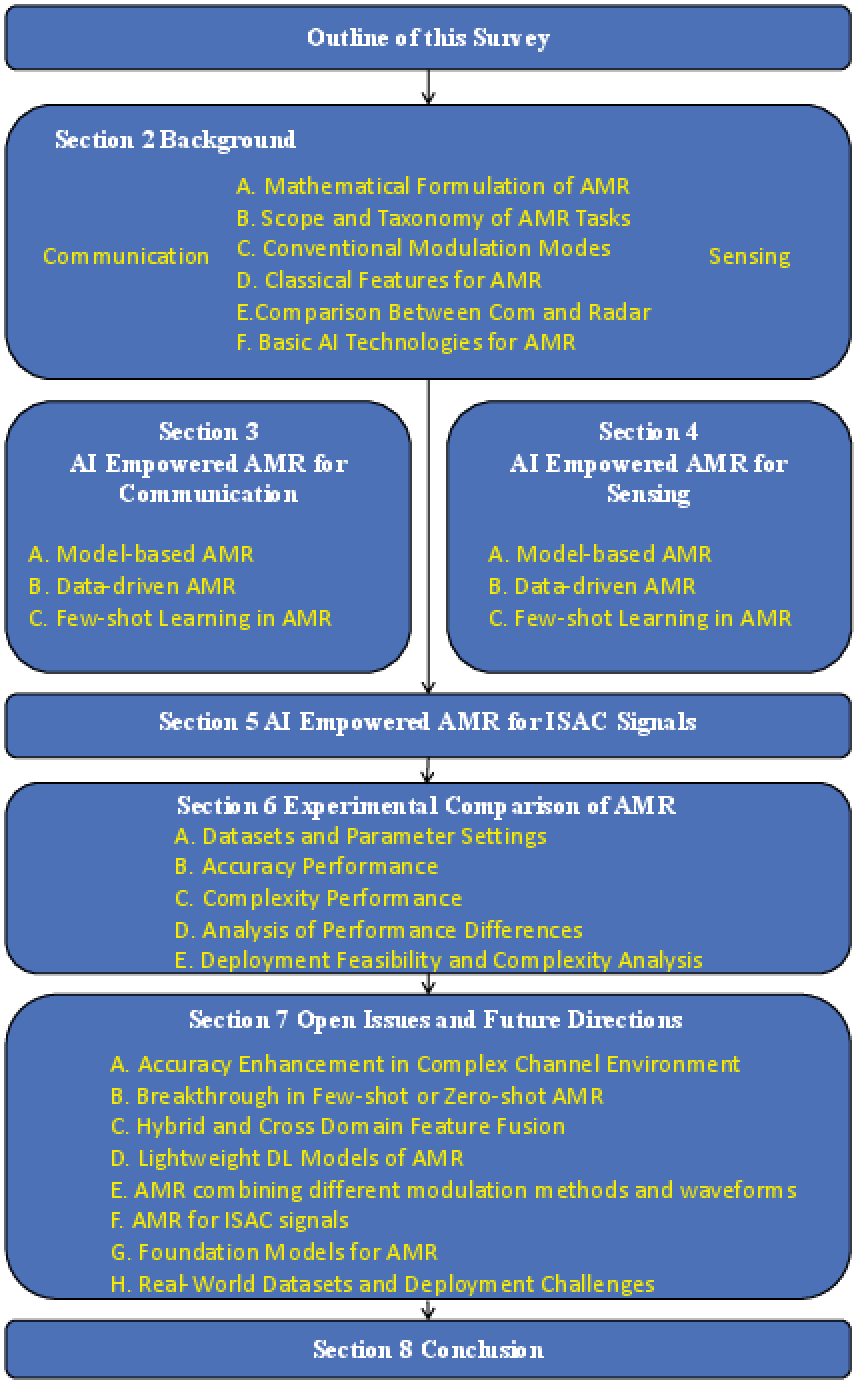}
	\caption{The outline of this paper.}
	\label{fig01}
\end{figure}

The content arrangement of this survey is outlined in Fig.~\ref{fig01}. Section \uppercase\expandafter{\romannumeral2} offers comparisons between AMR in communication and radar from perspectives of fundamental background of conventional modulation modes, features used in AMR, and basic AI technologies suitable for AMR. Section \uppercase\expandafter{\romannumeral3} delves into an in-depth analysis of existing studies for communication, including model-based methods and data-driven methods. While Section \uppercase\expandafter{\romannumeral4} makes a detail survey on AMR for sensing. Section \uppercase\expandafter{\romannumeral5} discusses recent ISAC signal recognition issues. Section \uppercase\expandafter{\romannumeral6} verifies the accuracy and efficiency performance of some classical models on four public datasets. Section \uppercase\expandafter{\romannumeral7} discusses open issues and provides some future directions, and the conclusion is drawn in \uppercase\expandafter{\romannumeral8}. 

\section{Background}
In this section, we first give a unified mathematical model of AMR, and then compare the differences between communication and radar tasks in detail from three aspects: modulation modes, recognition features, and recognition methods. The overall comparison between AMR in communication and radar tasks is given in Table~\ref{tab_compare}.
\begin{table*}[!h]
	\caption{Comparison between AMR in communication and radar tasks. \label{tab_compare}}
	\renewcommand{\arraystretch}{1}
	\centering
	\resizebox{1.0\linewidth}{!}{
	\begin{tabular}{|c|c|c|}
		\hline
		Dimension & Communication & Sensing \\
		\hline
		Mission objectives & Demodulation & Target parameter estimation, interference identification \\
		\hline
		Core performance indicators & Recognition accuracy & \makecell{Detection probability, false alarm probability,\\ parameter estimation error and recognition accuracy}  \\
		\hline
		Typical features & \makecell{I/Q sequence,\\ constellation diagrams,\\ high-order cumulants, etc.}& \makecell{Time-frequency spectrum, Doppler shift,\\ intra-pulse phase, frequency coding features, etc.} \\
		\hline
		Channel characteristics & AWGN, multipath fading & multipath fading, scattering, Doppler shift \\
		\hline
		Classical AI models & \makecell{Model-based methods (SVM, RF, etc.) \\Data-driven methods (CNN, RNN, Transformer \\ hybrid model, etc.)} & \makecell{Model-based methods (SVM, KNN, etc.) \\Data-driven methods (CNN, Transformer, Mamba, etc.)}\\
		\hline
	\end{tabular}}
\end{table*}
\begin{table*}[!h]
	\caption{Conventional modulation signal categories. \label{tab03}}
	\renewcommand{\arraystretch}{1}
	\centering
	\resizebox{1.0\linewidth}{!}{
	\begin{tabular}{|c|c|c|c|c|c|}
		\hline
		\multicolumn{6}{|c|}{Communication}\\
		\hline
		\multicolumn{2}{|c|}{Modulation types} & Representation & Advantage & Disadvantage & Typical application \\
		\hline
		\multicolumn{2}{|c|}{Amplitude Shift Keying (ASK)} & ASK, OASK & low cost, simple circuit & weak anti-interference & RFID \cite{ISO14443-2:2016,Goay2022RFID} \\
		\cline{2-6}
		\multicolumn{2}{|c|}{Frequency Shift Keying (FSK)} & FSK, GFSK & strong anti-interference & \makecell{inefficient utilization of\\ frequency band resources} &  \makecell{2G mobile communication \cite{Mouly1992},\\Bluetooth \cite{Bluetooth}, ZigBee \cite{Zigbee,Zigbee2}}\\
		\cline{2-6}
		\multicolumn{2}{|c|}{Phase Shift Keying (PSK)} & BPSK, QPSK & \makecell{strong anti-interference,\\high power utilization,\\high bandwidth utilization } & \makecell{decreased anti-interference\\ability during high-order\\ modulation} & \makecell{3G/4G mobile communication \cite{3GPP_TS_25213,3GPP_TS_36211},\\wifi \cite{IEEE_80211}, satellite communication \cite{gaudenzi2006turbo}} \\
		\cline{2-6}
		\multicolumn{2}{|c|}{Quadrature Amplitude Modulation (QAM)}  & 16/32/64QAM & \makecell{high bandwidth utilization,\\medium power utilization} & weak anti-interference & \makecell{4G/5G mobile communication \cite{3GPP_TS_36211,LTE_modulation_comparison,3GPP_TS_38211,Lin_2019},\\optical fiber communication \cite{agrawal2010fiber}} \\
		\hline
		\multicolumn{6}{|c|}{Radar Sensing}\\
		\hline
		\multicolumn{2}{|c|}{Modulation types} & Representation & Advantage & Disadvantage & Typical application \\
		\hline
		\multicolumn{2}{|c|}{Pulse Modulation (PLM)} & PLM & strong anti-interference & high peak to average power & Meteorological detection radar \cite{Gomi2017PLM} \\
		\hline
		\multicolumn{2}{|c|}{Continuous Wave (CW)} & Single Frequency CW & low peak to average power & vulnerability to narrowband interference & UAV obstacle avoidance radar \cite{Wessendorp2021LFM} \\
		\hline
		\multicolumn{2}{|c|}{Phase Modulation (PM)} & Phase Coding Modulation (PCM) & high distance resolution & complex signal processing & Synthetic Aperture Radar \cite{Sachidananda1998}\\
		\hline
		\multicolumn{2}{|c|}{Frequency Modualtion (FM)} & Linear FM & \makecell{high distance resolution,\\low peak to average power,\\strong anti-interference} & \makecell{complex signal processing,\\sensitive to Doppler effect} & Remote surveillance radar, SAR \cite{Sachidananda1998}\\
		\hline
	\end{tabular}}
\end{table*}

\subsection{Mathematical Formulation of AMR}
\subsubsection{Observation Models}
AMR can be formulated as a statistical inference and multi-hypothesis classification problem. Given a received signal observation, the objective of AMR is to determine the modulation format that most likely generated the observation from a predefined modulation set. Let $\mathcal{Q} = {Q_1,Q_2,...,Q_K}$ denote the set of candidate modulation types. For communication systems, $\mathcal{Q}$ may include ASK, FSK, PSK, QAM and OFDM signals, whereas for radar sensing systems it may contain LFM, FMCW, PCM, PLM and other radar waveforms.

The received observation can generally be represented as
\begin{equation}
    	\mathbf r = \mathcal{G} \left(Q_k,\boldsymbol{\eta}\right) + \mathbf n, \label{eq:unified_model}
\end{equation} 
where $Q_k$ denotes the modulation hypothesis, $\mathbf n$ represents additive noise, and $\boldsymbol{\eta}$ contains unknown nuisance parameters except for additive noise, such as channel fading, timing offset, frequency offset, phase offset, Doppler shift, clutter, and target scattering characteristic. Depending on the application scenario, the observation model can correspond to communication signals, radar sensing signals, or ISAC signals.

\subsubsection{AMR as a Multi-Hypothesis Test}
Based on the above signal models, AMR can be formulated as a multi-hypothesis testing problem
\begin{equation}
	H_k:\mathbf r \sim	p(\mathbf r|Q_k,\boldsymbol{\eta}),	\quad k=1,\ldots,K	\label{eq:hypothesis}
\end{equation}
where $p(\mathbf r|Q_k,\boldsymbol{\eta})$ denotes the conditional distribution of the observation under modulation hypothesis $Q_k$. The nuisance parameter vector is defined as
\begin{equation}
	\boldsymbol{\eta} =	\left\{	h,	\sigma^2,	\tau,	\Delta f,	\phi,	\nu,	\alpha	\right\},	\label{eq:nuisance}
\end{equation}
where $h$ is the channel response, $\sigma^2$ is the noise variance, $\tau$ is the timing offset, $\Delta f$ is the carrier frequency offset, $\phi$ is the phase offset, $\nu$ is the Doppler shift, and $\alpha$ denotes the target scattering parameters. For communication systems, the dominant nuisance parameters are usually channel fading, timing offset, carrier frequency offset, and phase offset. In radar sensing systems, Doppler shifts, target scattering characteristics, clutter, and multipath reflections further complicate the recognition process.

\subsubsection{Communication Signal Model}
For communication systems, assuming transceivers operate at a same center frequency, $s_{com}(t)$ is the time domain modulation signal carrying information, and the complex carrier mathematical model under hypothesis $Q_k$ can be expressed as:
\begin{equation}
	s_{com}(t) = \sum_{m=1}^{M}\sqrt{E} a_m g(t-mT_s)e^{j(\omega_ct+\theta_c)},
\end{equation}
where $\omega_c$ is the carrier frequency, $\theta_c$ is the carrier phase, $M$ is the total number of symbols, $E$ is the power of modulation signal, and $a_m$ denotes the sequence of symbol. After multipath channel transmission and noise interference, the signal arrives at the receiving end, which will be:
\begin{equation}
	r_{com}(t)	=	e^{j(2\pi\Delta f t+\phi_0)}\int h(\tau)s_{com}(t-\tau-\tau_0)d\tau + n(t)
	\label{eq:comm_rx}
\end{equation}
where $h()$ is the channel impulse response function considering wireless signal distortion caused by multiple factors, $n(t)$ is the additive white Gaussian noise with a mean of 0 and a variance of $\sigma^2$, and $\phi_0$, $\Delta f$, and $\tau$ are corresponding phase offset, frequency offset and timing offset, respectively. The received communication signal is affected by multipath fading, timing offset, carrier frequency offset, phase offset and additive noise, which constitute the primary nuisance parameters in communication AMR.

\subsubsection{Radar Signal Model}

Radar observations additionally include Doppler, scattering and clutter effects. The recognition of radar modulation signals is still largely focused on single channel modulation systems, except for a few works that default to OFDM signal form \cite{Wang2014OFDM,Du2021OFDM}. For radar sensing systems, the received signal can be represented as
\begin{equation}
	r_{radr}(t) = \sum_{p_t=1}^{P_t}\alpha_{p_t} s(t-\tau_{p_t})e^{j2\pi \nu_{p_t}t} + n(t),
\end{equation}
where $s(t)$ denotes the transmitted signal, $\alpha_{p_t}$ is the scattering coefficient, $\tau_{p_t} = 2 R_{p_t}/c$ is the target delay ($R_{p_t}$ is the range and $c$ is the signal propagation speed), $\nu_{p_t}=2v_{p_t}/\lambda$ is the Doppler factor ($v_{p_t}$ is the target velocity and $\lambda$ is the wave length). Unlike communication signals, radar observations are additionally affected by target-dependent factors including Doppler shifts, scattering coefficients, clutter, and multipath reflections. These factors become important nuisance parameters in radar modulation recognition.

\subsubsection{ISAC Signal Model}
With the increasing scarcity of spectrum resources, ISAC has become one of the important research topics for future 6G communication and will also become a key area of focus for AMR in the future. Assuming that a node is equipped with $J$ transmit antennas, as a general ISAC signal model for cooperative ISAC systems, the transmitted signal is expressed as:
\begin{equation}
	\boldsymbol{S} = \sqrt{E^c}\boldsymbol{W}^c\boldsymbol{S}^c + \sqrt{E^r}\boldsymbol{W}^r\boldsymbol{S}^r,
\end{equation}
where $\sqrt{E^c}$ and $\sqrt{E^r}$ denote the transmit powers allocated to communication and radar sensing signals, respectively. The matrices $\boldsymbol{W}^c \in \mathbb{C}^{J \times I}$ and $\boldsymbol{W}^r\in \mathbb{C}^{J \times J}$ represent the corresponding precoding matrices for $I$ users. $\boldsymbol{S}^c \in \mathbb{C}^{I \times O}$ and $\boldsymbol{S}^r \in \mathbb{C}^{J \times O}$ denote the modulated communication signals and radar sensing signals, respectively. The received signal can be expressed as
\begin{equation}
	r_{ISAC}(t) = H_c\left( \sqrt{E^c}\boldsymbol{W}^c\boldsymbol{S}^c \right) + H_r\left( \sqrt{E^r}\boldsymbol{W}^r\boldsymbol{S}^r \right) + n(t),
	\label{eq:isac_tx}
\end{equation}
where $H_c$ is the communication channel, and $H_r$ is the sensing echo channel. This received signal can be further written as
\begin{equation}
	r_{ISAC}(t)=	H_c s_{com}(t)	+	\sum_{p_t=1}^{P_t}	\alpha_{p_t} s_r(t-\tau_{p_t})	e^{j2\pi\nu_{p_t} t}+n(t).
	\label{eq:isac_rx}
\end{equation}
In \eqref{eq:isac_rx}, sensing signals $\sum_{p_t=1}^{P_t}	\alpha_{p_t} s_r(t-\tau_{p_t})	e^{j2\pi\nu_{p_t} t}$ actually become additional interference terms for communication modulation signals. In a conclusion, the received observation is generated by the superposition of communication and sensing waveforms in ISAC systems. Consequently, modulation recognition becomes more challenging due to the coexistence of communication symbols, sensing echoes, channel distortion, and target-dependent effects.

\subsubsection{Optimal Decision Rule}
The optimal Bayesian decision rule is given by
\begin{equation}
	\hat Q	=	\arg\max_{Q_k}	P(Q_k|\mathbf r),
	\label{eq:bayes_rule}
\end{equation}
where $P(Q_k|\mathbf r)$ denotes the posterior probability of modulation hypothesis $Q_k$. Applying Bayes' theorem yields
\begin{equation}
	\hat Q	=	\arg\max_{Q_k}	p(\mathbf r|Q_k)P(Q_k).	
\end{equation}
If all modulation classes are assumed equally likely, the decision rule reduces to the maximum likelihood criterion
\begin{equation}
	\hat Q	=	\arg\max_{Q_k}	p(\mathbf r|Q_k).	\label{eq:ml_rule}
\end{equation}
This formulation provides the theoretical optimum for AMR and serves as the foundation of likelihood-based recognition methods. 

Existing AMR techniques can be interpreted as different approximations to the decision rule in \eqref{eq:bayes_rule} to \eqref{eq:ml_rule}. Likelihood-based methods attempt to explicitly model the probability density function $p(\mathbf r|Q_k)$ and perform optimal hypothesis testing according to the maximum likelihood criterion. Model-based machine learning methods first extract discriminative features $\mathbf z=f(\mathbf r)$, such as high-order cumulants, cyclostationary features, entropy features, constellation features, and time-frequency representations. A classifier then performs
\begin{equation}
	\hat Q	=	g(\mathbf z),	\label{eq:ml_classifier}
\end{equation}
where $g()$ may denote SVM, KNN, RF, decision trees, or other machine learning models.

Deep learning methods directly learn a nonlinear mapping from observations to modulation labels
\begin{equation}
	\hat Q	=	F_{\beta}(\mathbf r),	\label{eq:dl_classifier}
\end{equation}
where $F_{\beta}()$ denotes a neural network parameterized by $\beta$. Typical architectures include CNNs, RNNs, Transformers, Mamba-based networks, and hybrid models. The network parameters are optimized by minimizing a classification loss function, commonly the cross-entropy loss
\begin{equation}
	\mathcal L	=	-\sum_{k=1}^{K}	y_k	\log(\hat y_k).	\label{eq:cross_entropy}
\end{equation}
Therefore, existing likelihood-based, model-based and DL methods can all be interpreted as different approximations of the Bayesian decision rule.

\subsection{Scope and Taxonomy of AMR Tasks}

Communication AMR and radar signal recognition differ in observation models, prior information, objectives, and evaluation metrics. Communication AMR identifies the modulation format of received signals without prior knowledge, whereas radar signal recognition includes waveform recognition, passive radar signal classification, and jamming/interference identification. In mono-static radar, the transmitted waveform is usually known, making interference identification the primary task, while passive radar requires waveform recognition due to unknown transmit signals. ISAC signal recognition is considered separately because the received signal contains superimposed communication and sensing components. Table~\ref{tab:taxonomy} gives the taxonomy of signal recognition tasks related to AMR.

\begin{table*}[t]
	\caption{Taxonomy of Signal Recognition Tasks Related to AMR}
	\label{tab:taxonomy}
	\centering
	\renewcommand{\arraystretch}{1.2}
	\begin{tabular}{p{3.2cm}|p{3.3cm}|p{3.5cm}|p{3cm}|p{2.5cm}}
		\hline
		\textbf{Task} &
		\textbf{Observation} &
		\textbf{Objective} &
		\textbf{Prior Information} &
		\textbf{Typical Metrics}
		\\
		\hline
		
		Communication AMR &
		Received communication signal &
		Modulation classification &
		Unknown modulation format &
		Recognition accuracy
		\\
		\hline
		
		Radar waveform recognition &
		Radar pulse or echo signal &
		Waveform classification &
		Waveform unknown or partially known &
		Recognition accuracy
		\\
		\hline
		
		Passive radar signal classification &
		Signals from non-cooperative emitters &
		Signal type identification &
		Limited prior information &
		Recognition accuracy
		\\
		\hline
		
		Deception jamming / interference identification &
		Radar echo and interference signals &
		Jamming detection and classification &
		Radar waveform known &
		Detection probability ($P_d$),
		false alarm probability ($P_{fa}$)
		\\
		\hline
		
		RF emitter identification &
		RF fingerprint features &
		Transmitter identification &
		Signal type usually known &
		Identification accuracy
		\\
		\hline
		
		ISAC signal recognition &
		Mixed communication and sensing signals &
		Signal separation and classification &
		Partial knowledge of communication/sensing signals &
		Classification accuracy,
		separation accuracy
		\\
		\hline
		
	\end{tabular}
\end{table*}

\subsection{Conventional Modulation Modes}
Communication systems typically employ standardized modulation schemes, such as BPSK, QPSK, for reliable and spectrally efficient information transmission, whereas radar sensing adopts flexible waveforms like chirp, pulse, and phase-coded signals to enhance detection and resolution. Representative communication and radar modulation schemes are summarized in Table~\ref{tab03}.

Analog modulation is vulnerable to noise and distortion, whereas digital modulation has become the mainstream because of its higher reliability and processing capability. Advanced multicarrier waveforms, such as orthogonal frequency division multiplexing (OFDM) signals and orthogonal time-frequency spatial (OTFS) signals, further improve throughput and robustness in multipath and high-mobility environments. Overall, communication modulation has evolved from analog to digital and from single-carrier to multicarrier schemes.

Most communication modulation types for SISO systems have been included in five popular datasets, including RML2016.04c (11 modulation classes), RML2016.10a (11 modulation classes), RML2016.10b \cite{Timothy2016RML} (10 modulation classes), RML2018.01a (24 modulation classes) \cite{Timothy2018RML}, and HisarMod2019.1 (26 modulation classes) \cite{Tekb2020Data}. RadChar \cite{Huang2023RadChar} is a publicly available dataset that covers 5 types of radar modulation signals, while most radar modulation datasets are rarely made public.

\subsection{Classical Features for AMR}

Communication and radar have different AMR objectives and evaluation metrics. Communication AMR mainly focuses on signal demodulation and spectrum monitoring, where recognition accuracy is the primary metric. In contrast, radar AMR aims at target sensing and additionally considers detection probability, false alarm probability, and parameter estimation accuracy. In mono-static radar, the transmitted waveform is usually known, making interference or deception jamming detection the primary task, whereas bi-static radar generally requires waveform recognition because the transmitted waveform is unknown or variable.

Communication and radar AMR rely on different features owing to their distinct signal characteristics. Communication signals follow deterministic modulation structures, whereas radar signals are dominated by target-dependent Doppler and scattering effects \cite{Clemente2015}. Consequently, communication AMR is primarily affected by noise and multipath fading, while radar AMR is more sensitive to Doppler and scattering interference. As a result, features such as constellation diagrams \cite{Marcelo2022Com} and high-order cumulants \cite{Swami2000HOC}, which are effective for communication AMR, may degrade significantly in radar applications. Representative AMR features are briefly reviewed below.

\subsubsection{Amplitude histogram}
\begin{figure}[!h]
	\centering
	\subfloat[]{\includegraphics[width=0.2\textwidth]{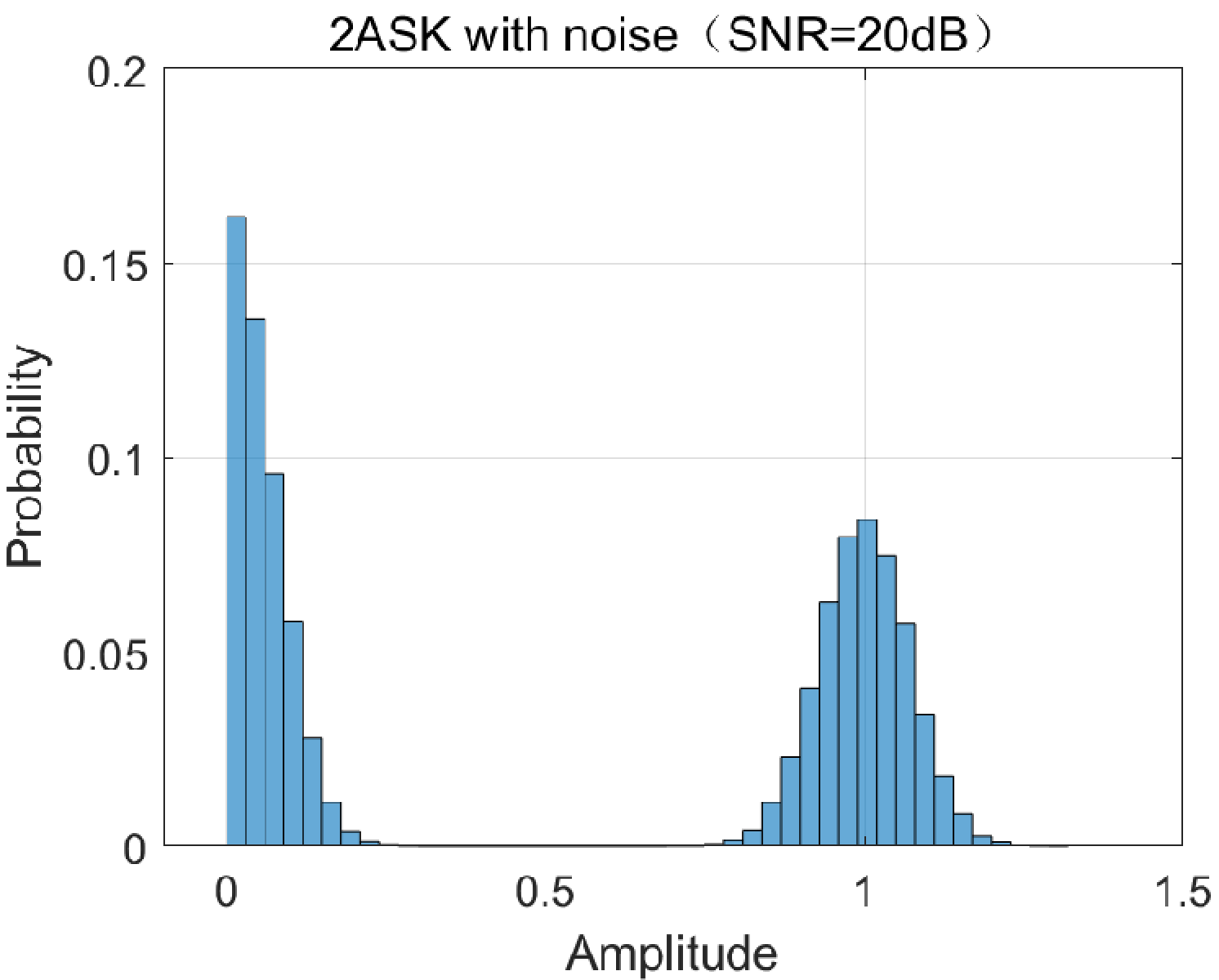}}
	\label{fig2a}
	\subfloat[]{\includegraphics[width=0.2\textwidth]{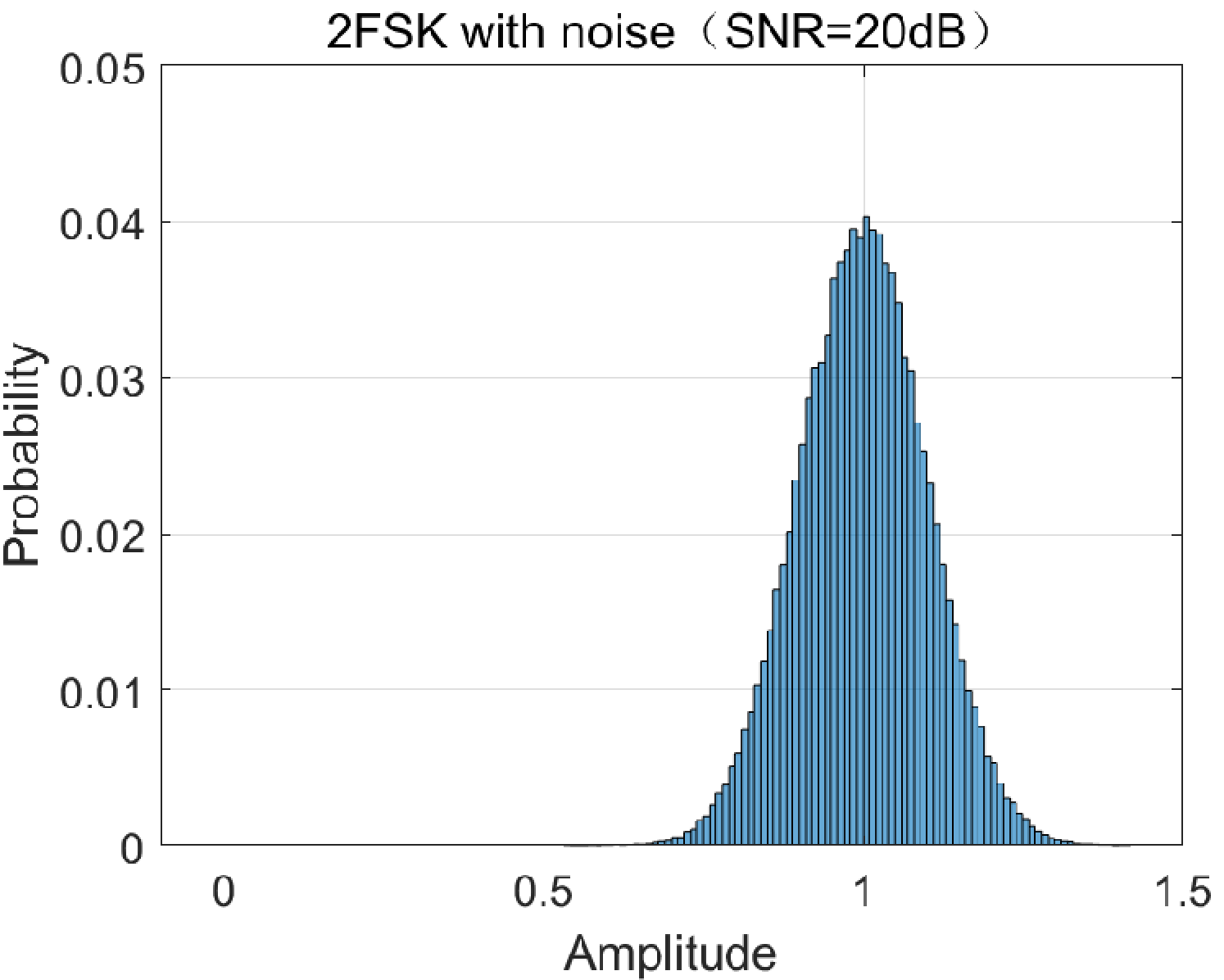}}
	\label{fig2b}\\
	\subfloat[]{\includegraphics[width=0.2\textwidth]{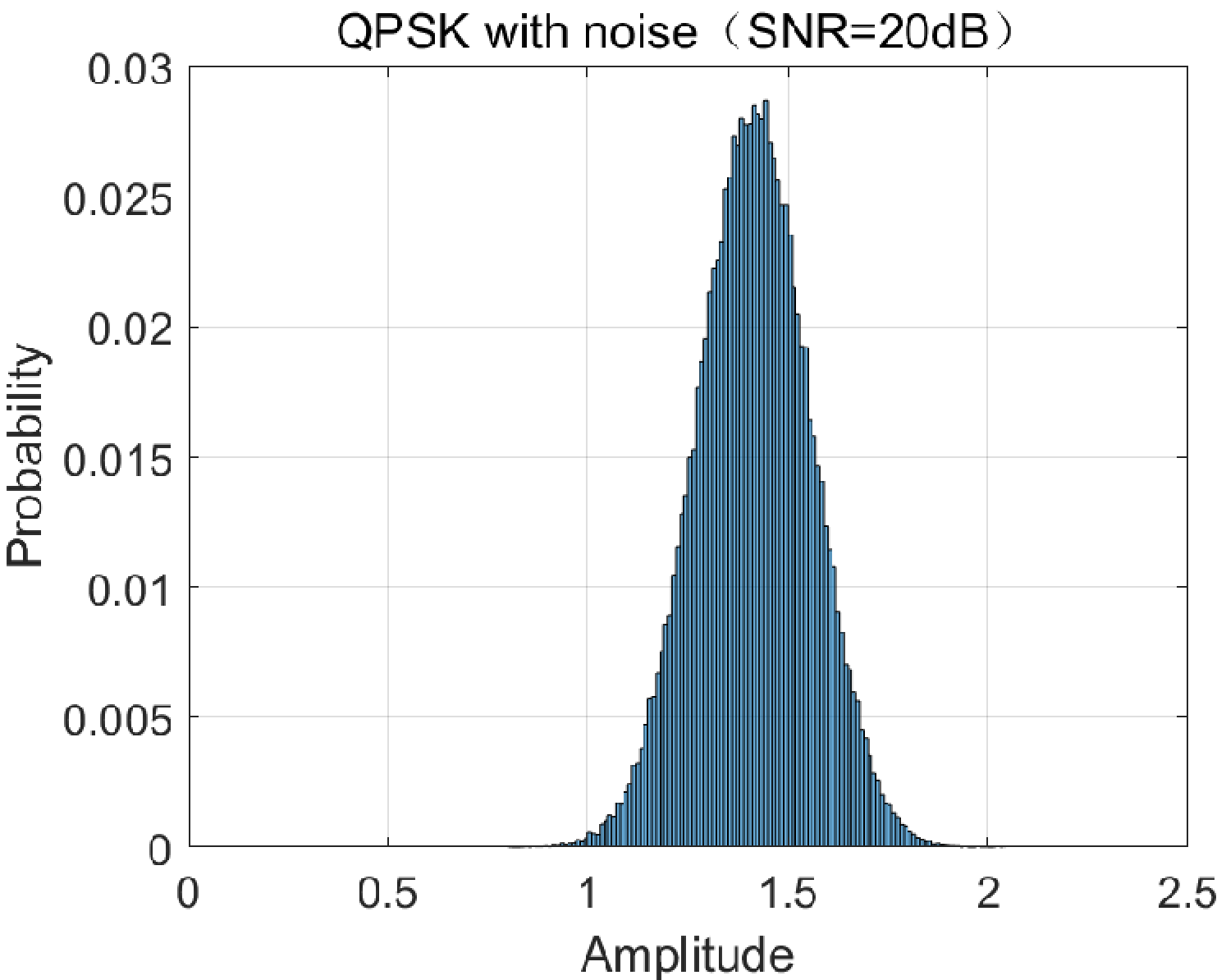}}
	\label{fig2c}
	\subfloat[]{\includegraphics[width=0.2\textwidth]{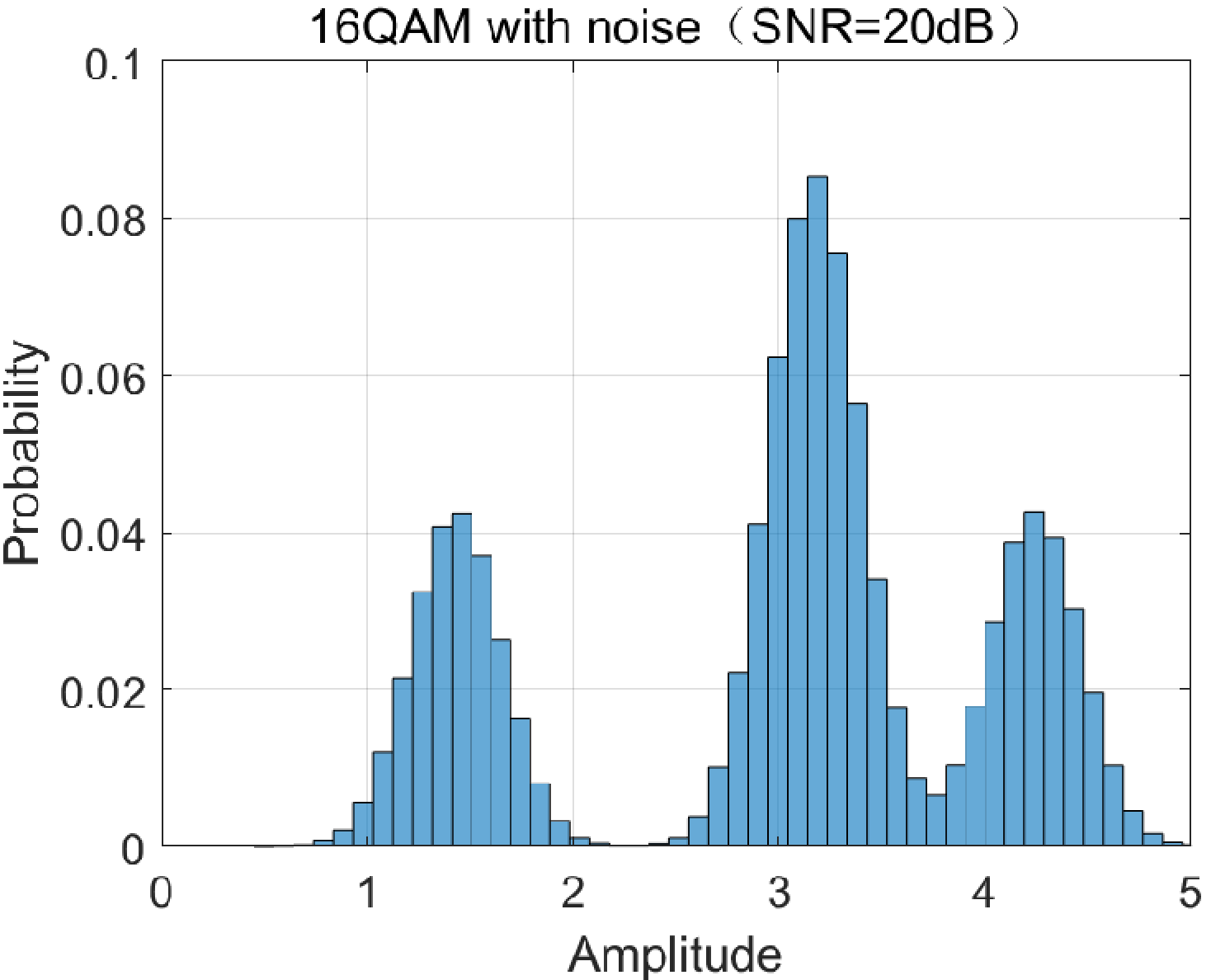}}
	\label{fig2d}
	\caption{Amplitude histogram for 2ASK, 2FSK, QPSK and 16QAM.}
	\label{fig02}
\end{figure}
Amplitude histogram is an early statistical feature based on signal envelope, which is shown in Fig.~\ref{fig02}. The basic idea is to distinguish different modulation methods by statistically analyzing the probability distribution characteristics of signal amplitude. This feature is not suitable for distinguishing modulation modes with insignificant envelope changes such as 2FSK and QPSK.


\subsubsection{Sequence Features}
The I/Q sequence preserves the complete temporal structure of the received signal and therefore contains all information theoretically available for modulation recognition. However, it is highly sensitive to channel impairments, synchronization errors, carrier frequency offsets, and noise. Consequently, effective feature extraction often relies on deep learning models capable of learning robust representations directly from raw observations.

In contrast, high-order cumulants \cite{Swami2000HOC,Han2004HOCSVM,Wang2009HOCSVM,Aslam2012HOCKNN,Han2012HOC} are widely used in AMR because Gaussian noise has zero cumulants above second order. Consequently, the cumulants of a received signal are theoretically determined only by the modulated signal itself under the assumptions of additive Gaussian noise, statistical independence between signal and noise, and sufficiently large observation length. However, finite-sample estimation errors, carrier frequency offsets, phase offsets, synchronization errors, and multipath fading may distort the cumulant estimates and reduce their discriminative capability.

\begin{figure*}[!h]
	\centering
	\subfloat[]{\includegraphics[width=0.2\textwidth]{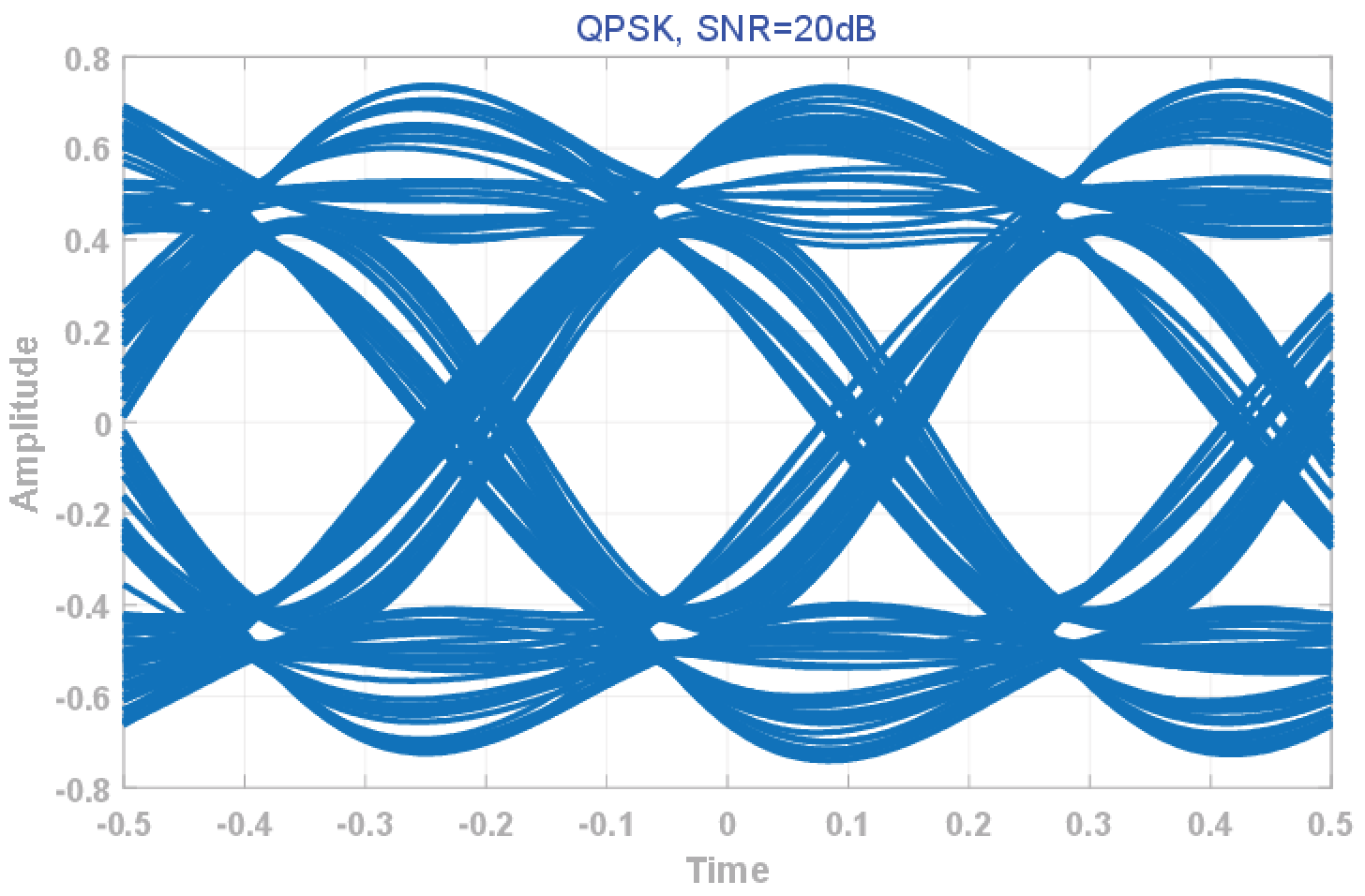}}
	\label{fig3a}
	\subfloat[]{\includegraphics[width=0.2\textwidth]{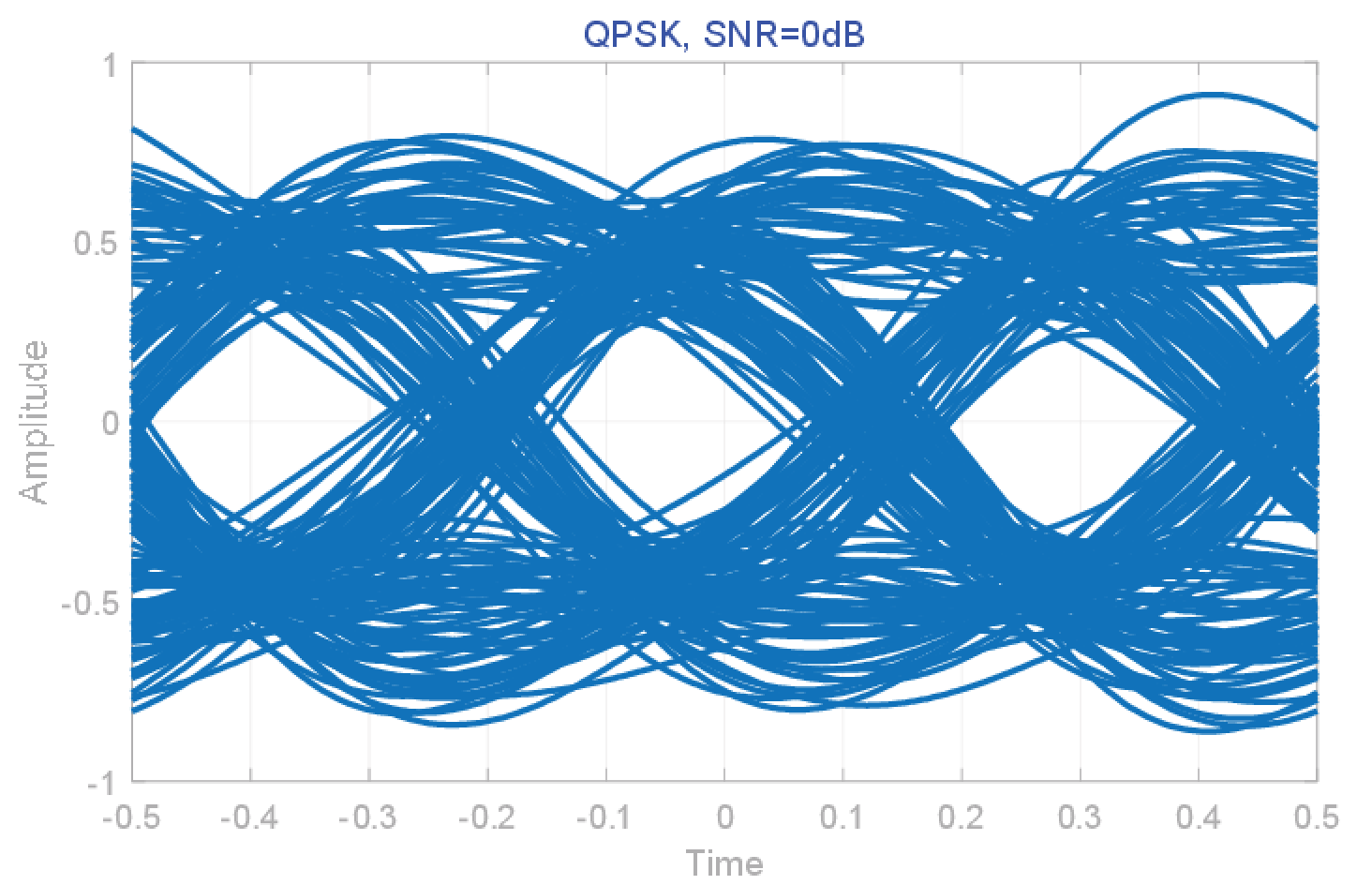}}
	\label{fig3b}
	\subfloat[]{\includegraphics[width=0.2\textwidth]{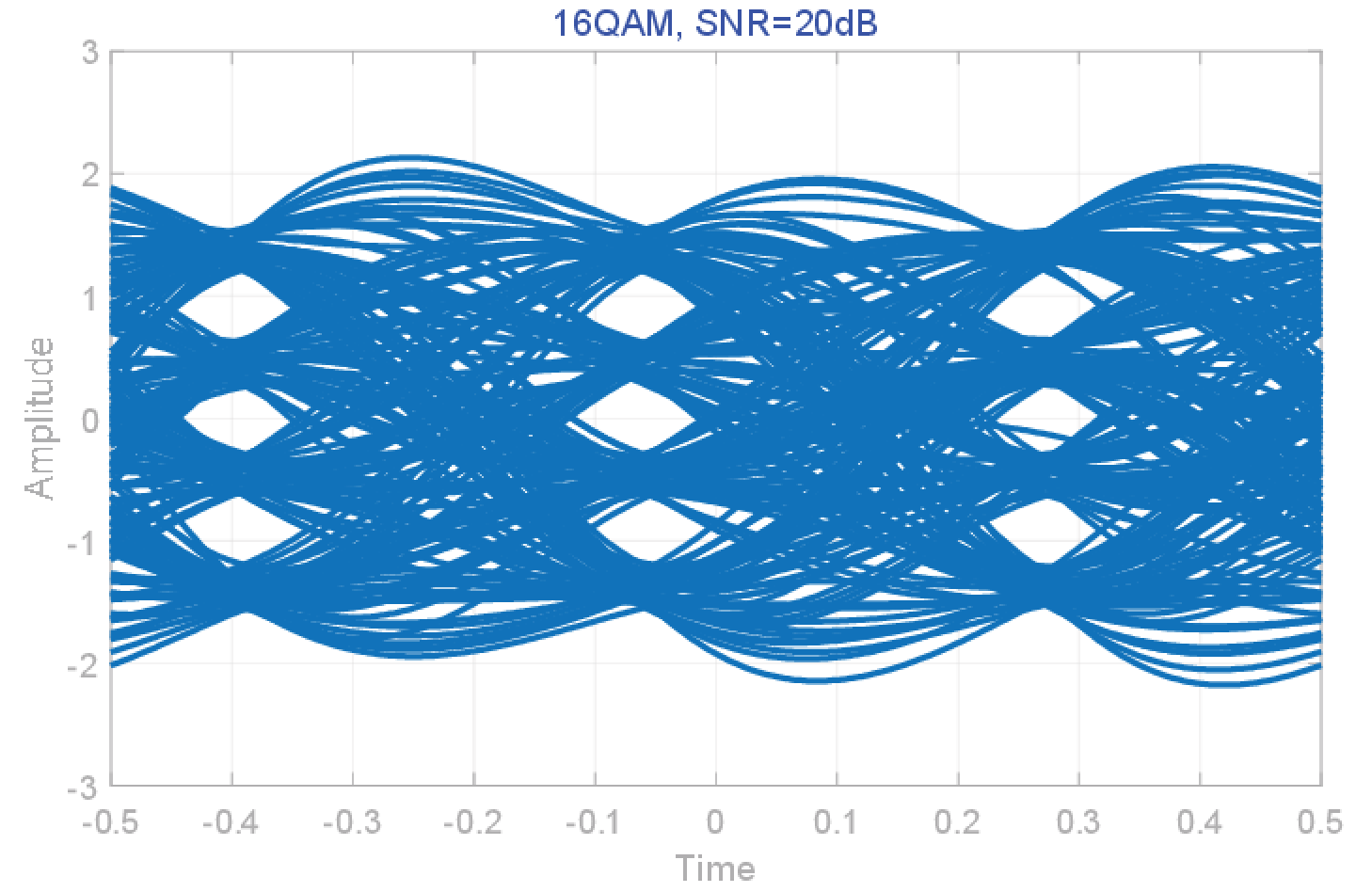}}
	\label{fig3c}
	\subfloat[]{\includegraphics[width=0.2\textwidth]{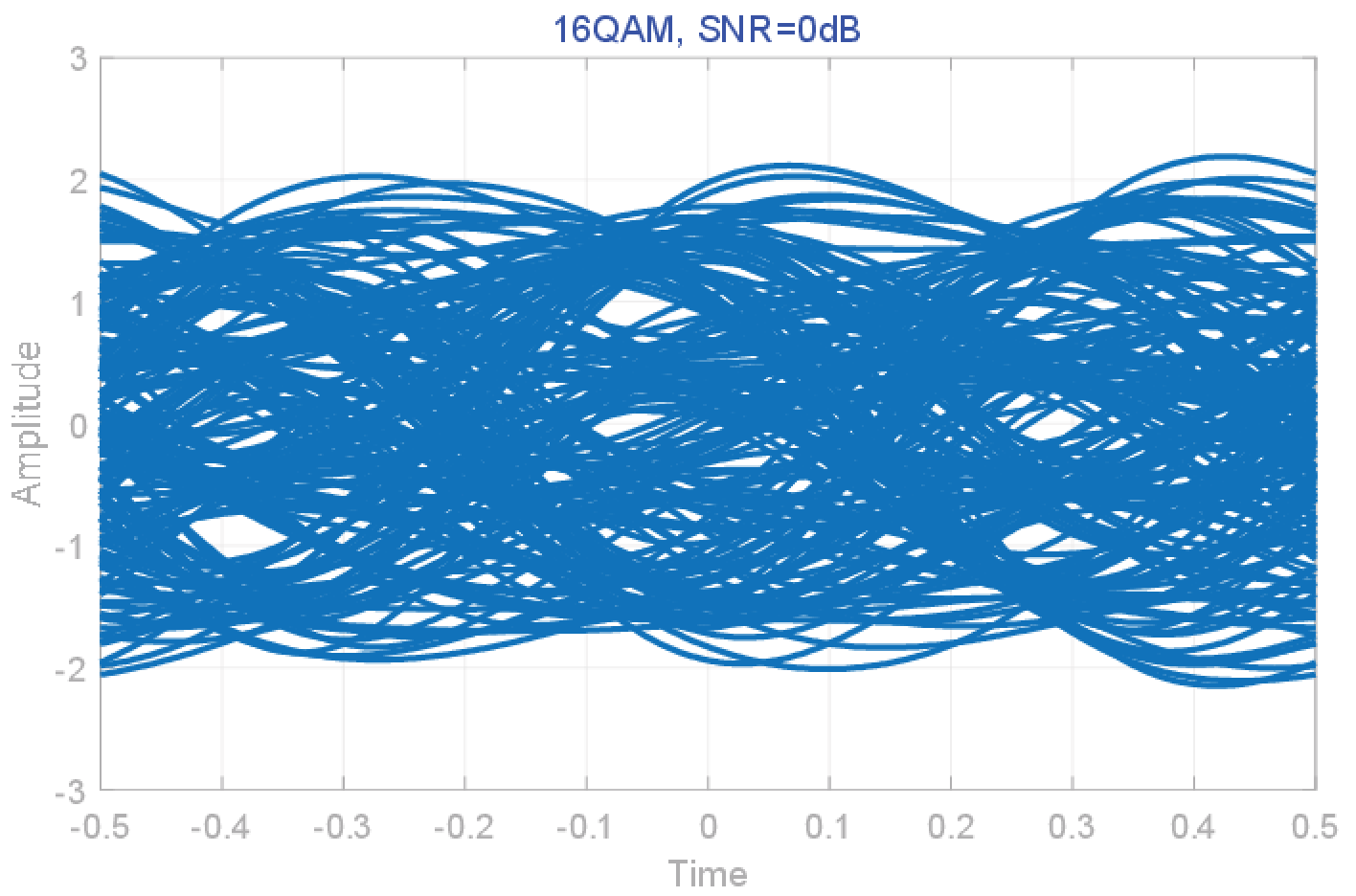}}
	\label{fig3d}
	\caption{Eye diagram of QPSK and 16QAM with different SNR.}
	\label{fig03}
\end{figure*}
\begin{figure*}[!h]
	\centering
	\subfloat[]{\includegraphics[width=0.2\textwidth]{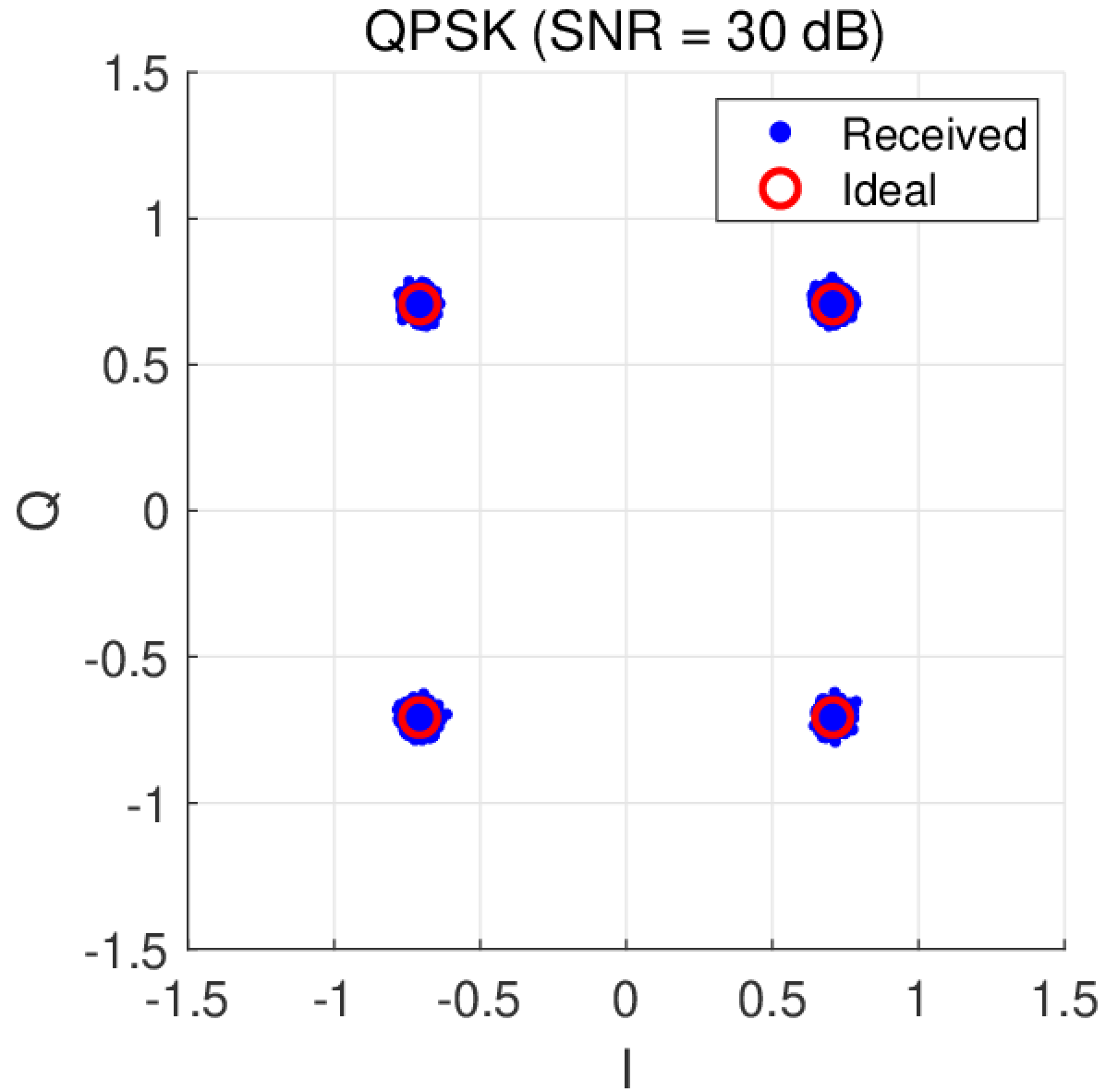}}
	\label{fig4a}
	\subfloat[]{\includegraphics[width=0.2\textwidth]{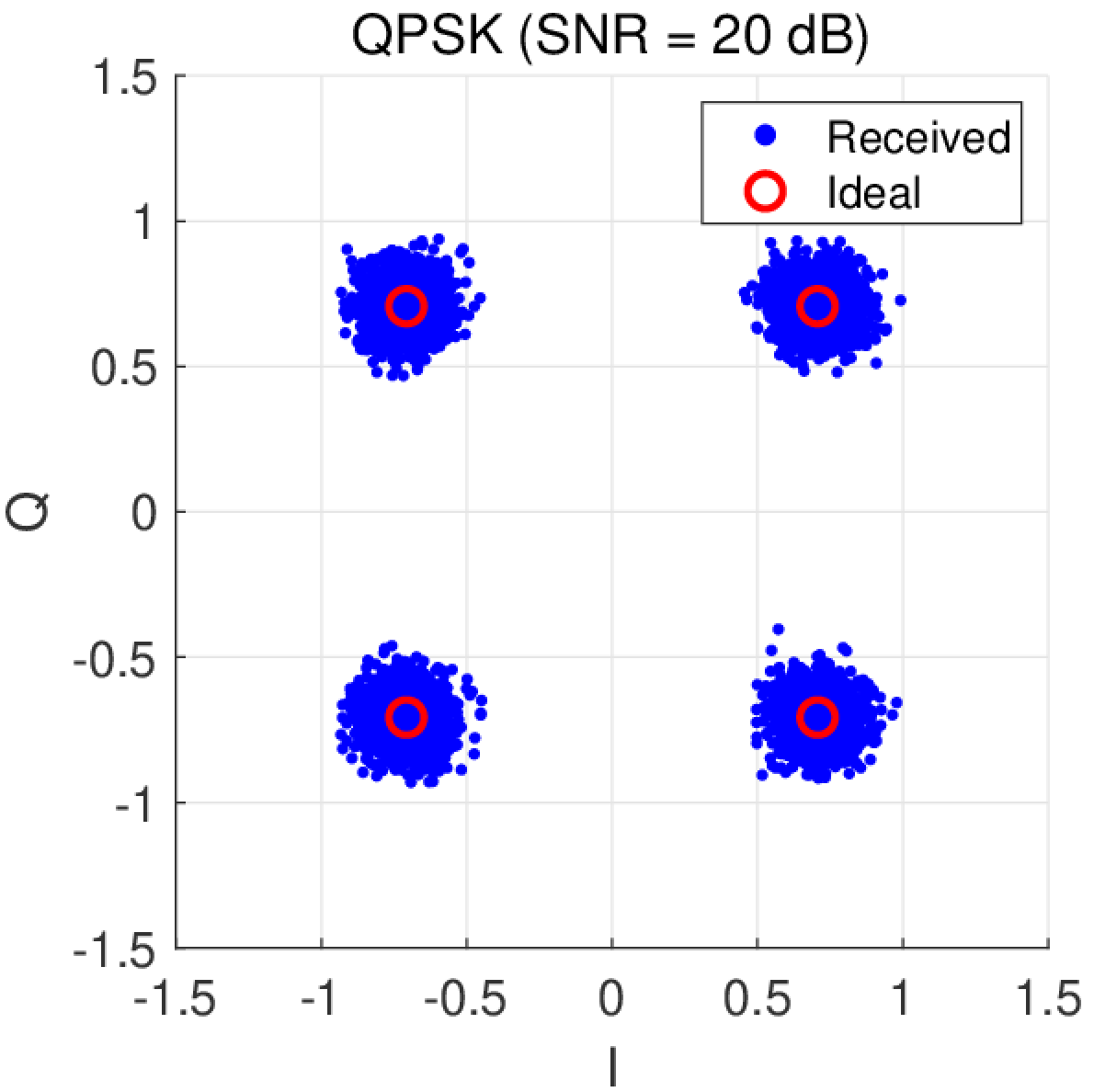}}
	\label{fig4b}
	\subfloat[]{\includegraphics[width=0.2\textwidth]{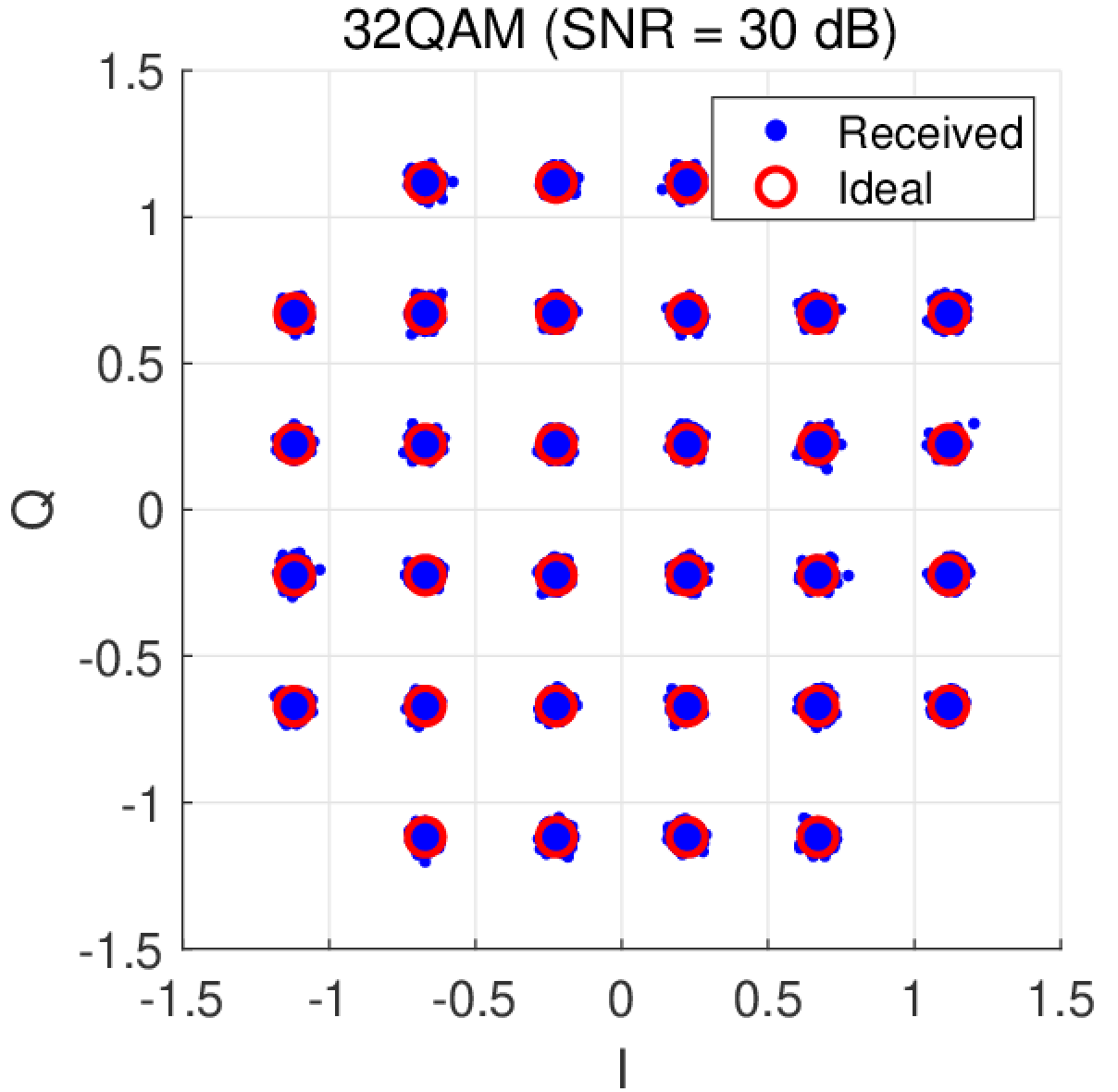}}
	\label{fig4c}
	\subfloat[]{\includegraphics[width=0.2\textwidth]{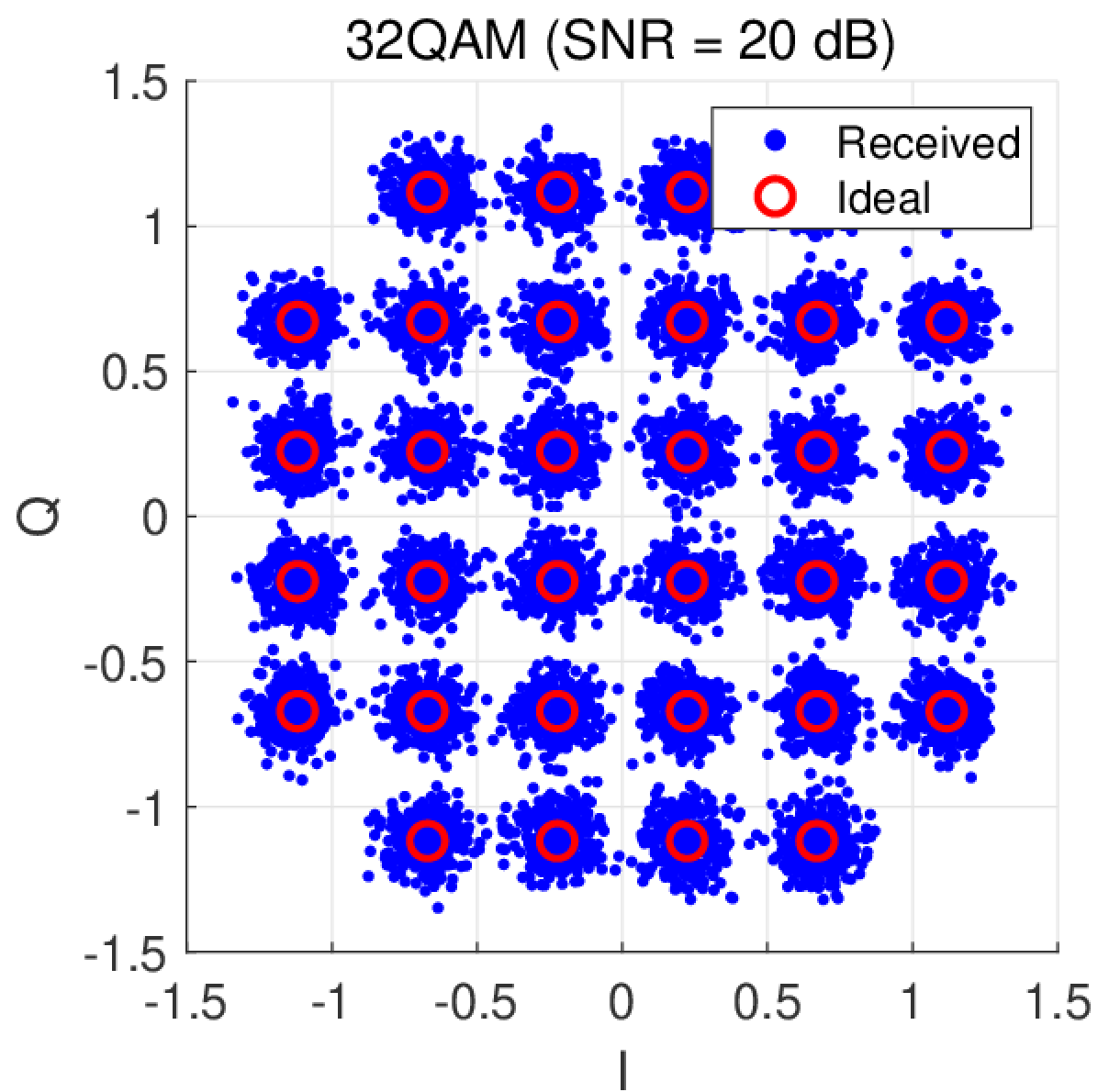}}
	\label{fig4d}
	\caption{Constellation diagram of QPSK and 32QAM with different SNR, (a) QPSK 30 dB SNR, (b) QPSK 20 dB SNR, (c) 32QAM 30 dB SNR,(d) 32QAM 20 dB SNR.}
	\label{fig04}
\end{figure*}

\subsubsection{Image Features}
Eye diagram, as a classic tool for evaluating signal integrity in digital communication systems, is formed by synchronously superimposing the baseband waveform of the received signal within the symbol period as shown in Fig.~\ref{fig03}. 
In modulation recognition applications, Eye diagrams represent symbol transition characteristics and are particularly useful for evaluating inter-symbol interference and synchronization quality. Their discriminative capability relies on accurate symbol timing recovery. Timing offsets, multipath propagation, and severe noise can reduce eye opening and blur the underlying modulation structure, thereby degrading recognition performance.

Constellation diagrams characterize the geometric distribution of modulation symbols in the complex plane. They provide strong discriminative capability for PSK and QAM signals under accurate synchronization and high SNR conditions. However, phase rotation causes global constellation rotation, carrier frequency offsets introduce time-varying phase drift, and nonlinear amplifier distortion may deform the constellation geometry. In radar and high-mobility scenarios, Doppler shifts can further distort constellation trajectories, significantly degrading recognition performance as shown in Fig.~\ref{fig04}.

\begin{figure}[!h]
	\centering
	\subfloat[]{\includegraphics[width=0.8\linewidth]{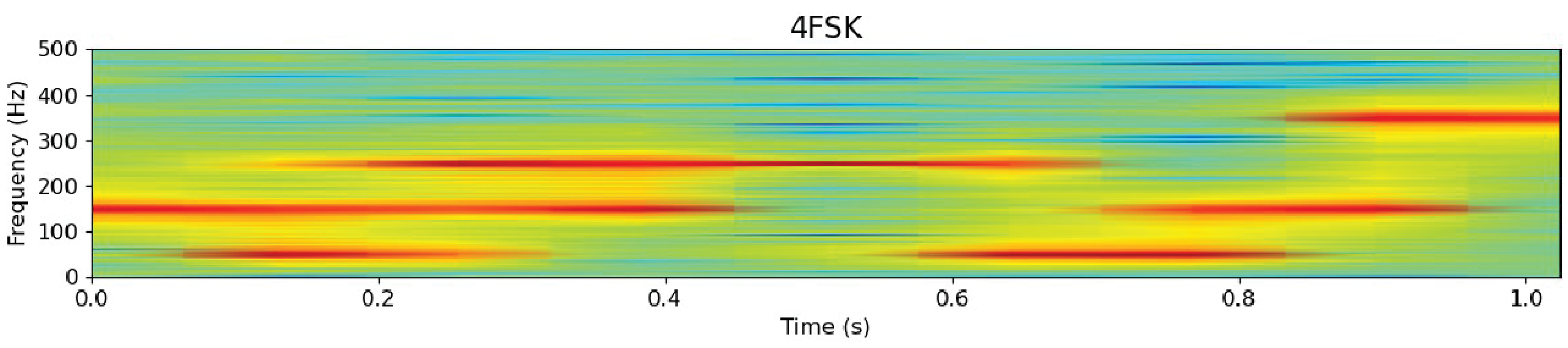}}
	\label{fig5a}
	\subfloat[]{\includegraphics[width=0.8\linewidth]{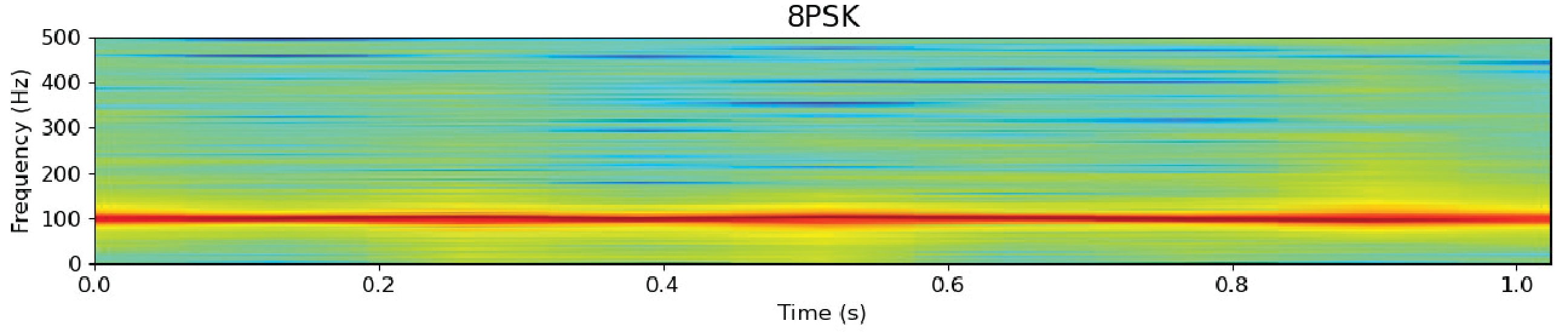}}
	\label{fig5b}
	\subfloat[]{\includegraphics[width=0.8\linewidth]{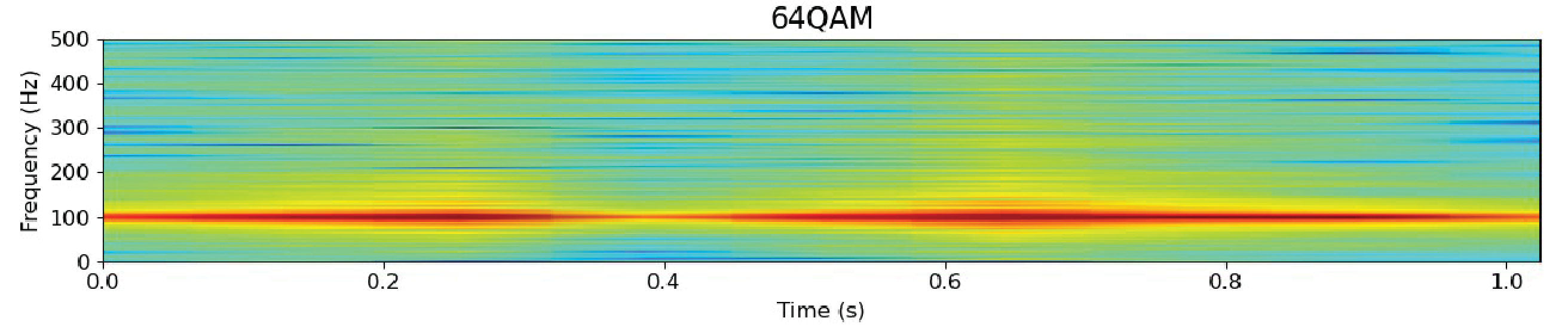}}
	\label{fig5c}
	\caption{Time-frequency feature of 4FSK, 8PSK and 64QAM.}
	\label{fig05}
\end{figure}

Time-frequency representations describe the joint distribution of signal energy in both time and frequency domains and are particularly effective for non-stationary radar waveforms. Their performance depends strongly on the selected transform, window length, overlap ratio, and resolution parameters. According to the time-frequency uncertainty principle, increasing time resolution generally reduces frequency resolution and vice versa. Therefore, the effectiveness of time-frequency features is closely related to the time-bandwidth structure of the target waveform.
\subsection{Comparison Between Communication and Radar Signals}
The differences in task objectives, signal models, prior information, and evaluation metrics lead to distinct AMR algorithm designs for communication and radar. Communication signals are typically stationary or cyclostationary with standardized modulation formats, such as PSK and QAM, and sufficient prior information, making recognition accuracy the primary objective. In contrast, radar signals are generally non-stationary, transient, and target-dependent, with limited or uncertain prior information. Accordingly, radar AMR additionally emphasizes detection probability, false alarm probability, and parameter estimation accuracy. These differences also influence algorithm selection: communication AMR mainly benefits from deep feature learning and statistical classification, whereas radar AMR often requires time-frequency analysis, sparse representation, or few-shot learning to address target-dependent and weak-prior scenarios.

\subsection{Basic AI Technologies for AMR}
\begin{figure}[!h]
	\centering
	\includegraphics[width=0.7\linewidth]{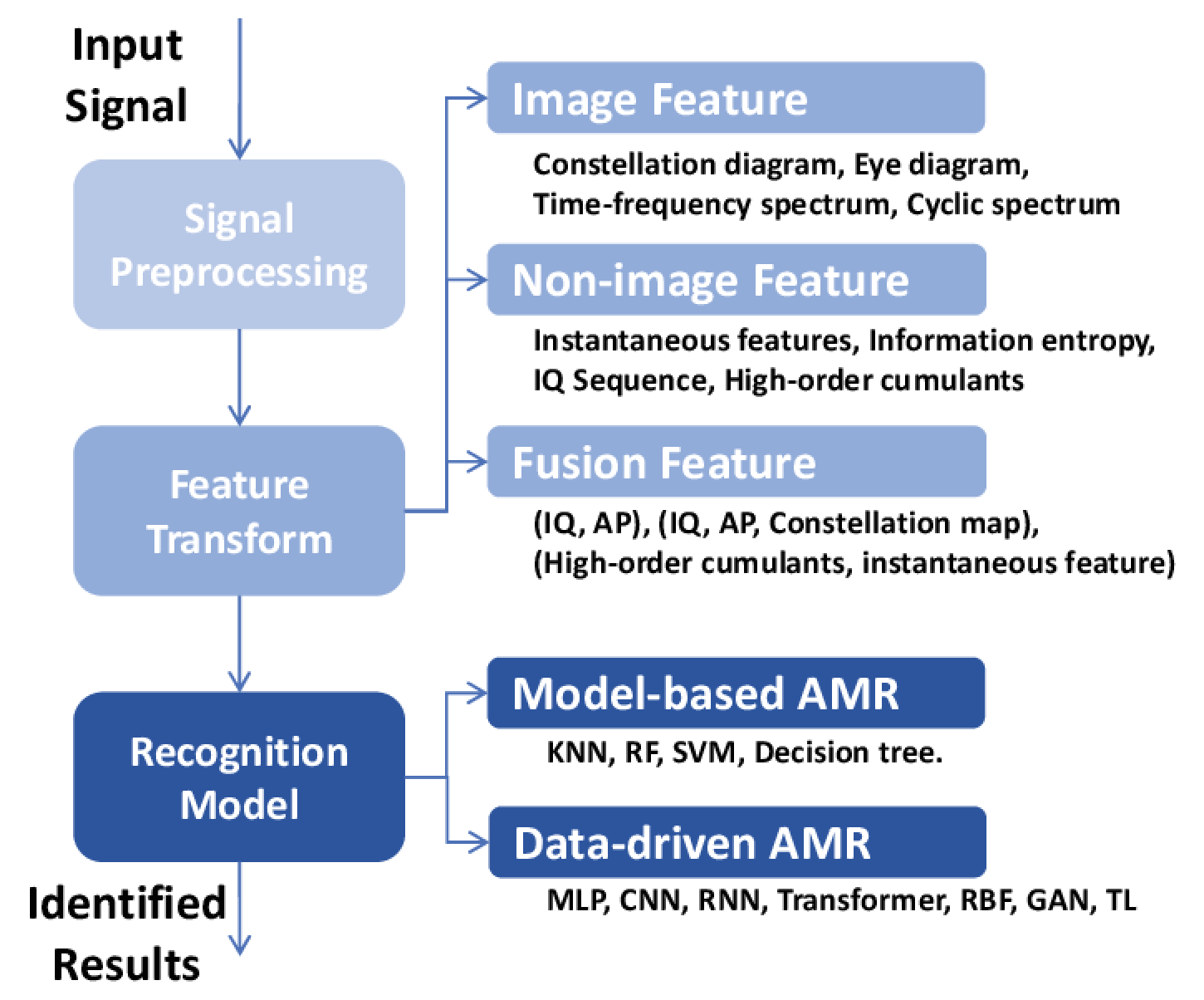}
	\caption{AMR structure based on AI technology.}
	\label{fig06}
\end{figure}

Communication and radar AMR have different feature extraction requirements. Communication AMR generally employs lightweight ML or DL models, such as SVM \cite{Dong2020CF}, RF \cite{Zhang2017RF}, CNN \cite{Gao2019CNN}, and RNN \cite{Ghasemzadeh2022GSQRNN}. By contrast, radar AMR often requires more powerful representation learning, including deep CNNs \cite{Jiang2024Radar}, Transformers \cite{Ren2023Radar}, and Mamba \cite{Hou2025Radar}, to capture target-dependent time-frequency characteristics. Representative model-based and data-driven AMR methods are reviewed below, and the overall AI-based framework is shown in Fig.~\ref{fig06}.

From a representation learning perspective, different AI models can be interpreted as imposing different inductive biases on the AMR problem. An inductive bias determines which structures of the received signal are assumed to be informative and therefore guides feature extraction and decision making.

\subsubsection{Model-based Methods}
Traditional model-based ML methods typically rely on manually designed and extracted signal features, such as high-order cumulants, cyclic spectral features, amplitude histograms, instantaneous frequency features. 

SVM is one of the most widely used models for AMR. SVM achieves signal classification by constructing the optimal hyperplane, and uses kernel functions to map nonlinear features to high-dimensional space, effectively distinguishing different modulation methods. Model-based classifiers such as SVM rely on the assumption that modulation classes can be separated in an appropriately designed feature space. Their performance therefore depends strongly on the discriminative power of hand-crafted features and on the validity of the underlying channel assumptions.

KNN is a non-parametric distance-based classifier that assigns a signal to the class of its nearest neighbors. It is simple and effective for low-dimensional features but is sensitive to noise and incurs high computational complexity for large datasets.

Model-based AMR methods are theoretically mature, interpretable, and computationally efficient. However, their performance strongly depends on handcrafted features and accurate channel models, and degrades significantly under multipath fading, frequency offsets, and low-SNR conditions.

\subsubsection{Data-driven Methods}
In recent years, data-driven DL methods have demonstrated strong feature learning and generalization capabilities in the field of AMR. 

CNNs are based on the assumption that informative patterns are locally correlated in the input space. Through local receptive fields and weight sharing, CNNs efficiently capture spatial structures such as constellation geometry, eye-diagram morphology, and localized time-frequency patterns. Consequently, CNNs are particularly suitable when modulation-specific information is encoded in local neighborhoods of an image-like representation.

RNN-based models assume that the underlying signal contains meaningful temporal dependencies. This inductive bias is well matched to raw I/Q sequences because modulation information is often embedded in symbol transitions, phase evolution, and temporal correlations. Therefore, RNN and its improved structures, such as LSTM and GRU, can effectively model the dynamic relationship of signals over time and are suitable for identifying modulation signals with temporal correlation, such as FSK, MSK, or underwater communication signals. RNN can capture temporal dependencies of phase and amplitude, but there are issues of gradient vanishing and high computational complexity during the training process.

Transformers assume that relevant information may be distributed across the entire observation sequence rather than being confined to local neighborhoods. By means of self-attention, they can model long-range interactions among symbols and waveform segments. This property is especially beneficial for long I/Q sequences and complex radar waveforms. However, the weaker structural prior of Transformers generally requires larger training datasets and higher computational resources.

As the complexity of DL models increases, a large amount of data is required for training, which is difficult to meet in some application scenarios. Few-shot AMR often suffers from overfitting due to limited labeled data. To address this issue, GANs and TL are widely adopted. GANs alleviate data scarcity by learning the underlying data distribution and generating synthetic samples, whereas TL improves recognition by transferring knowledge learned from related domains.

Beyond CNNs, RNNs, Transformers, and Mamba architectures, recent studies have begun exploring foundation-model-based wireless intelligence frameworks. Unlike task-specific architectures, foundation models aim to learn transferable signal representations from large-scale heterogeneous datasets and may become an important direction for future AMR systems.

\section{AI Empowered AMR for Communication}
This section surveys AI-based methods for communication AMR, including model-based methods, data-driven methods, and few-shot learning approaches. These methods are categorized by recognition model and further grouped according to the adopted feature types.
\subsection{Model-based AMR for Communication}
Model-based AMR mainly includes SVM, decision tree, and clustering methods, using features such as instantaneous, high-order cumulant, time-frequency, and entropy features.
\subsubsection{AMR for Communication based on Decision Tree}
\paragraph{Instantaneous Features}
Pattern recognition for AMR was first introduced by Weaver et al. \cite{Weaver1969PR}, while the model-based ML framework was established by Liedtke in 1984 \cite{LIEDTKE1984AMR}. Liedtke extracted statistical features, such as amplitude, frequency, and phase histograms, and combined them with a nearest-neighbor classifier for digital modulation recognition. Building on this work, Nandi and Azzouz further investigated instantaneous features (e.g., amplitude, frequency, and phase) and employed decision trees for analog \cite{NANDI1995Ana} and digital \cite{AZZOUZ1995Digi} modulation recognition. These handcrafted features have become the foundation of model-based AMR despite their limited performance under low-SNR conditions.

\paragraph{High-order cumulants}
Feature extraction is critical to model-based AMR, and high-order cumulants (HOCs) are among the most widely used handcrafted features because Gaussian noise has zero cumulants above the second order \cite{Swami2000HOC,Han2004HOCSVM,Wang2009HOCSVM,Aslam2012HOCKNN,Han2012HOC}. Swami et al. \cite{Swami2000HOC} and Han et al. \cite{Han2012HOC} employed fourth-order cumulants with decision trees for modulation recognition, achieving high accuracy for common modulation schemes. However, HOCs provide limited discrimination for high-order constellations, such as 16QAM and 64QAM.

\paragraph{Entropy Features}
Classical modulation features characterize the statistical distribution and geometric structure of signal amplitude, phase, and frequency, while entropy features characterize the disorder, complexity, and information uncertainty of signal timing and spectral domain. Entropy is a nonlinear upgrade and supplement of classical features at the information theory level. The entropy feature can effectively remove the random interference of noise and preserve the structural differences of the modulated signal itself by quantifying the distribution uncertainty of the signal. Zhang et al. \cite{Zhang2017RF} adopted a combination of power spectrum Shannon entropy, wavelet energy spectrum entropy, and Rényi entropy to establish a low complexity RF model, effectively removing Gaussian channel noise interference and achieving recognition accuracy of over 90\% even under SNR conditions below 0 dB. However, only 7 modulation modes were considered in this paper, thus the universality of the model was not validated.
 
\subsubsection{AMR for Communication based on SVM}
SVM is a very popular modulation signal recognition model that mainly utilizes high-order statistical features and time-frequency spectrum features.
\paragraph{High-order cumulants}
Utilizing the high-dimensional spatial feature mapping capability of SVM, Han et al. \cite{Han2004HOCSVM} added sixth-order cumulants to enhance the resolution of QAM signals, but it fails to distinguish 2ASK and 2PSK. To tackle this problem, Wang et al. \cite{Wang2009HOCSVM} further adopted sixth-order cumulants and combinations of high and low order cumulants to distinguish 2ASK signals from the others. However, both \cite{Han2004HOCSVM} and \cite{Wang2009HOCSVM} require manual design of high-order cumulative features, which limits their generality.

\paragraph{Time-frequency Spectrum}
Modulation signals often contain transient components and rapidly changing fluctuations. Wavelet transform is widely used in model-based AMR because of its ability to characterize transient and non-stationary signals in both the time and frequency domains \cite{Ho1995WT,Wu2005SVM,Inan2006WT,AVCI2008WT,Hassan2009WT}. Representative studies combined wavelet features with SVM or other classifiers to improve modulation recognition over a wide SNR range \cite{Wu2005SVM,Inan2006WT,AVCI2008WT}. However, its performance strongly depends on the selection of wavelet bases and scale parameters, which generally requires prior knowledge of signal characteristics.

Cyclic spectrum is widely used in AMR because it jointly characterizes the spectral and periodic statistical properties of cyclostationary signals. Dong et al. \cite{Dong2020CF} combined cyclostationary features with an SVM classifier for non-cooperative modulation recognition. However, reliable cyclic spectrum estimation requires a large number of samples, limiting its performance in few-shot and low-SNR scenarios.

\subsubsection{AMR for Communication based on Clustering}
\paragraph{High-order cumulants}
To address the problem of manually designing features in existing methods, Aslam et al. \cite{Aslam2012HOCKNN} combined genetic programming algorithm (GP) with KNN for AMR, in which GP automatically generates new features composed of high-order cumulants through evolutionary progresses, without relying on prior knowledge to design features. However, the article only verified the applicability of four digital modulations (BPSK, QPSK, 16QAM, 64QAM) and did not involve analog modulation or other common digital modulations. 

Most existing model-based methods are supervised and require sufficient labeled samples. To address unlabeled scenarios, Spooner et al. \cite{Spooner2017HOCC} proposed an unsupervised AMR method based on higher-order cyclic cumulants. However, its clustering performance degrades under low-SNR conditions and pulse distortions.

\paragraph{Entropy Features}
Unlike early clustering methods based on high-order statistics, Huang et al. \cite{Huang2022OAE} combined optimized autoencoder (OAE) and enhanced KNN (EKNN) evaluation for modulation recognition in underwater acoustic communication systems. The autoencoder was used to extract features and reduce dimensionality of the data, and the EKNN was used for classification. Although the accuracy was over 99\%, but they only covered several low-order modulation signals.

\begin{table*}[!h]
	\caption{Model-based AMR for communication. \label{tab04}}
	\renewcommand{\arraystretch}{1}
	\centering
	\resizebox{1.0\linewidth}{!}{
	\begin{tabular}{|c|c|c|c|c|c|}
		\hline
		Core Input Features & Ref. & Year & Model & Modulation Type & Recognition Accuracy \\
		\hline
		\multirow{3}{*}{Instantaneous Features} & \cite{LIEDTKE1984AMR} & 1984 & Decision Tree & AM, 2ASK, 2FSK, 2PSK & $\geq 90\%$ , 18 dB, AWGN \\
		\cline{2-6}
		& \cite{NANDI1995Ana} & 1995 & Decision Tree & AM, DSB, VSB, LSB, USB, FM & $\geq 90\%$ , 10 dB, AWGN \\
		\cline{2-6}
		& \cite{AZZOUZ1995Digi} & 1995 & Decision Tree & 2ASK, 4ASK, 2FSK, 4FSK, 2PSK, 4PSK & $\geq 89\%$ , 10 dB, AWGN \\
		\hline
		\multirow{7}{*}{High-order cumulants} & \cite{Swami2000HOC} & 2000 & Decision Tree & BPSK, 4PAM, 16QAM, 8PSK & $\geq 98\%$ , 5 dB, AWGN \\
		\cline{2-6}
		& \cite{Han2004HOCSVM} & 2004 & SVM & 2ASK/2PSK, 4ASK, 8PSK, 4PSK, 16QAM & $\geq 98\%$ , 4 dB, AWGN \\
		\cline{2-6}
		& \cite{Wang2009HOCSVM} & 2009 & SVM & 2ASK, 4ASK, QPSK, 2FSK, 4FSK & $\geq 97\%$ , 10 dB, AWGN \\
		\cline{2-6}
		& \cite{Han2012HOC} & 2012 & Decision Tree & \makecell{BPSK, QPSK, OQPSK, $\pi$/4DQPSK, 8PSK, \\ 16QAM, 16APK, 64QAM}  & $\geq 90\%$ , 10 dB, AWGN \\
		\cline{2-6}
		& \cite{Aslam2012HOCKNN} & 2012 & GP-KNN & BPSK, QPSK, 16QAM, 64QAM & $\geq 97\%$ , 5 dB, AWGN \\
		\cline{2-6}
		& \cite{Spooner2017HOCC} & 2017 & Unsupervised clustering & QPSK, 8PSK, 32PSK, 32QAM, 32APSK, 8QAM & $\geq 91\%$ , 20 dB, AWGN \\
		\hline
		\multirow{2}{*}{Time-frequency Spectrum} & \cite{Wu2005SVM} & 2005 & Wavelet SVM & \makecell{2ASK, 4ASK, 2FSK, 4FSK, 2PSK, 4PSK,\\ 16QAM, $\pi$/4QPSK, OQPSK} & $\geq 96.5\%$ , 3 dB, AWGN \\
		\cline{2-6}
		& \cite{Dong2020CF} & 2020 & SVM & FSK, ASK, MSK, BPSK, QPSK & $\geq 93\%$ , 7 dB, AWGN \\
		\hline
		\multirow{2}{*}{Entropy Features} & \cite{Zhang2017RF} & 2017 & Random Forest & 2FSK, 4FSK, 8FSK, BPSK, QPSK, 16QAM, MSK & $\geq 95\%$ , 6 dB, AWGN \\
		\cline{2-6}
		& \cite{Huang2022OAE} & 2022 & OAE-EEKNN & 2/4FSK, 2/4PSK, 16/64QAM, OFDM, DSSS & $\geq 99\%$ , real  data \\
		\hline
	\end{tabular}}
\end{table*}

Table~\ref{tab04} summarizes representative model-based AMR methods. Despite their effectiveness, several challenges remain. Most studies assume Gaussian noise channels, and their robustness under multipath propagation and low-SNR conditions remains limited. In addition, existing methods mainly focus on single-carrier signals and rely on a single handcrafted feature, restricting their applicability to emerging multicarrier systems. Therefore, multi-feature fusion for low-SNR and multicarrier AMR deserves further investigation.

\subsection{Data-driven AMR for Communication}
Data-driven DL models have the advantages in extracting complex nonlinear features from input data, reducing the difficulty of designing features, which provides new solutions for AMR. In this section, we will make a comprehensive study on data-driven AMR methods. 

\subsubsection{Application of MLP for AMR}
MLP was the earliest data-driven DL model applied to AMR \cite{Hinton1986BP}. Early studies employed MLP with handcrafted features, such as cyclostationary and high-order cumulant features, demonstrating improved nonlinear classification capability over traditional classifiers \cite{Lu1996MLP,Xie2019MLP}. Subsequent works extended MLP-based AMR to multipath fading and low-SNR scenarios \cite{Eric2008MLP} and further enhanced high-order MQAM recognition using heuristic optimization \cite{Zhang2021BSAMLP}. However, MLPs have limited representation learning capability and generally rely on handcrafted features, making them less effective for high-order or multicarrier modulation recognition while increasing computational complexity when combined with feature engineering or optimization algorithms.

\subsubsection{Application of CNN for AMR}
\begin{figure}[!h]
	\centering
	\includegraphics[width=0.7\linewidth]{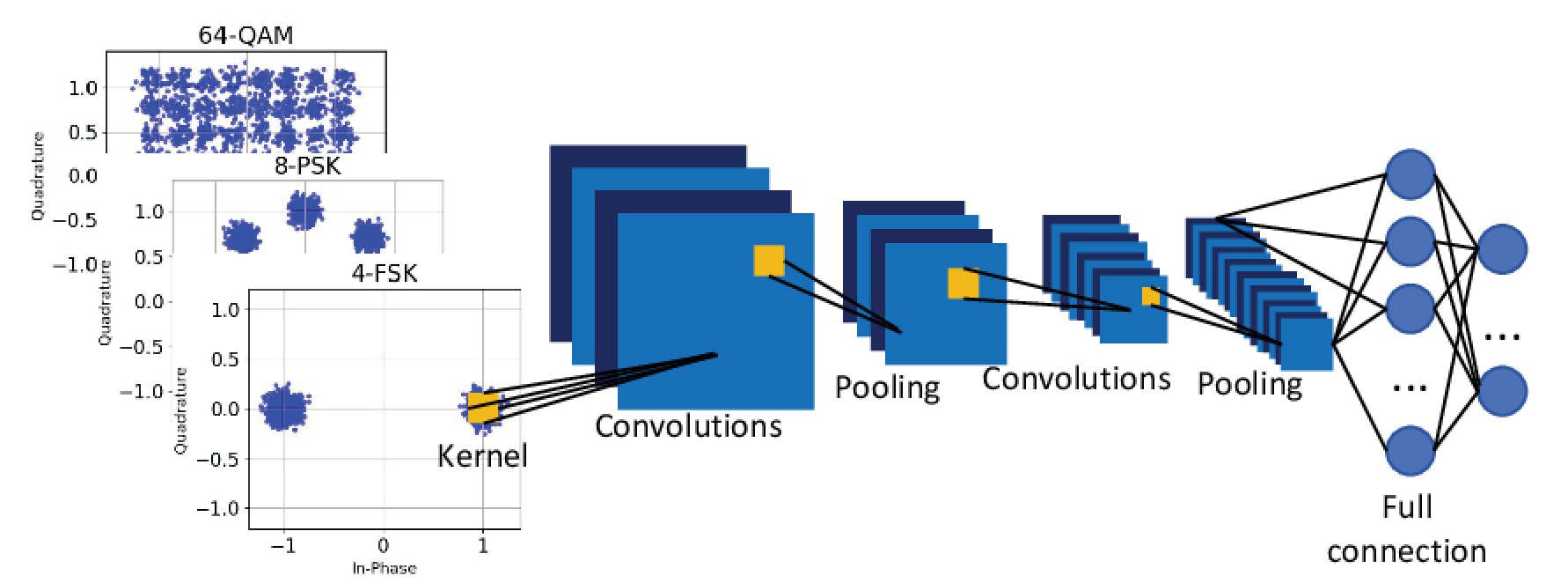}
	\caption{A typical CNN-based AMR model.}
	\label{fig07}
\end{figure}

CNN \cite{Lecun1998CNN}, with its efficient ability to extract spatial features, has shown outstanding performance in using image features such as constellation maps and time-frequency maps for AMR, and has thus become the mainstream basic AMR model for a long time. Typical CNN-based AMR model is shown in Fig.~\ref{fig07}.

\paragraph{I/Q Sequence}
O'Shea et al. \cite{Shea2016CNN} first introduced CNNs into AMR, demonstrating their effectiveness in learning discriminative representations directly from complex-valued time-domain signals. Subsequent studies enhanced CNNs with attention mechanisms to improve high-order modulation recognition \cite{Kong2021CTDNN} and extended their application to multicarrier OFDM signals \cite{Hong2019OFDM}. Nevertheless, CNN-based methods still suffer from performance degradation under low-SNR conditions and confusion among similar modulation formats.

To extend CNN-based AMR to MIMO systems, Wang et al. \cite{Wang2020MIMO,Wang2020CNN} employed zero-forcing equalization followed by CNN-based classification. Subsequently, Thien et al. \cite{Thien20223DCNN} proposed a 3D CNN to jointly exploit sample, antenna, and I/Q correlations for MIMO multicarrier AMR. However, both studies were evaluated on only a few modulation formats and lacked validation on high-order modulations.

Recent waveform reuse techniques, such as OTFS and affine frequency division multiplexing (AFDM), have introduced new challenges for AMR. Deep learning has been successfully applied to OTFS modulation recognition, demonstrating improved performance over conventional CNN- and RNN-based models \cite{Wang2024OTFS,Liu2025OTFS}. Meanwhile, deep neural network receivers have been developed for AFDM, providing a foundation for future AMR research \cite{HUANG2025AFDM,Yin2026AFDM}. However, existing studies remain limited to a few modulation formats and lack validation using real-world signals.

\paragraph{Constellation map}
Constellation diagrams \cite{Marcelo2022Com} provide intuitive amplitude and phase representations, making them suitable for CNN-based AMR. Wang et al. \cite{Wang2017CNN} first demonstrated high-accuracy modulation recognition using constellation images. Subsequently, Tian et al. \cite{Tian2019CNN} compared several CNN architectures and showed that residual networks achieved the best performance by facilitating deeper feature learning.

Building upon constellation images, Peng et al. \cite{Peng2019CNN} further enhanced input representations and demonstrated the effectiveness of CNN-based AMR using different network architectures. To improve deployment efficiency, Wang et al. \cite{Wang2024EL} proposed a lightweight CNN that significantly reduced model complexity while maintaining comparable recognition performance.

\paragraph{Eye diagram}
Eye diagrams provide intuitive visual representations of signal quality and symbol transitions, making them suitable for CNN-based AMR \cite{Marcelo2022Com}. Wang et al. \cite{Wang2017CNNEye} employed eye diagrams as CNN inputs for modulation recognition, demonstrating promising performance. However, the method was evaluated on only a limited number of modulation formats.

Zha et al. \cite{Zha2019SSDCNN} introduced ResNet \cite{He2016DRL} into AMR, dealing with the problem of vanishing gradients in deep networks. For low order modulation methods, a recognition rate of over 98\% was achieved with 3 dB AWGN, but for high-order modulation methods such as 64QAM, the recognition accuracy deteriorated significantly. 

\paragraph{Time-frequency Spectrum}
Time-frequency representations have also been adopted as CNN inputs to jointly exploit temporal and spectral characteristics. Guan et al. \cite{Guan2018CNN} demonstrated their effectiveness for low-SNR modulation recognition, while Gao et al. \cite{Gao2019CNN} further proposed a lightweight CNN to reduce computational complexity. However, existing studies mainly focus on single-carrier signals and a limited range of modulation formats. Different from most works, Tian et al. \cite{Tian2019nGCNN} addressed AMR under non-Gaussian impulsive noise by combining the fractional low-order Choi--Williams distribution with a CNN for time-frequency feature learning. Although effective under impulsive noise, the method was validated on only a few binary modulation formats.

Liu et al. \cite{Liu2020CNN} proposed a modulation recognition scheme based on CNN and cyclic spectrum of signals to address the challenge of identifying mixed data of primary and secondary modulation signals in communication systems. The author compared the performance of four classic architectures (VGG16 \cite{Simonyan2015VGG}, VGG19 \cite{Simonyan2015VGG}, AlexNet \cite{Alex2017AlexNet}, and ResNet18 \cite{He2016DRL}), among which Resnet performed the best. However, the model was prone to confusion with secondary modulation signals due to the reason that external PM modulated secondary signals (BFSK-PM, BPSK-PM, QPSK-PM) shared PM features.

\paragraph{Hybrid Features}
To enrich feature representations, multimodal AMR methods have been proposed by fusing complementary signal features. Representative studies combined IQ signals with eye diagrams \cite{Li2019IQEyeCNN}, integrated handcrafted features with CNN-GRU \cite{Liu2022CNNGRU}, and fused sequence and constellation features \cite{An2022SC}, demonstrating improved robustness under low-SNR and channel-impaired conditions. However, existing methods still lack comprehensive validation across diverse modulation formats and practical communication scenarios. Khan et al. \cite{Khan20213DCNN} and Hasnaine et al. \cite{Hasnaine2025CNN} further investigated CNN-based modulation classification under noisy channel conditions and Khan et al. \cite{Khan20213DCNN} jointly exploited spatial and temporal signal representations. Their results demonstrated improved robustness against channel noise and highlighted the potential of deep CNN architectures for low-SNR AMR scenarios. The comparison of CNN-based AMR Methods for communication is given in Table~\ref{tab06}.
\begin{table*}[!h]
	\caption{CNN-based AMR Methods for communication. \label{tab06}}
	\renewcommand{\arraystretch}{1}
	\centering
	\resizebox{1.0\linewidth}{!}{
		\begin{tabular}{|c|c|c|c|c|c|}
			\hline
			Core Input Features & Ref. & Year & Model & Modulation Type & Recognition Accuracy \\
			\hline
			\multirow{10}{*}{I/Q Sequence} & \cite{Shea2016CNN} & 2016 & CNN & \makecell{8PSK, DSB, SSB, BPSK, GFSK, CPFSK,\\ PAM, 16QAM, 64QAM, QPSK, WBFM} & $\geq$90\%, 0 dB, AWGN\\
			\cline{2-6}
			& \cite{Hong2019OFDM} & 2019 & CNN & BPSK, QPSK, 8PSK, 16QAM, 64QAM & $\geq$90\%, 15 dB, AWGN\\
			\cline{2-6}
			& \cite{Wang2020MIMO} & 2020 & CNN & BPSK, QPSK, 8PSK, 16QAM & $\geq$92\%, 5 dB, MIMO-AWGN\\
			\cline{2-6}
			& \cite{Wang2020CNN} & 2020 & CNN & BPSK, QPSK, 8PSK, 16QAM & $\geq$90\%, 0 dB, MIMO-AWGN\\
			\cline{2-6}
			& \cite{Kong2021CTDNN} & 2021 & Attention CNN & \makecell{8PSK, AM-DSB, BPSK, CPFSK, \\GFSK, 4PAM, 16QAM, 64QAM, QPSK, WBFM} & $\geq$90\%, 0 dB, AWGN\\
			\cline{2-6}
			& \cite{Thien20223DCNN} & 2022 & 3DCNN & BPSK, QPSK, 8PSK, 16QAM & $\geq$90\%, 0 dB, AWGN\\
			\cline{2-6}
			& \cite{Chen2023OpenSet} & 2023 & ResNet18 & \makecell{OOK, 4ASK, 8ASK, BPSK, QPSK, \\ 8PSK, 16PSK, 32PSK, 16APSK, 32APSK} & $\geq$90\%, 8 dB, AWGN\\
			\cline{2-6}
			& \cite{Zhang2023SAtt} & 2023 & Attention CNN & \makecell{8PSK, AM-DSB, BPSK, CPFSK, \\GFSK, 4PAM, 16QAM, 64QAM, QPSK, WBFM} & $\geq$86\%, 0 dB, AWGN\\
			\cline{2-6}
			& \cite{Wang2024ETE} & 2024 & Gated-IQNet & \makecell{8PSK, AM-DSB, BPSK, CPFSK, \\GFSK, 4PAM, 16QAM, 64QAM, QPSK, WBFM} & $\geq$90\%, 6 dB, AWGN\\
			\cline{2-6}
			& \cite{Wang2024OTFS} & 2024 & \makecell{CNN, RESNET,\\ LSTM, LCDNN} & \makecell{OTFS, 4QAM, 16QAM,\\ 64QAM, 256QAM} & $\geq$80\%, 5 dB, AWGN\\
			\cline{2-6}
			& \cite{Liu2025OTFS} & 2025 & CNN & \makecell{OFTS, 8PSK, BPSK, QPSK, \\ 16QAM, 64QAM, 256QAM} & $\geq$90\%, 5 dB, EVA\\
			\cline{2-6}
			& \cite{Hasnaine2025CNN} & 2025 & Improved VT-CNN2 &  \makecell{AM-DSB, AM-SSB, BPSK, QPSK, QAM16,64,\\CPFSK, GFSK, PAM4, 8PSK, WBFM} & $\geq$60\%, -3 dB, AWGN \\
			\hline
			\multirow{4}{*}{Constellation map} & \cite{Wang2017CNN} & 2017 & CNN & QPSK, 8PSK, 8QAM, 16QAM, 32QAM, 64QAM & $\geq$95\%, 20 dB, AWGN\\
			\cline{2-6}
			& \cite{Tian2019CNN} & 2019 & ResNet50 & 2PSK, 4PSK, 8PSK, 16QAM, 32QAM, 64QAM & $\geq$95\%, 14 dB, AWGN\\
			\cline{2-6}
			& \cite{Peng2019CNN} & 2019 & \makecell{AlexNet,\\ GoogLeNet} & \makecell{BPSK, 4ASK, QPSK, OQPSK, \\ 8PSK, 16QAM, 32QAM, 64QAM} & $\geq$90\%, 4 dB, AWGN\\
			\cline{2-6}
			& \cite{Wang2024EL} & 2024 & Lightwight-CaffeNet & 2/4/8PSK, 2/4FSK, 16QAM & $\geq$81.8\%, 5 dB, AWGN\\
			\hline
			\multirow{2}{*}{Eye diagram} & \cite{Wang2017CNNEye} & 2017 & CNN & 4PAM, DPSK, RZ, NRZ & $\geq$95.2\%, 10 dB, AWGN \\
			\cline{2-6}
			& \cite{Zha2019SSDCNN} & 2019 & ResCNN & \makecell{BPSK, QPSK, OQPSK, 8PSK, \\ 16QAM, 16APSK, 32APSK, 64QAM} & $\geq$69\%, 3 dB, AWGN \\
			\hline
			\multirow{3}{*}{Time-frequency Spectrum} & \cite{Guan2018CNN} & 2018 & CNN & BPSK, QPSK, 8PSK, 16QAM, 64QAM & $\geq$90\%, -5 dB, AWGN \\
			\cline{2-6}
			& \cite{Gao2019CNN} & 2019 & Lightweight CNN & \makecell{2ASK, 4ASK, 2FSK, 4FSK,\\ BPSK, QPSK, 16QAM, 64QAM} & $\geq$90.4\%, -2 dB, AWGN \\
			\cline{2-6}
			& \cite{Tian2019nGCNN} & 2019 & CNN & 2ASK, 2FSK, 2PSK & $\geq$98\%, 2 dB, Impulsive noise \\
			\hline
			\makecell{I/Q Sequence, \\Eye Diagram}& \cite{Li2019IQEyeCNN} & 2019 & ResNet & \makecell{BPSK, QPSK, OQPSK, 8PSK, \\ 16QAM, 16APSK, 32APSK, 64QAM} & $\geq$97\%, 5 dB, AWGN \\
			
			\hline
			\makecell{I-Q constellation image, \\Spectrum-based image}& \cite{Khan20213DCNN} & 2021 & 3DCNN &  BPSK, QPSK, 16QAM, 64QAM & $\geq$96.97\%, SNR not given, AWGN \\
			\hline
			\makecell{I/Q Sequence, \\Constellation Map}& \cite{An2022SC} & 2022 & CNN & 2/4/8PSK, 16/64QAM & $\geq$90\%, -1 dB, AWGN \\
			\hline
			\makecell{High-order cumulants, \\Wavelet Feature, \\Instantaneous Feature,\\ SNR Estimation,\\Cyclic Spectral Features}& \cite{Liu2022CNNGRU} & 2022 & CNN-GRU &  \makecell{2PSK, 2ASK, 2FSK, 4PSK, \\ 4ASK, 4FSK, 16QAM, 64QAM} & $\geq$80\%, -10 dB, AWGN \\
			\hline
			\makecell{Wavelet Feature, \\PSD,\\Cyclic Spectral Features}& \cite{Wang2024HN} & 2024 & S\&SEFM &  2/4FSK, 2/4PSK, DSSS, OFDM & $\geq$95\%, 4 dB, AWGN \\
			\hline
	\end{tabular}}
\end{table*}

\subsubsection{Application of RNN for AMR}
\begin{figure}[!h]
	\centering
	\includegraphics[width=0.7\linewidth]{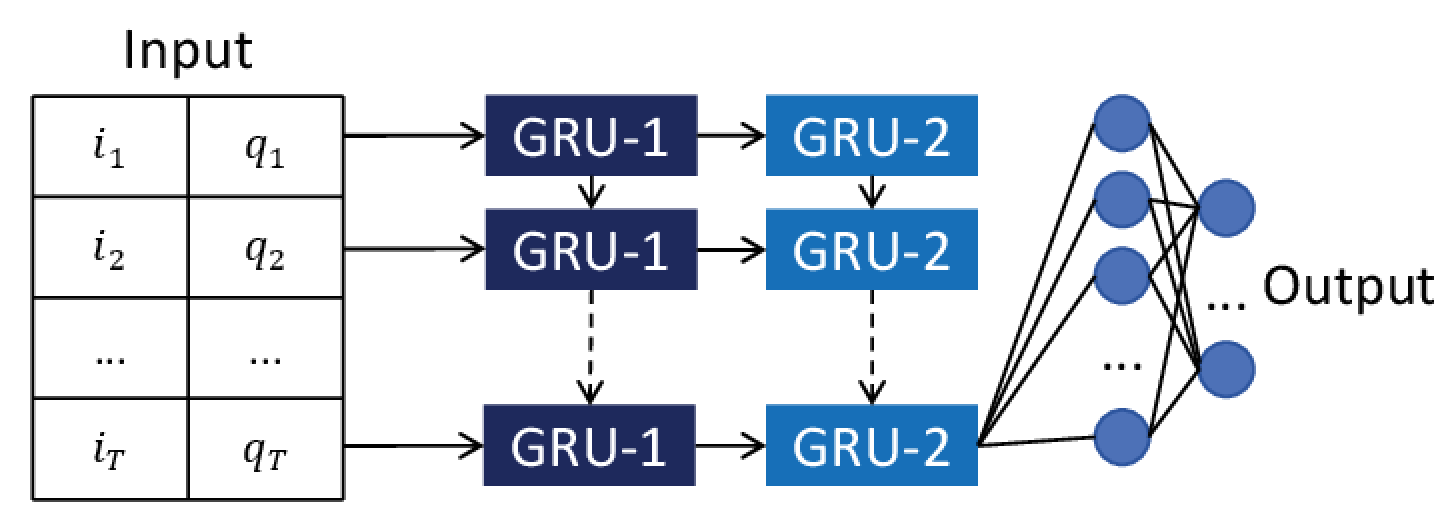}
	\caption{A typical AMR model dealing with sequence data.}
	\label{fig08}
\end{figure}
The modulation signal itself is a temporal sequence, and image features cannot represent its temporal details. Therefore, researchers have proposed AMR methods based on RNN \cite{ELMAN1990RNN} to fully extract the temporal sequence features of modulation signals as shown in Fig.\ref{fig08}. 

Hong et al. \cite{Hong2017RNN} introduced an RNN-based framework to exploit the temporal dependencies of communication signals, achieving higher accuracy than CNN-based AMR \cite{Shea2016CNN}. They further compared RNN, GRU, and LSTM models, showing that GRU achieved the best performance. However, confusion among similar modulation formats remained, even under high-SNR conditions.

In order to achieve a balance between classification accuracy and computational complexity, and meet the real-time spectrum sensing needs of IoT devices, Ghasemzadeh et al. \cite{Ghasemzadeh2022GSQRNN} proposed a gated stacked quasi RNN (GS-QRNN) model. This work achieved a significant improvement in execution efficiency by sacrificing a certain level of precision performance.

To fully utilize temporal and spatial features, Ke et al. \cite{Ke2022LSTM} proposed a LSTM-based model for AMR that extracted amplitude/phase features of I/Q sequence and power spectral density (PSD). Under high SNR conditions, the recognition accuracy was improved by 8\% compared to traditional models, but it still cannot solve the confusion problem between 16QAM and 64QAM, AM-DSB and WBFM.

\subsubsection{Hybrid Neural Networks for AMR}

\begin{figure}[!h]
	\centering
	\includegraphics[width=0.7\linewidth]{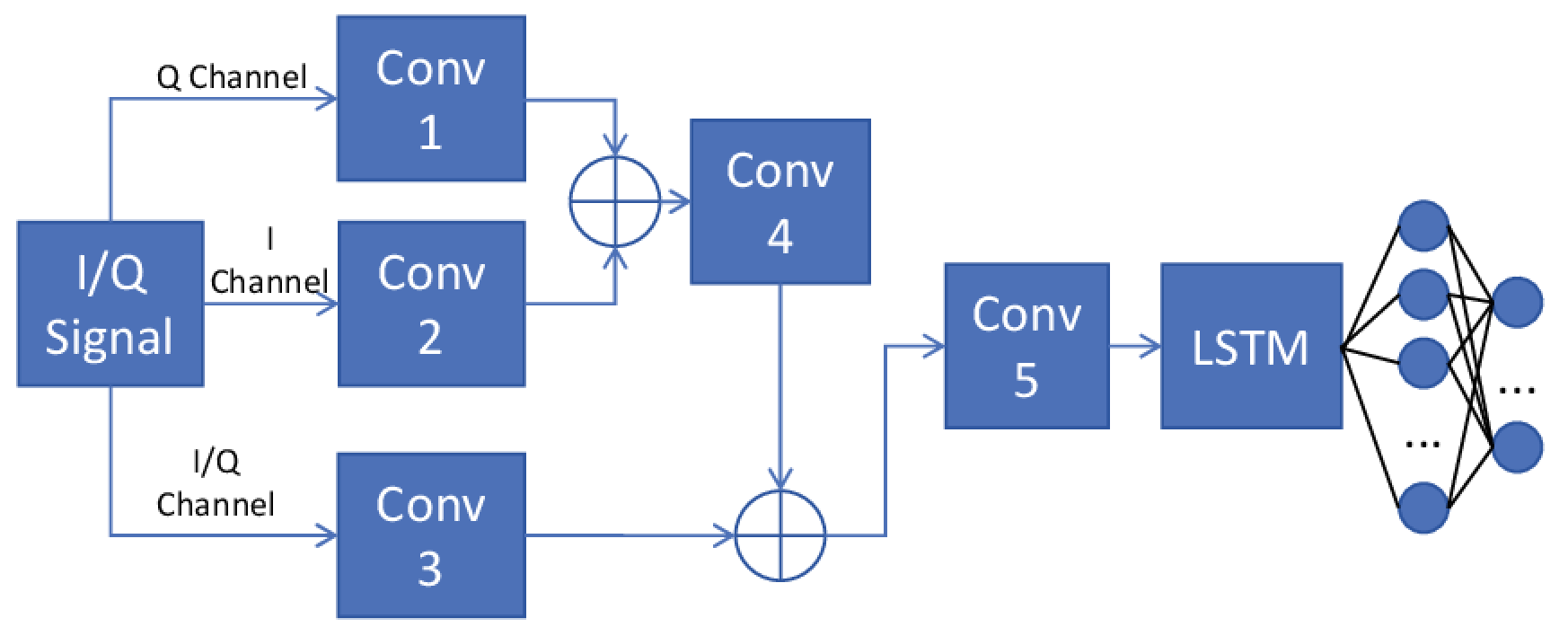}
	\caption{A hybrid model combining CNN and LSTM for AMR.}
	\label{fig09}
\end{figure}

To jointly exploit spatial and temporal information, hybrid CNN–RNN architectures have been proposed for AMR. Xu et al. \cite{Xu2020CNNLSTM} developed a multi-channel convolutional long short-term deep neural network (MCLDNN) that learns complementary features from I, Q, and I/Q channels (Fig.~\ref{fig09}). Although MCLDNN improves recognition performance, confusion among similar modulation formats remains under low-SNR conditions.

Subsequent studies further improved hybrid AMR models by integrating residual learning, recurrent networks, multimodal fusion, and Transformers. Representative methods include IRLNet \cite{Yang2021IRLNet}, the multi-cue fusion (MCF) model \cite{Wang2021MCF}, and the dual-flow convolutional Transformer network (DFCTNet) \cite{SHEN2025DFCT}, which enhance spatial-temporal feature representation. However, confusion among similar modulation formats and high computational complexity remain major challenges. Representative hybrid models are summarized in Table~\ref{tab08}.
\begin{table*}[!h]
	\caption{Hybrid Neural Networks for AMR in communication. \label{tab08}}
	\renewcommand{\arraystretch}{0.6}
	\centering
	\resizebox{1.0\linewidth}{!}{
	\begin{tabular}{|c|c|c|c|c|c|}
		\hline
		Core Input Features & Ref. & Year & Model & Modulation Type & Recognition Accuracy \\
		\hline
		\multirow{4}{*}{I/Q Sequence} & \cite{Xu2020CNNLSTM} & 2020 & MCLDNN (CNN \& LSTM ) &  \makecell{8PSK, DSB, SSB, BPSK, GFSK, CPFSK,\\ PAM, 16QAM, 64QAM, QPSK, WBFM} & $\geq$92\%, 0 dB, AWGN \\
		\cline{2-6}
		& \cite{Yang2021IRLNet} & 2021 & IRLNet & \makecell{2/4/8/16FSK, 4/8/16PAM, 2/4/8/16/32/64PSK\\ 4/8/16/32/64/128/256QAM, AM-DSB, \\ AM-SSB, AM-SC, AM-USB, AM-LSB, FM, PM} & $\geq$82\%, 0 dB, AWGN\\
		\cline{2-6}
		& \cite{Zhang2022RCNN} & 2022 & R\&CNN & \makecell{2/4FSK,2/4PSK\\ 16/64QAM, OFDM, DSSS} & $\geq$98.2\%, Real data\\
		\hline
		\makecell{I/Q Sequence,\\constellation maps} & \cite{Wang2021MCF} & 2021 & MCF &  \makecell{8PSK, BPSK, CPFSK, GFSK, PAM4, 16/64QAM,\\ QPSK, AM-DSB, AM-SSB, WBFM} & $\geq$94.5\%, 10 dB, AWGN \\
		\hline
		\makecell{I/Q Sequence,\\Time-frequency Spectrum} & \cite{SHEN2025DFCT} & 2025 & DFCTNet &  \makecell{2/4/8/16/32/64PSK, 4/8/16/32/64/128/256QAM,\\ 2/4/8/16FSK, 4/8/16PAM, DSB,\\DSB-SC, USB, LSB, FM, PM} & $\geq$95\%, 0 dB, AWGN \\
		\hline
	\end{tabular}}
\end{table*}

\subsubsection{Other Neural Networks for AMR}
Different from mainstream AMR models, Ahmed and Ergun \cite{Ahmed2021RBF} combined principal component analysis (PCA) with radial basis function (RBF) networks for AMR, reducing training complexity while improving low-SNR recognition performance compared with MLP. However, the computational complexity of RBF increases rapidly with the number of hidden nodes, limiting its scalability to large modulation sets.

Recent studies have proposed specialized architectures for challenging AMR scenarios. SCSNN improves robustness under low-SNR conditions \cite{Wei2023Shrinkage}, TSTR exploits dual-stream Transformers for shallow-water multipath channels \cite{Li2024Trans}, and IQFormer jointly learns I/Q sequences and time-frequency features \cite{Shao2025Trans}. However, existing methods still exhibit limited performance for high-order and structurally similar modulation formats. Representative specialized models are summarized in Table~\ref{tab09}.
\begin{table*}[!h]
	\caption{Other Neural Networks for AMR in communication. \label{tab09}}
	\renewcommand{\arraystretch}{0.6}
	\centering
	\resizebox{1.0\linewidth}{!}{
	\begin{tabular}{|c|c|c|c|c|c|}
		\hline
		Core Input Features & Ref. & Year & Model & Modulation Type & Recognition Accuracy \\
		\hline
		High-order cumulants & \cite{Ahmed2021RBF} & 2021 & PCA-RBF & \makecell{BPSK, QPSK, 8PSK, 16/32/64APSK,\\ 16/32/64/256QAM} & $\geq$96\%, 0 dB, AWGN \\
		\hline
		I/Q Sequence & \cite{Wei2023Shrinkage} & 2023 & SCSNN &  \makecell{8PSK, BPSK, CPFSK, GFSK, PAM4, 16/64QAM,\\ QPSK, AM-DSB, AM-SSB, WBFM} & $\geq$83\%, -2 dB, AWGN \\
		\hline
		Time-Frequency Spectrum & \cite{Li2024AttRes} & 2024 & EfficientDet &  \makecell{2/4FSK, 2/4PSK,\\ CW, DSSS, LFM} & $\geq$91\%, -4 dB, AWGN \\
		\hline
		\multirow{2}{*}{\makecell{I/Q Sequence,\\Time-Frequency Spectrum}} & \cite{Li2024Trans} & 2024 & TSTR &  \makecell{2/4/8FSK, 2/4/8PSK,\\ 16/64QAM, OFDM} & $\geq$80\%, 5 dB, Impulse Noise \\
		\cline{2-6}
		& \cite{Shao2025Trans} & 2025 & IQFormer & \makecell{8PSK, BPSK, CPFSK, GFSK, PAM4, 16/64QAM,\\ QPSK, AM-DSB, AM-SSB, WBFM} & $\geq$92.7\%, 10 dB, AWGN \\
		\hline
	\end{tabular}}
\end{table*}

\subsection{Few-shot Learning in AMR for Communication}
Data-driven methods have excellent feature extraction capabilities, as the parameter size grows, the required training data also increases, which is difficult to meet in practical application scenarios. Therefore, researchers have proposed some improvement measures for few-shot learning scenarios. The term few-shot AMR is often used broadly in the literature to describe situations in which sufficient labeled training data are unavailable. However, data scarcity may arise from different sources, leading to substantially different learning problems.

In this survey, we distinguish four representative data-limited AMR scenarios:
\textit{Few-sample learning}, only a small number of labeled samples are available for each modulation class, while the modulation categories in training and testing remain identical. \textit{Novel-class recognition}, previously unseen modulation types or waveform classes appear during testing. This problem is closely related to open-set recognition and zero-shot learning. \textit{Cross-domain adaptation}, the modulation classes remain unchanged, but the channel environment, propagation conditions, or hardware platforms differ between training and testing. \textit{Distribution-shift learning}, the statistical properties of observations change due to factors such as SNR variation, Doppler effects, interference, or target dynamics.

These scenarios impose different requirements on AMR algorithms and therefore motivate different solutions such as data augmentation, transfer learning, meta-learning, domain adaptation, and generative modeling. For example, GAN mainly addresses few-sample scarcity, while TL mainly addresses cross-domain adaptation.

Tang et al. \cite{Tang2018GAN} first introduced auxiliary classifier generative adversarial network (ACGAN) for AMR data augmentation, demonstrating improved recognition performance, particularly under low-SNR conditions. Subsequently, Zhou et al. \cite{Zhou2020GAN,Zhou2022FL} combined GANs with semi-supervised learning to exploit unlabeled data and further improve recognition accuracy. However, GAN-based methods remain limited by unstable adversarial training and mode collapse.

To enhance the feature extraction ability, Chen et al. \cite{Chen2020AttCRGAN} combined ACGAN with CNN and bidirectional RNNs to enhance spatiotemporal feature learning for few-shot AMR. Wang et al. \cite{Wang2022IAFNet} further proposed IAFNet by integrating pulse-noise preprocessing, attention mechanisms, and few-shot learning for underwater communication. However, existing methods remain sensitive to the distribution mismatch between generated and real samples as well as complex channel impairments such as multipath propagation.

\begin{figure}[!h]
	\centering
	\includegraphics[width=0.7\linewidth]{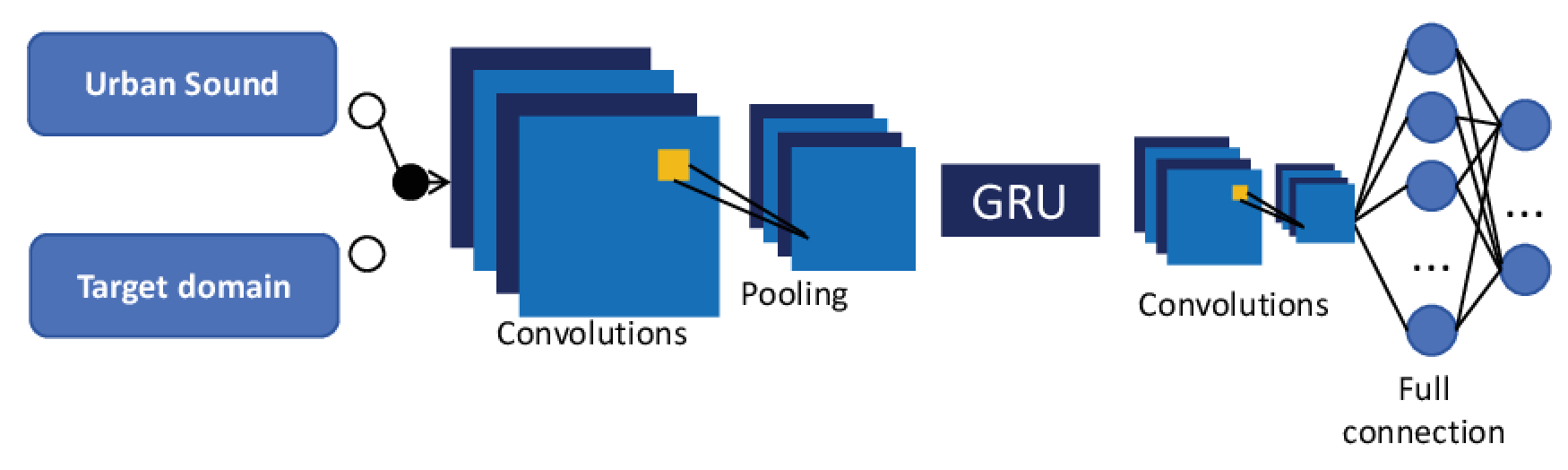}
	\caption{A domain transfer learning structure for AMR.}
	\label{fig11}
\end{figure}

Transfer learning (TL) has been widely adopted to alleviate data scarcity in AMR. Jiang et al. \cite{Jiang2019TL} first introduced TL into AMR using stacked autoencoders, while Bu et al. \cite{Bu2020ATL} proposed an adversarial transfer learning framework for cross-domain knowledge transfer between different signal domains. Although TL significantly improves data efficiency, its performance remains sensitive to channel mismatch and severe multipath propagation (Fig.~\ref{fig11}).

To further improve feature transfer, Liang et al. \cite{Liang2021AttRes} incorporated attention mechanisms into ResNeXt for cross-domain AMR, whereas Li et al. \cite{Li2021CN} investigated lightweight capsule networks to improve parameter efficiency. Nevertheless, both methods still exhibit degraded performance under low-SNR conditions.

Subsequent studies explored more label-efficient learning paradigms. Chang et al. \cite{Chang2022ML} proposed a multi-task learning framework by jointly exploiting I/Q and amplitude-phase features, while Chen et al. \cite{Chen2025ZS} introduced zero-shot learning for recognizing unseen modulation types through semantic attribute learning. However, these methods still rely on domain consistency or auxiliary knowledge, limiting their generalization across heterogeneous communication environments. Representative few-shot AMR methods are summarized in Table~\ref{tab10}.

\begin{table*}[!h]
	\caption{Data-driven AMR in Few-shot Learning Scenario. \label{tab10}}
	\renewcommand{\arraystretch}{1}
	\centering
	\resizebox{1.0\linewidth}{!}{
	\begin{tabular}{|c|c|c|c|c|c|}
		\hline
		Core Input Features & Ref. & Year & Model & Modulation Type & Recognition Accuracy \\
		\hline
		Contour Stellar Image & \cite{Tang2018GAN} & 2018 & ACGAN \& AlexNet & \makecell{4ASK, BPSK, QPSK, OQPSK, \\8PSK, 16QAM, 32QAM, 64QAM} & \makecell{$\geq$98\%, 0 dB, AWGN\\$\geq$62\%, -6 dB, AWGN}\\
		\hline
		Power Spectrum & \cite{Jiang2019TL} & 2019 & SAE & 2FSK, 4FSK, 8FSK, BPSK, OFDM, LFM & $\geq$88\%, 22 dB, AWGN \\
		\hline
		\multirow{9}{*}{I/Q Sequence} & \cite{Zhou2020GAN} & 2020 & E3SGAN & \makecell{BPSK, QPSK, 8PSK, 16QAM, 64QAM, GFSK, \\CPFSK, 4PAM, WBFM, AM-SSB, AM-DSB} & $\geq$91.8\%, 12 dB, AWGN \\	
		\cline{2-6} & \cite{Zhou2022FL} & 2022 & CNN-GAN & \makecell{BPSK, QPSK, 8PSK, 16QAM, 64QAM, GFSK, \\CPFSK, 4PAM, WBFM, AM-SSB, AM-DSB} & $\geq$80\%, 0 dB, AWGN \\	
		\cline{2-6}
		& \cite{Chen2020AttCRGAN} & 2020 & \makecell{ACGAN,\\Bidirectional RNN,\\Cyclic connected CNN} & \makecell{BPSK, QPSK, 8PSK, 16QAM, 64QAM, GFSK, \\CPFSK, 4PAM, WBFM, AM-SSB, AM-DSB} & $\geq$90.1\%, 0 dB, AWGN \\	
		\cline{2-6}
		& \cite{Bu2020ATL} & 2020 & ATLA & \makecell{BPSK, QPSK, 8PSK, 16QAM, 64QAM, GFSK, \\CPFSK, 4PAM, WBFM, AM-SSB, AM-DSB}	& $\geq$80\%, 5 dB, AWGN\\
		\cline{2-6}
		& \cite{Li2021CN} & 2021 & CapsNet & \makecell{BPSK, QPSK, 8PSK, 16QAM, 64QAM, GFSK, \\CPFSK, 4PAM, WBFM, AM-SSB, AM-DSB}	& $\geq$90\%, 12 dB, AWGN\\
		\cline{2-6}
		& \cite{Wang2022IAFNet} & 2022 & IAFNet & \makecell{2/4/8FSK, BPSK, QPSK, \\LFM, OFDM, FH}	& \makecell{ $\geq$93.3\%, 0 dB, AWGN,\\ Pulse noise}\\
		\cline{2-6}
		& \cite{Lin2023TL} & 2023 & CNN-GRU &  \makecell{BPSK, QPSK, 8PSK, 16QAM, 64QAM, GFSK, \\CPFSK, 4PAM, WBFM, AM-SSB, AM-DSB}	& $\geq$90\%, 0 dB, AWGN\\
		\cline{2-6}
		& \cite{Chen2025ZS} & 2025 & GCN &  4/8PSK, 16/32/64QAM, 16APSK	& $\geq$90\%, 0 dB, AWGN\\
		\hline
		\makecell{I/Q Sequence,\\A/P Feature} & \cite{Chang2022ML} & 2022 & MLDNN & \makecell{BPSK, QPSK, 8PSK, 16QAM, 64QAM, GFSK, \\CPFSK, 4PAM, WBFM, AM-SSB, AM-DSB} & $\geq$90\%, 0 dB, AWGN \\
		\hline		
		Time-frequency Spectrum & \cite{Liang2021AttRes} & 2021 & Att-ResNeXt & 2/4/8/16PSK, 32/64APSK, 16/32/64QAM, OQPSK & $\geq$80\%, 2 dB, AWGN \\
		\hline		
	\end{tabular}}
\end{table*}

\section{AI Empowered AMR for Sensing}
The signal modulation method is an important intra pulse feature of radar signals, and radar signal waveform recognition is the process of identifying the signal modulation method, based on which the radar types can be further identified. Early radar waveform recognition methods extracted parameters from received signals and relied on prior knowledge for recognition, but due to the lack of reasoning, they are hard to be suitable for the increasingly complex electromagnetic environment. The AMR of radar signals can be traced back to the 1980s \cite{Danielsen1988ESM,Hsue1990IET}, which proposed a modulation recognizer based on zero-crossing technology. Then, in 2004 and 2007, Kawalec et al. \cite{Kawalec2024Radar} and Lunden et al. \cite{Lunden2007radar} systematically propose automatic recognition methods for radar pulse modulation, respectively. Modern radar signal waveform recognition can also be divided into model driven and data-driven, and also faces the problem of few-shot learning. We will provide a survey of recent popular AI models for AMR of radar signals based on the three classifications mentioned above, and further divide them according to the types of features used, which is shown in Table~\ref{tab11}.

\begin{table*}[!h]
	\caption{AI-based AMR for Radar Sensing. \label{tab11}}
	\renewcommand{\arraystretch}{1}
	\centering
	\resizebox{1.0\linewidth}{!}{
		\begin{tabular}{|c|c|c|c|c|c|}
			\hline
			Core Input Features & Ref. & Year & Model & Modulation Type & Recognition Accuracy \\
			\hline
			\multicolumn{6}{|c|}{Model-based AMR for radar signal} \\
			\hline
			\multirow{2}{*}{Entropy Features} & \cite{Zhang2004SVM} & 2004 & SVM & \makecell{BPSK, QPSK, MPSK, LFM, NLFM, \\CW, FD, FSK, IPFE, CSF} & $\geq$87.4\%, 5 dB, AWGN \\
			\cline{2-6}
			& \cite{Zhou2024RadarKNN} & 2019 & KNN & \makecell{Sine AMFS, NAM, NFM, \\ BPSK, LFMP, Sine FM, Saw FM} & $\geq$98.8\%, -10 dB, AWGN \\
			\hline
			\multirow{2}{*}{Time-frequency Feature}& \cite{Gao2019Radar} & 2019 & SVM & \makecell{BPSK, COSTAS, NLFM, FRANK, \\ LFM, P1-P4} & $\geq$91.5\%, -2 dB, AWGN \\
			\cline{2-6}
			& \cite{Liu2024RF} & 2019 & Random Forest & \makecell{LFM, NLFM, BPSK, QPSK, \\ MPSK, LFMCW, FSK, NS} & $\geq$93.0\%, -2 dB, AWGN \\
			\hline
			\multicolumn{6}{|c|}{Data-driven AMR for radar signal} \\
			\hline
			\multirow{2}{*}{Time-domain Feature} & \cite{Liao2022Radar} & 2022 & MLP & \makecell{LFM, Yark-42, \\ Cessna, An-26} & $\geq$92.37\%, K=L, AWGN \\
			\cline{2-6}
			& \cite{Wu2025Radar} & 2025 & MMFAR(CNN-based) & \makecell{Costas, Barker, Frank, LFM, \\SFM, T1-T4} & $\geq$74.2\%, -20 dB, AWGN \\
			\hline
			\multirow{13}{*}{Time-frequency Spectrum} & \cite{Xie2021Radar} & 2021 & MLP, RBF & LFM, Frank, P1-P4 & $\geq$94.17\%, 0 dB, AWGN \\
			\cline{2-6}
			& \cite{Wei2020Radar} & 2020 & ACSE(CNN-based) & \makecell{2ASK, 2FSK, BPSK, CW, \\EXP, LFM, SFW, SIN} & $\geq$93\%, -10 dB, AWGN \\
			\cline{2-6}
			& \cite{Qu2020Radar} & 2020 & CNN, DQN & \makecell{LFM, SFM, BPSK, \\2FSK, 4FSK, EQFM, Frank} & $\geq$83.05\%, -8 dB, AWGN \\
			\cline{2-6}
			& \cite{Pan2020Radar} & 2020 & MIML-DCNN & LFM, Barker, Frank, Costas & $\geq$83\%, 6 dB, AWGN \\
			\cline{2-6}
			& \cite{Si2021Radar} & 2021 & DCNN & \makecell{2FSK, 4FSK, BPSK, EQFM, \\Frank, LFM, SFM} & $\geq$83.4\%, -10 dB, AWGN \\
			\cline{2-6}
			& \cite{Yu2021Radar} & 2021 & GoogLeNet \& ResNet-18 & LFM, SFM, PFM, Single Carrier, FSK & $\geq$90\%, -5 dB, AWGN \\
			\cline{2-6}
			& \cite{Jiang2022LPI} & 2022 & Unet, DCNN & \makecell{BPSK, Costas, Frank, LFM, \\P1-P4, T1-T4} & $\geq$91.17\%, -10 dB, AWGN \\
			\cline{2-6}
			& \cite{Ren2023Radar} & 2023 & Swin-Transformer & \makecell{CW, LFM, SIN-FM, 2FSK, 4FSK, \\ BPSK, QPSK, P1-P4, Frank} & $\geq$80\%, -12 dB, AWGN \\
			\cline{2-6}
			& \cite{Chen2024Radar} & 2024 & Improved ResNet & \makecell{BPSK, Costas, Frank, LFM \\ T1-T4, P1-P4} & $\geq$94.3\%, -12 dB, AWGN \\
			\cline{2-6}
			& \cite{Jiang2024Radar} & 2024 & Improved CNN & \makecell{BPSK, Costas, Frank, LFM \\ T1-T4, P1-P4} & $\geq$85.17\%, -12 dB, AWGN \\
			\cline{2-6}
			& \cite{Gao2024FS} & 2024 & BPL(CNN-based) & LFM, T1-T4 & $\geq$97.29\%, 0 dB, AWGN \\
			\cline{2-6}
			& \cite{Hou2025Radar} & 2025 & CNN \& Mamba (DCMNet) &\makecell{LFM, SFM, EQFM, BPSK, \\FSK, 4FSK, Frank} & $\geq$90\%, -8 dB, AWGN \\
			\cline{2-6}
			& \cite{Xu2025Radar} & 2025 & YOLOv7 &LFM, BPSK, Frank, Costas & $\geq$91\%, -4 dB, AWGN \\
			\hline
	\end{tabular}}
\end{table*}

\subsection{Model-based AMR for Radar Sensing}
\subsubsection{AMR for Radar Sensing based on SVM}
The existing SVM based models for radar modulation recognition mainly utilize entropy features and signal time-frequency features. Zhang et al. \cite{Zhang2004SVM} first identified radar radiation sources using SVM with entropy features. The experiment proved that the SVM classifier is applicable in the field of radar radiation source recognition. Gao et al. \cite{Gao2019Radar} combined three types of Wigner Ville distribution images, trained them using a non negative matrix factorization network and different CNNs, classified the results using SVM, and used heuristic algorithm for ensemble learning.

\subsubsection{AMR for Radar Sensing based on KNN}
Cai et al. \cite{Cai2022CL} and Zhou et al. \cite{Zhou2024RadarKNN} have implemented KNN for radar modulation signal recognition, however, multiple signal aliasing and real-time deployment constraints were not considered, and there is still room for optimization in the inference efficiency of KNN under high-dimensional and large sample conditions.

\subsubsection{AMR for Radar Sensing based on Decision Tree}
Decision trees (DT) and random forests (RF) have the advantages of high interpretability and low computational cost, and are used for radar intra pulse modulation recognition. Tian et al. \cite{Tian2022DT} and Liu et al. \cite{Liu2024RF} proposed radar signal recognition models based on decision tree and random forest, respectively. Decision trees have weak noise resistance and generalization. Although random forests are more robust, the inference is slow under high-dimensional features, and recognition rates still significantly decline under low signal-to-noise ratios.

In general, SVM has the advantages of simple structure and strong generalization ability in recognition problems, but it is sensitive to kernel function selection and difficult to solve multi classification problems. KNN algorithm is simple, easy to understand, and robust in denoising data through the selection of K. However, it requires a large amount of space to store all known instances, and its time complexity is high due to the need to compare the instances to be classified with all known instances. The decision tree is intuitive and effective for small-scale datasets, but it is inconvenient to handle continuous variables and when there are many categories, errors increase quickly and scalability is average.

\subsection{Data-driven AMR for Radar Sensing}
\subsubsection{MLP-based AMR for Radar Sensing}
MLP is an early emerged deep learning model that was quickly used for radar modulation signal recognition. Liao et al. \cite{Liao2022Radar} proposed an interpretable depth probability model based on the high-dimensional nature of high range resolution (HRR) radar signals and the unexplainability of traditional methods, leveraging time-domain features. The model has high interpretability, but due to the need to build an independent inference network for each azimuth frame, the parameter scale is large. Xie et al. \cite{Xie2021Radar} focused on the difficult problem of LFM radar waveform recognition and proposed an algorithm based on fractional Fourier transform (FrFT) and time-frequency analysis. The core was to extract three key features and achieve efficient recognition by assembling a neural network classifier. Under low SNR, Frank code was prone to be classified as P3 code, while P4 code was prone to be classified as LFM. Moreover, noise can significantly affect the accuracy of extracting the standard deviation of the target component width.

\subsubsection{CNN-based AMR for Radar Sensing}
Convolutional based models are the most widely used deep learning models in radar modulation recognition, such as CNN, ResNet, GoogLeNet, and Yolo models. In this type of method, the input features utilized by the model are mainly time-domain features and time-frequency spectrum features.

\paragraph{Time-domain Feature}
To improve recognition performance in low SNR and complex environments, Wu et al. \cite{Wu2025Radar} proposed a radar signal AMR method based on multi-level and multi-scale feature learning, which combined mask autoencoder and contrastive learning. The model needs to fine tune multiple hyperparameters, such as mask ratio, contrast temperature, loss weight, etc., which increases optimization complexity and practical deployment difficulty.

\paragraph{Time-frequency Spectrum}
Time frequency analysis is an important means of signal modulation recognition, as it contains both time and frequency information. Recognition accuracy is the most important evaluation indicator for radar modulation signal recognition, and the main goal of most work is to achieve high signal recognition accuracy.

Wei et al. \cite{Wei2020Radar} proposed a radar signal AMR method based on multi-branch asymmetric convolution squeeze-and-excitation (ACSE) network and multi-domain feature fusion, which solved the problems of low accuracy and high complexity of traditional methods under low SNR. However, only 8 common radar signals were recognized in their work. To enhance the ability to resist noise, Jiang et al. \cite{Jiang2022LPI} proposed a two-stage framework combining feature enhancement and classification recognition. Firstly, the radar signal was converted into an image through time-frequency analysis, and then the features were denoised through Unet. Finally, CNN was used to identify 12 typical modulation signals. However, the model was prone to confusion when dealing with multi-phase modulation signals such as T1-T2 codes. Chen et al. \cite{Chen2024Radar} addressed the problem of AMR of 15 types of radar signals in complex electromagnetic environments with low SNR, colored noise, and multipath fading. When the SNR is as low as -12 dB, the average recognition accuracy of 15 types of signals reaches 94.93\%. However, for signals with similar time-frequency characteristics such as Frank and P1-P4, there was still a misjudgment rate of about 5\%. Jiang et al. \cite{Jiang2024Radar} designed a recognition framework that combined multi-layer decomposition denoising and deep learning. When the SNR was as low as -14 dB, the recognition accuracy of 12 signals reached 75.3\%, which was about 11\% higher than direct recognition of the original signal. However, the recognition performance of multi-phase code signals such as P1 and P4 was not good, with about 27\% of P1 codes being misclassified as P4 codes. Due to the high bandwidth occupation and frequency jitter of these signals, the denoising model was difficult to capture details.

In response to the three major problems of saturation, dependence on test sets, and inability to explain the impact of SNR on CNN performance in traditional methods of intra pulse modulation recognition of radar signals, Yu et al. \cite{Yu2021Radar} proposed a Grad-CAM Position Score (GCPS) quantitative evaluation criterion. Signal parameters are only randomly adjusted within a fixed range, but in actual environments, radar signal types and parameters are more complex, so further validation of generalization was needed for the model. 

Different perspective from the above methods, Hou et al. \cite{Hou2025Radar} addressed the problems of difficulty in feature extraction and insufficient capture of long sequence dependencies in complex electromagnetic environments using traditional methods. By integrating deformable convolution and Mamba architecture, a balance between high precision and lightweight was achieved through multi-view feature extraction and cross gate feature fusion. The performance of the model under different interferences has not been tested, and its generalization ability can be further improved.

The above work mainly focuses on the situation of a single signal, and in fact, there may be multiple signal superposition. How to simultaneously identify multiple signals is also one of the urgent problems to be solved. Aiming at the problem of overlapping time and frequency domains of multiple radar signals in electromagnetic environments, Pan et al. \cite{Pan2020Radar} proposed an AMR framework based on multi-instance multi-label learning (MIML) and CNN. The framework can accurately identify the modulation types of each component in overlapping signals by training only with a single type of signal, while also considering robustness under changes in SNR and power ratio. However, only the overlapping scenarios of 4 modulation types have been verified, and the generalization performance needs further validation. Qu et al. \cite{Qu2020Radar} extended the recognition types of radar signals to 8 and proposed an intra-pulse modulation recognition method suitable for single component and dual component radar signals, without the need to know the number of signal components in advance. The core was to solve the problem of identifying pulse overlap signals under low SNR by combining multi-core time-frequency analysis, deep learning feature extraction, and reinforcement learning classification. Similarly for low SNR scenarios, Si et al. \cite{Si2021Radar} proposed a multi-class learning framework, which improved accuracy compared to traditional CNN, but sacrificed execution efficiency. More importantly, 

\subsubsection{Transformer-based AMR for Radar Sensing}
Transformer is a widely used deep learning architecture in recent years. Ren et al. \cite{Ren2023Radar} proposed the ResSwinT network framework, which is the first method to use Swin-Transformer for the joint task of denoising and recognition of time-frequency maps of radar dual component signals. By adaptively triggering denoising through SNR, significant performance improvement can be achieved at extremely low signal-to-noise ratios while maintaining computational efficiency. However, for the above deep learning methods, the performance of these models depends on training with a large amount of data. Therefore, the design of network structures and the improvement of signal type adaptability under limited datasets are still unresolved issues.

\subsection{Few-shot Learning in AMR for Radar Sensing}
To address the issue of insufficient training samples, Xu et al. \cite{Xu2025Radar} introduced the idea of target detection, and proposed the concept of time-frequency ridge feature matching unit. By training the model with only a single signal sample, modulation recognition of dual overlapping signals can be achieved. The article also constructed an overlapping signal dataset based on time-frequency representation. For signals with highly similar time-frequency ridge shapes (such as low slope LFM and NS), the model was prone to confusion and difficult to achieve accurate classification. Gao et al. \cite{Gao2024FS} focused on the intra-pulse modulation recognition of few-shot radar signals, and proposed a Bayesian prototype learning (BPL) method that significantly improved generalization and robustness. The core conclusion was that through BPL, deep and shallow feature fusion, and class covariance measurement, basic class knowledge can be quickly transferred to new category radar signal recognition. The method still relies on a small number of manually annotated samples for model fine-tuning and new category adaptation, and cannot escape the dependence on manually annotated resources.

\section{AI Empowered AMR for ISAC Signals}
ISAC is a key enabler for next-generation wireless systems (6G), where a single waveform and hardware platform simultaneously support data transmission and radar-like sensing. From a modulation recognition perspective, ISAC introduces new challenges: the received signal is a superposition of communication symbols and target echoes, often with the same or tightly coupled modulation formats. Automatic modulation recognition (AMR) for ISAC must therefore not only identify the underlying modulation type (e.g., QPSK, OFDM, LFM) but also distinguish whether a signal originates from a communication transmitter, a sensing reflection, or a mixed ISAC waveform. Addressing these challenges through advanced machine learning and multi-domain feature fusion is essential for realizing agile, spectrum-efficient ISAC systems.
 
The automatic recognition of ISAC signals is an emerging field with only a few cutting-edge works. Zhang et al. \cite{Zhang2022ISAC} completed a representative work for ISAC signal recognition. In \cite{Zhang2022ISAC}, the authors proposed an automatic modulation recognition method based on neural architecture search for ISAC systems that achieves better recognition accuracy with fewer parameters and FLOPs than fixed-structure networks, especially under low SNR, however, only eight modulation types in simulated Rician fading channels were tested, lacking real-world signal validation. Similarly, Li et al. \cite{Li2025ISAC} proposed a cooperative spectrum detection scheme for ISAC services by combining an improved denoising auto-encoder with CNN and federated learning, which effectively reduces communication overhead, suppresses noise interference, and maintains high sensing accuracy under large compression ratios, yet it only validated performance in simulated Rayleigh fading channels, and did not test in real ISAC deployment environments.

AMR for ISAC signals is an important trend for future development. Currently, there are relatively few modulation methods involved in the work, and it is still in the simulation stage, lacking measured datasets and real-world experiments.

\section{Comparison of Representative Data-Driven AMR Models}
Although this survey reviews both model-based and data-driven AMR approaches, the experimental comparison in this section focuses on representative deep learning architectures. This choice is motivated by the fact that most recent AMR research has shifted toward data-driven methods and that a fair experimental comparison between model-based and deep learning approaches would require unified implementations, feature extraction pipelines, and benchmark settings, which are often unavailable in the literature.

\subsection{Datasets and Parameter Settings}
The experiment was conducted on four popular datasets, including three communication modulation datasets RML 2016.10a, RML 2016.10b, and RML 2016.04c, and one radar modulation dataset RadChar. We divided each RML dataset and the RadChar dataset into training set, validation set and test set with a ratio of 6:2:2. The categorical cross-entropy was set as the loss function and the Adam algorithm was adopted as the optimizer. The initial learning rate was set to be 0.001 and the batch size was set as 500. The experiments were implemented using  NVIDIA GeForce RTX 5060 GPU and Pytorch as the platform. The comparison of these datasets are given in Table~\ref{tab:evidence_strength}.
\begin{table*}[t]
	\caption{Evidence Strength Across AMR Research Domains}
	\label{tab:evidence_strength}
	\centering
	\renewcommand{\arraystretch}{1.2}
	\begin{tabular}{p{3cm}|p{3cm}|p{3cm}|p{2.5cm}|p{3cm}}
		\hline
		\textbf{Domain}
		&
		\textbf{Typical Dataset}
		&
		\textbf{Channel Assumptions}
		&
		\textbf{SNR Coverage}
		&
		\textbf{Validation Type}
		\\
		\hline
		
		Communication AMR
		&
		RadioML
		&
		AWGN, Rayleigh, Rician
		&
		-20 to 30 dB
		&
		Mostly simulation, some OTA
		\\
		\hline
		
		Radar Recognition
		&
		RadChar, proprietary datasets
		&
		Simplified target models
		&
		0 to 20 dB
		&
		Limited measured data
		\\
		\hline
		
		ISAC Recognition
		&
		Self-generated datasets
		&
		Synthetic ISAC channels
		&
		-10 to 20 dB
		&
		Predominantly simulation
		\\
		\hline
	\end{tabular}
\end{table*}

\subsection{Accuracy Performance}
\begin{figure}[!h]
	\centering
	\subfloat[]{\includegraphics[width=0.4\linewidth]{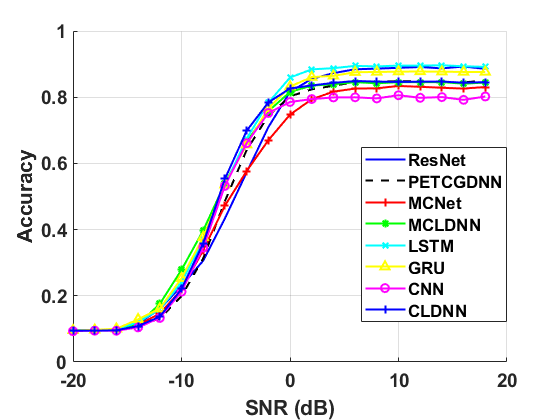}}
	\label{fig12a}
	\subfloat[]{\includegraphics[width=0.4\linewidth]{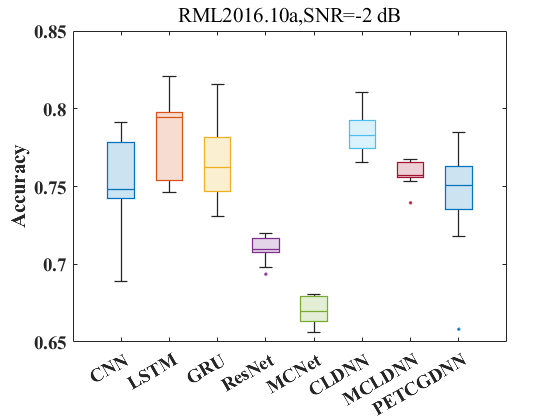}}
	\label{fig12b}\\
	\subfloat[]{\includegraphics[width=0.4\linewidth]{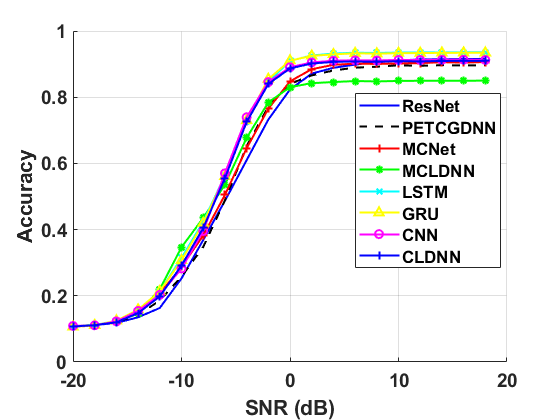}}
	\label{fig12c}
	\subfloat[]{\includegraphics[width=0.4\linewidth]{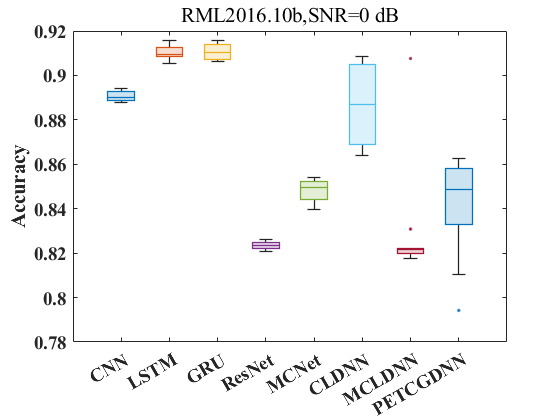}}
	\label{fig12d}\\
	\subfloat[]{\includegraphics[width=0.4\linewidth]{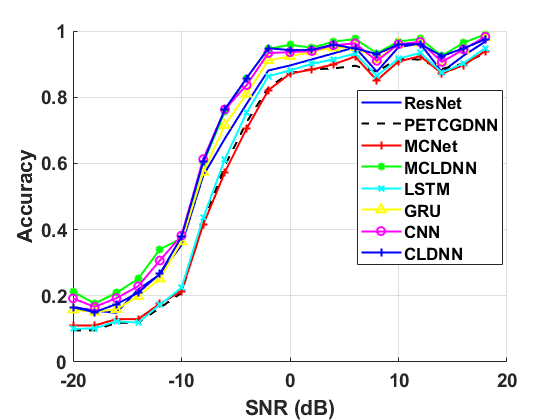}}
	\label{fig12e}
	\subfloat[]{\includegraphics[width=0.4\linewidth]{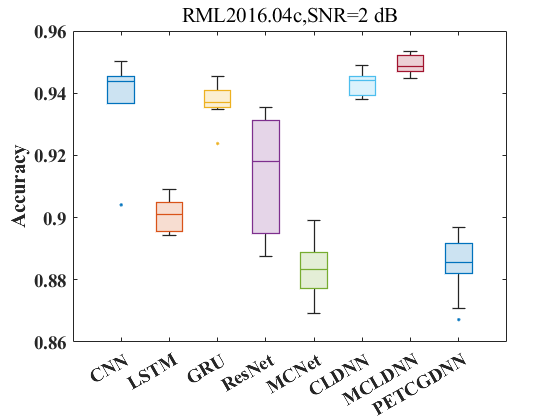}}
	\label{fig12f}
	\caption{Accuracy versus SNR and the fluctuation of accuracy variance. (a)(b) RML2016.10a, (c)(d) RML2016.10b, and (e)(f) RML2016.04c.}
	\label{fig12}
\end{figure}
Fig.~\ref{fig12} shows the average recognition accuracy of 8 typical deep learning models as a function of SNR on three datasets: RML2016.10a, RML2016.10b, and \textcolor{blue}{RML2016.04c}. These results are obtained by 10 times of experiments with different random seeds (42,40,20,123,5,7,8,256,338,15). Under the same experimental parameter settings, the highest recognition accuracies achieved on the three datasets were 91.59\% (obtained by the LSTM model at SNR=12 dB), 93.45\% (obtained by the MCLDNN model at SNR=18 dB), and 99.07\% (also obtained by the MCLDNN model at SNR=18 dB). Fig.~\ref{fig12} (b), (d), and (f) illustrate the statistical distributions of the recognition accuracies at representative SNR values over multiple independent runs. Compared with other methods, PETCGDNN achieves not only competitive median accuracy but also a smaller interquartile range and fewer outliers, demonstrating more stable recognition performance and better repeatability. These results indicate that the proposed model is less affected by the randomness introduced by different initialization seeds and provides consistently reliable classification performance.

From the overall experimental results, whether using I/Q signals or amplitude/phase signals as inputs, various AMR models based on deep learning, such as LSTM, GRU, MCLDNN, and PET-CGDNN, have shown significant advantages in recognition accuracy. Among these methods, RNN and its fusion structure model can effectively capture the temporal characteristics of signals and have stronger discriminative ability in medium to high SNR regions. In contrast, although ResNet-based networks perform well in fields such as computer vision, their recognition performance on the RML series dataset does not meet expectations due to their large parameter scale and complex structure. However, CNN models still have strong spatial feature extraction ability under relatively ideal channel conditions, and can achieve relatively stable recognition performance in high SNR regions.

\begin{figure*}[!h]
	\centering
	\subfloat[]{\includegraphics[width=0.2\textwidth]{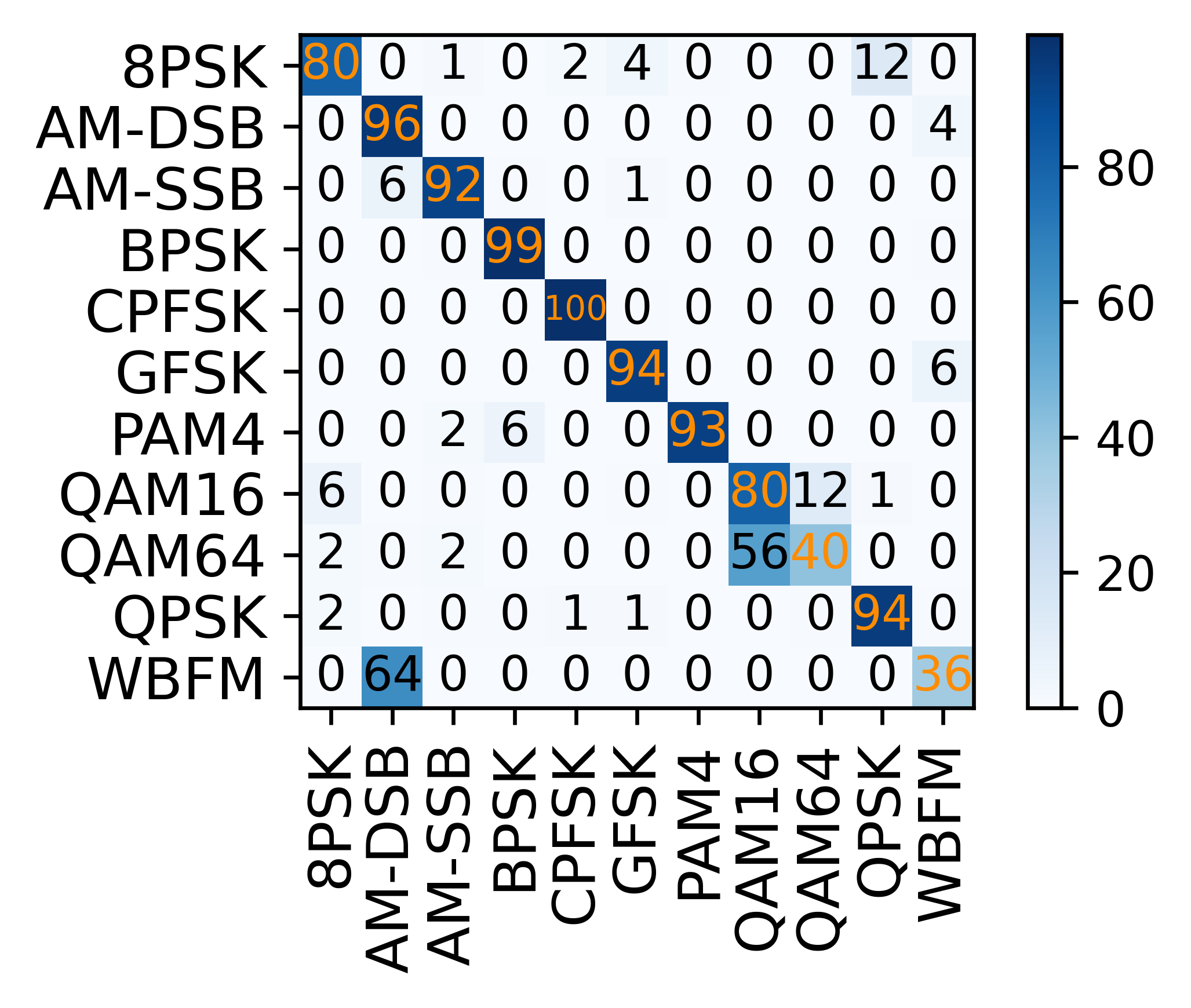}}
	\label{fig13a}
	\subfloat[]{\includegraphics[width=0.2\textwidth]{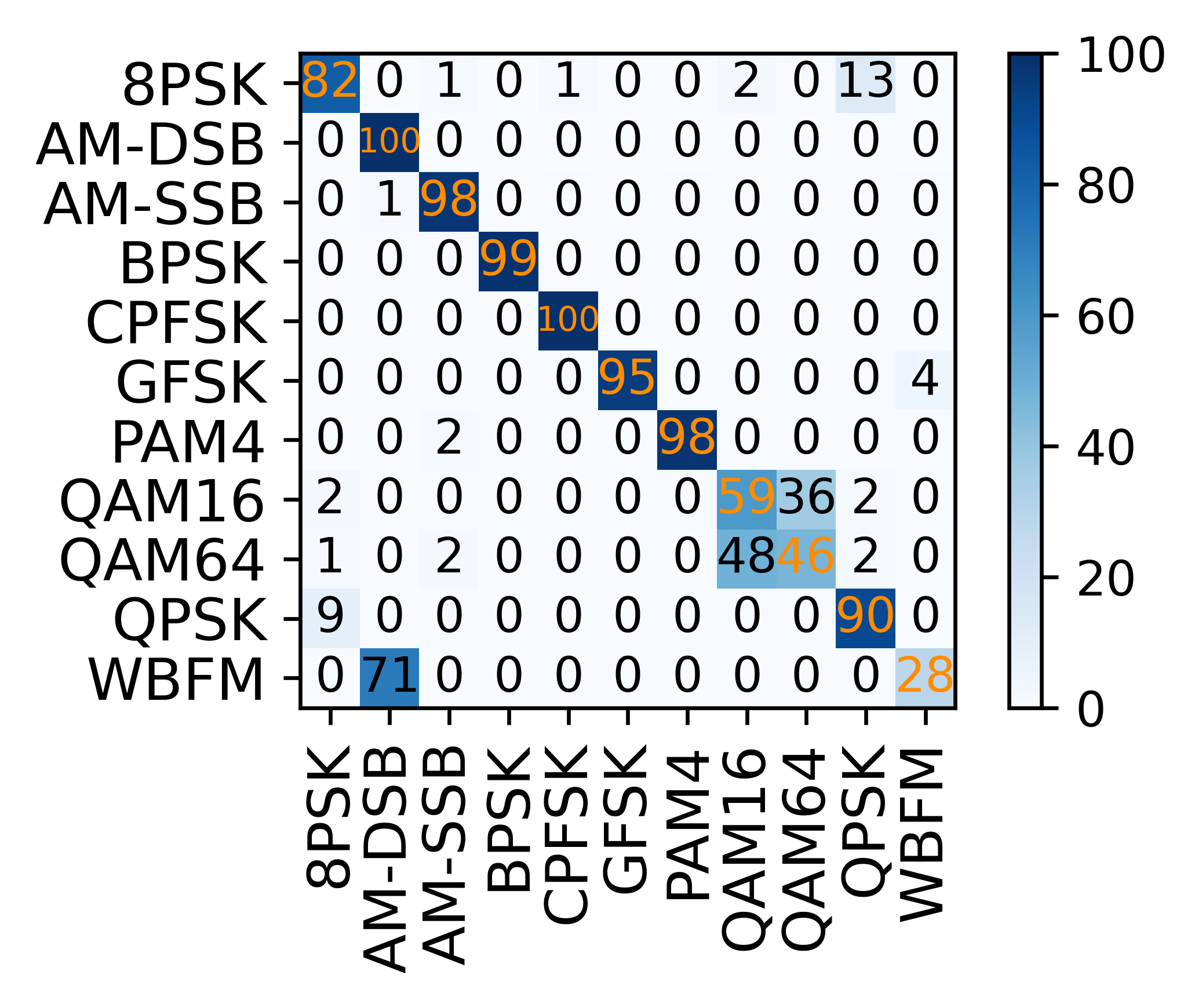}}
	\label{fig13b}
	\subfloat[]{\includegraphics[width=0.2\textwidth]{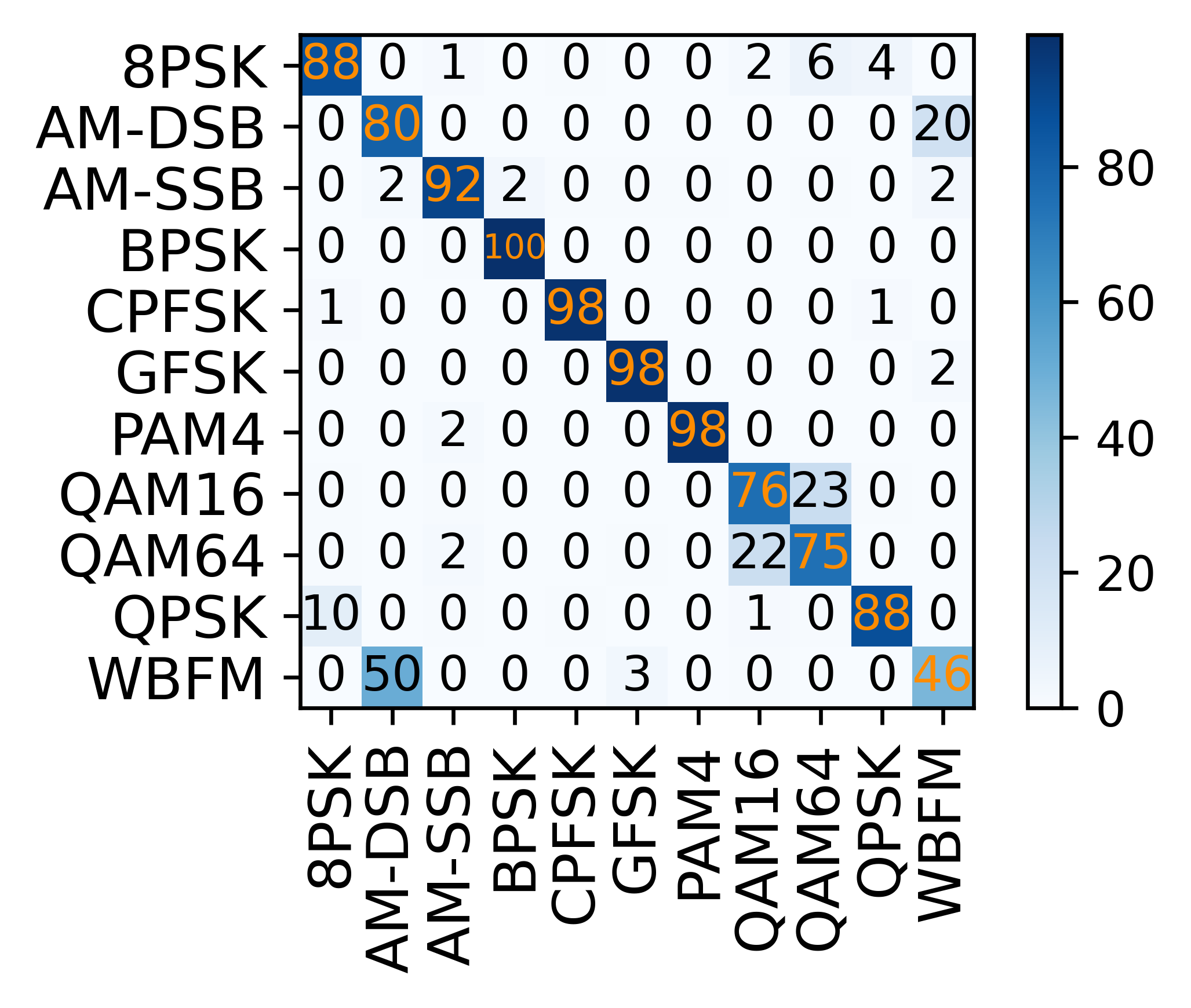}}
	\label{fig13c}
	\subfloat[]{\includegraphics[width=0.2\textwidth]{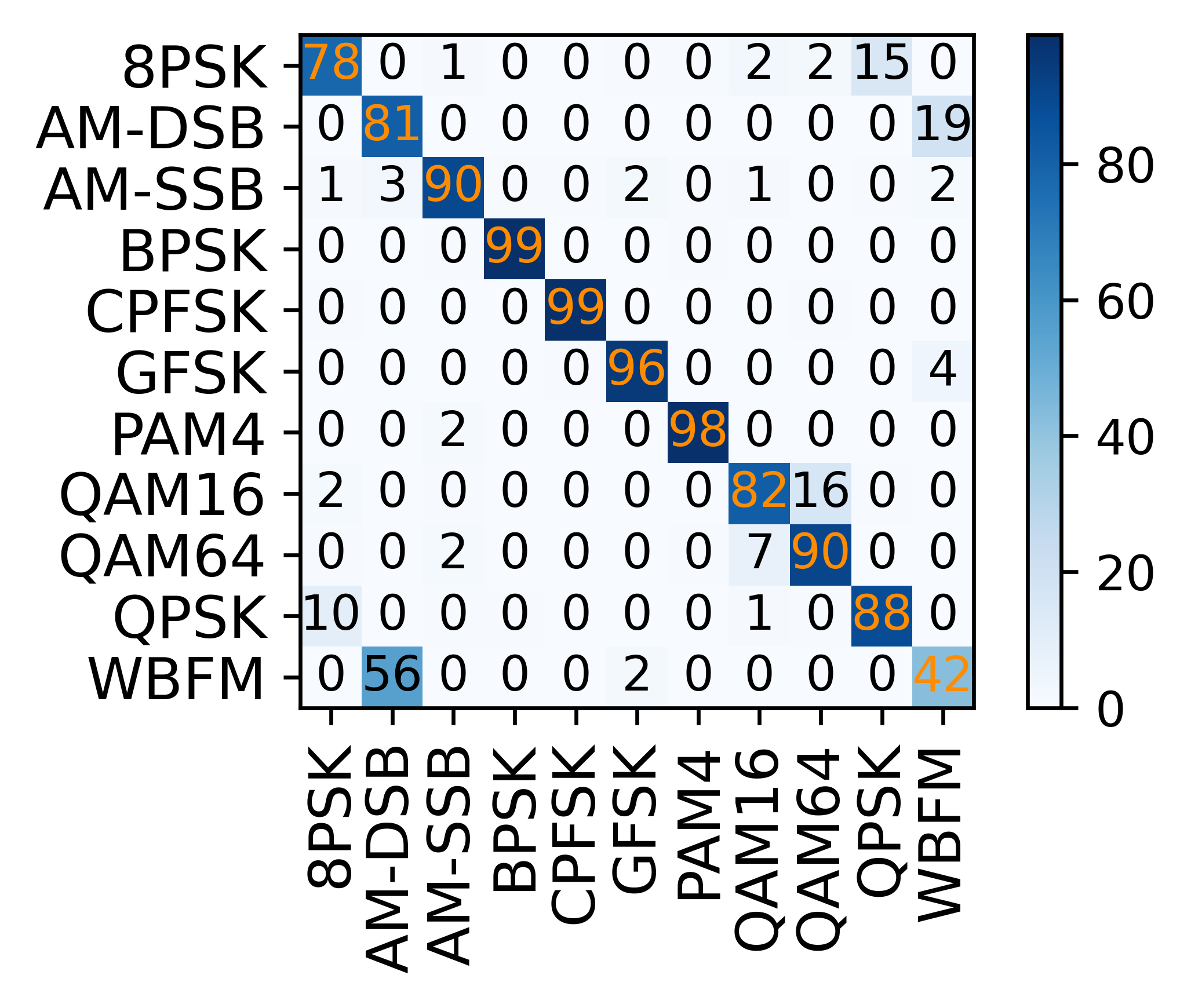}}
	\label{fig13d}\\
	\subfloat[]{\includegraphics[width=0.2\textwidth]{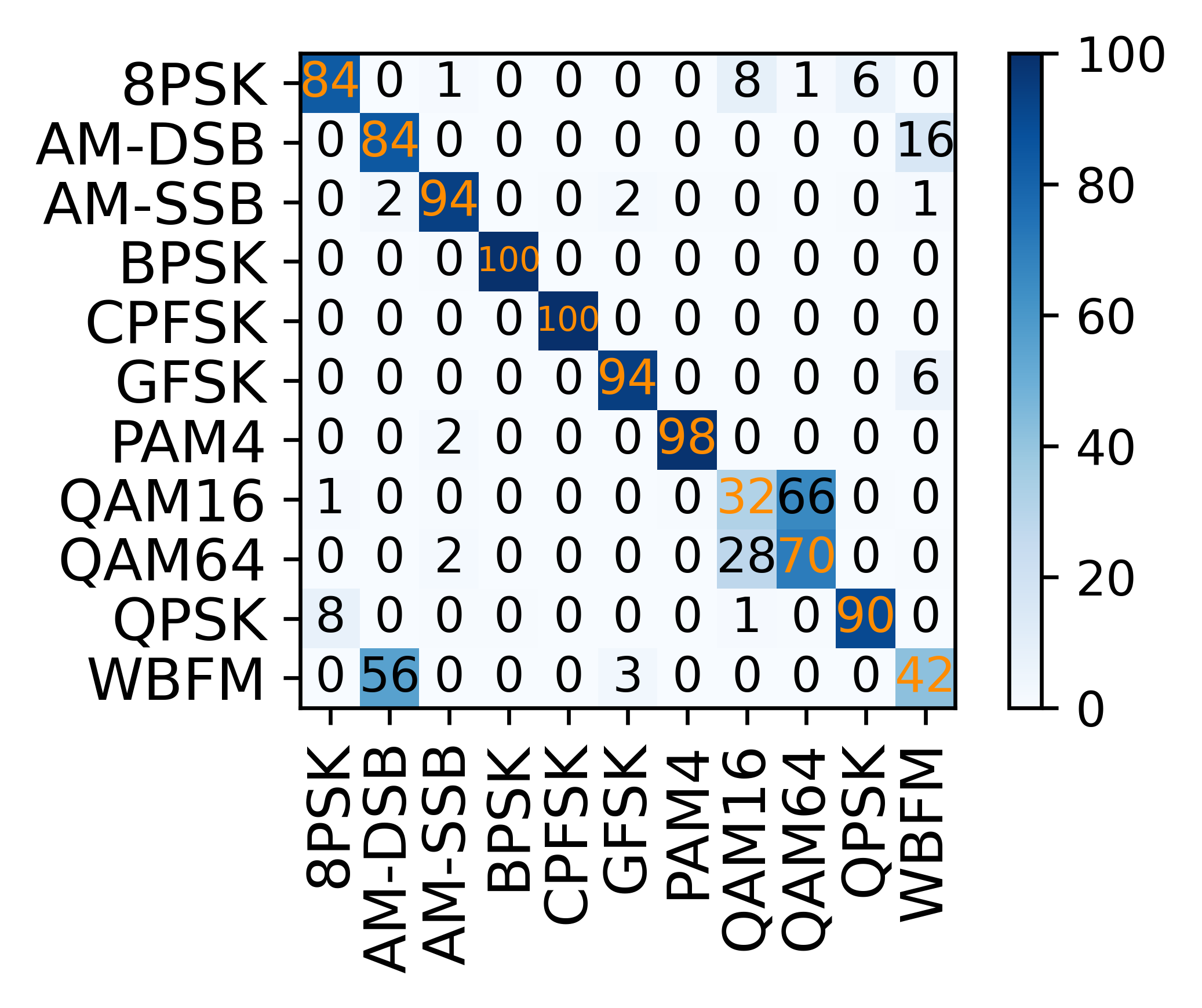}}
	\label{fig13i}
	\subfloat[]{\includegraphics[width=0.2\textwidth]{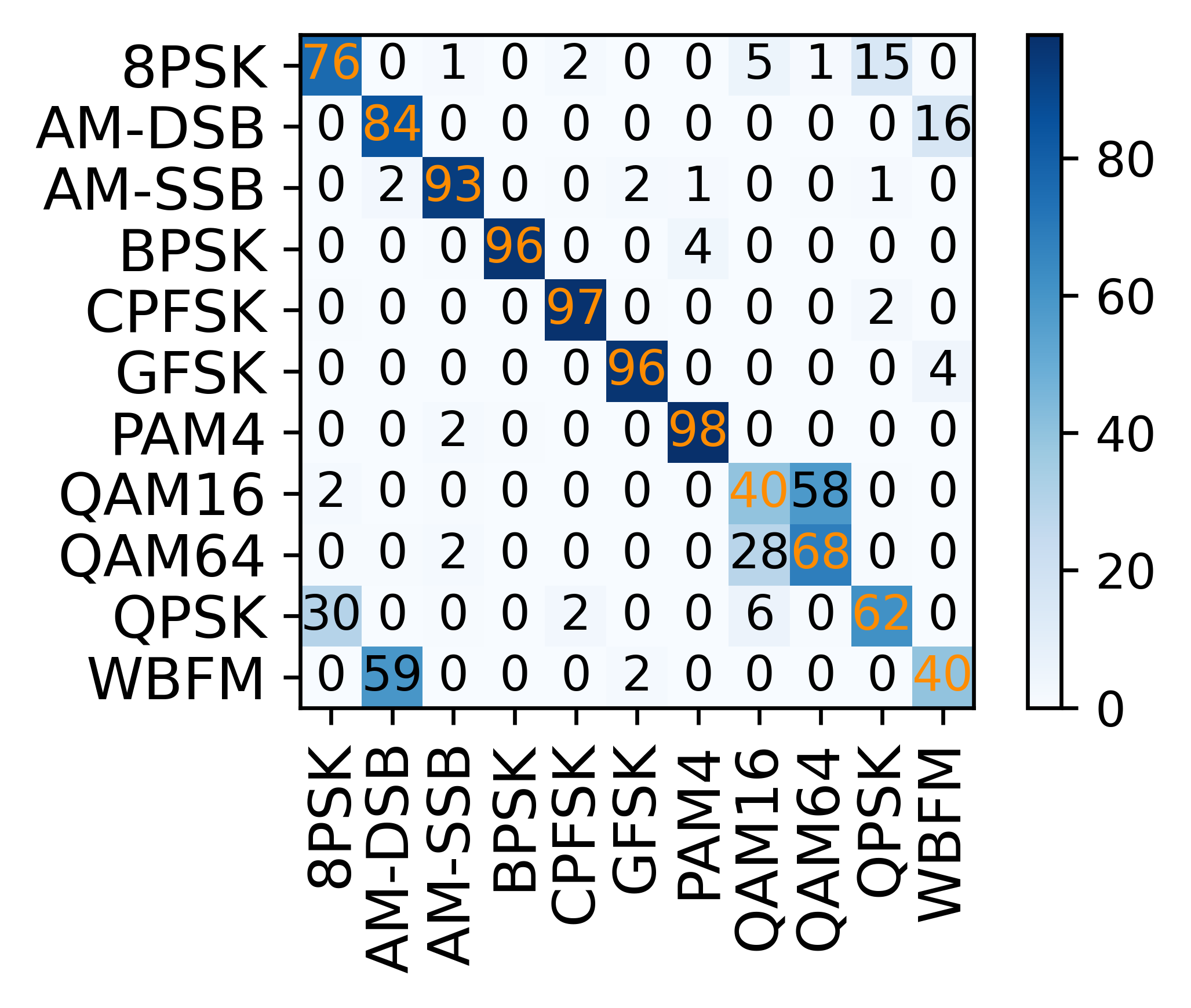}}
	\label{fig13j}
	\subfloat[]{\includegraphics[width=0.2\textwidth]{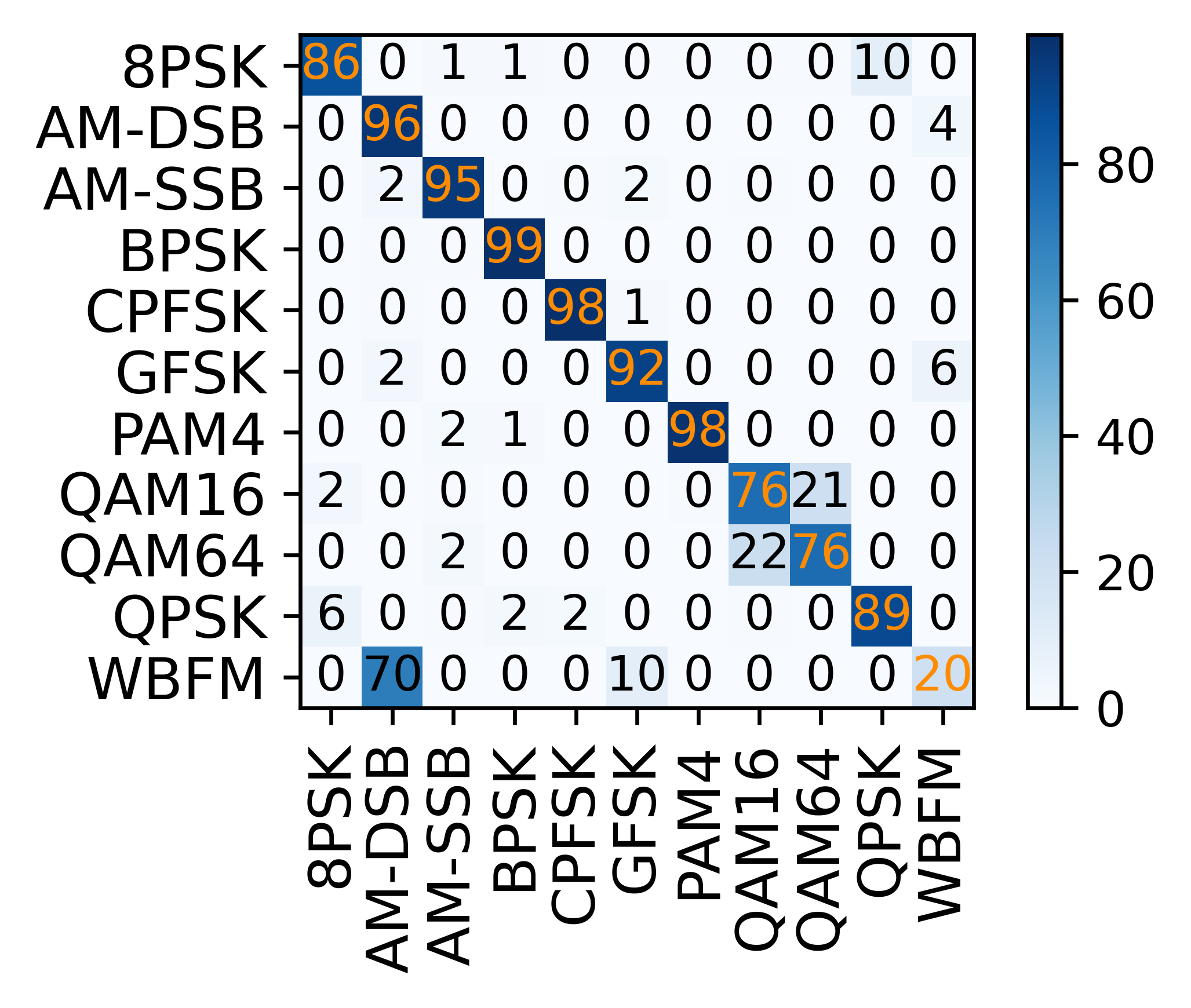}}
	\label{fig13k}
	\subfloat[]{\includegraphics[width=0.2\textwidth]{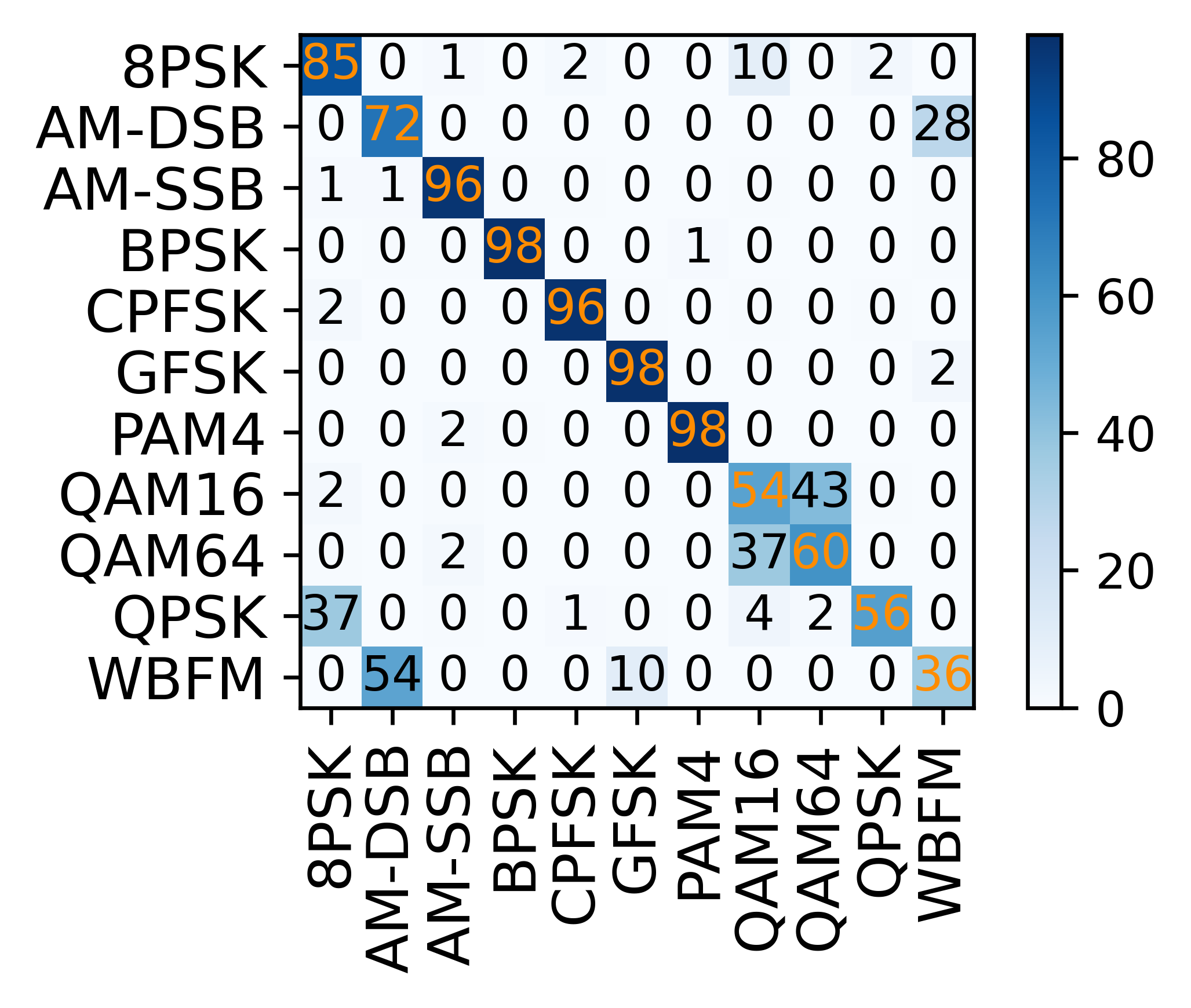}}
	\label{fig13l}\\
	\caption{Confusion matrix for communication modulation signals. (a) CLDNN 0 dB, (b) CNN 0 dB, (c) GRU 0 dB, (d) LSTM 0 dB, (e) MCLDNN 0 dB, (f) MCNet 0 dB, (g) PET-CGDNN 0 dB, (h) ResNet 0 dB.}
	\label{fig13}
\end{figure*}
Fig.~\ref{fig13} shows the confusion matrix results of 8 models under 0 dB SNR conditions, used to analyze the main sources of classification errors and the confusion characteristics between various modulation types. We select the typical confusion matrices of 8 models on the RML2016.10a dataset with SNR of 0 dB as representative results for analysis. The vertical axis represents the true modulation type label, and the horizontal axis represents the prediction result of the model.

The confusion between 16QAM and 64QAM modulation types is most evident. All models experienced misjudgments of these two types of modulation on a small number of samples, mainly due to the high similarity in the constellation distribution between the two, with some constellation points overlapping, making it difficult for the models to distinguish between them during discrimination. In contrast, the LSTM model performs the best, thanks to its multi-channel feature extraction structure that can more effectively distinguish constellation point distributions that are not completely overlapping.

There is significant confusion between WBFM and AM-DSB modulation types. Almost all models exhibit misclassification of WBFM signals as AM-DSB. This misjudgment mainly stems from the fact that both types of signal samples come from modulated signals with speech segments, which both contain pause and pause features in the time domain, resulting in the model extracting similar time patterns when learning features, leading to misclassification.
\begin{figure*}[!h]
	\centering
	\subfloat[]{\includegraphics[width=0.15\textwidth]{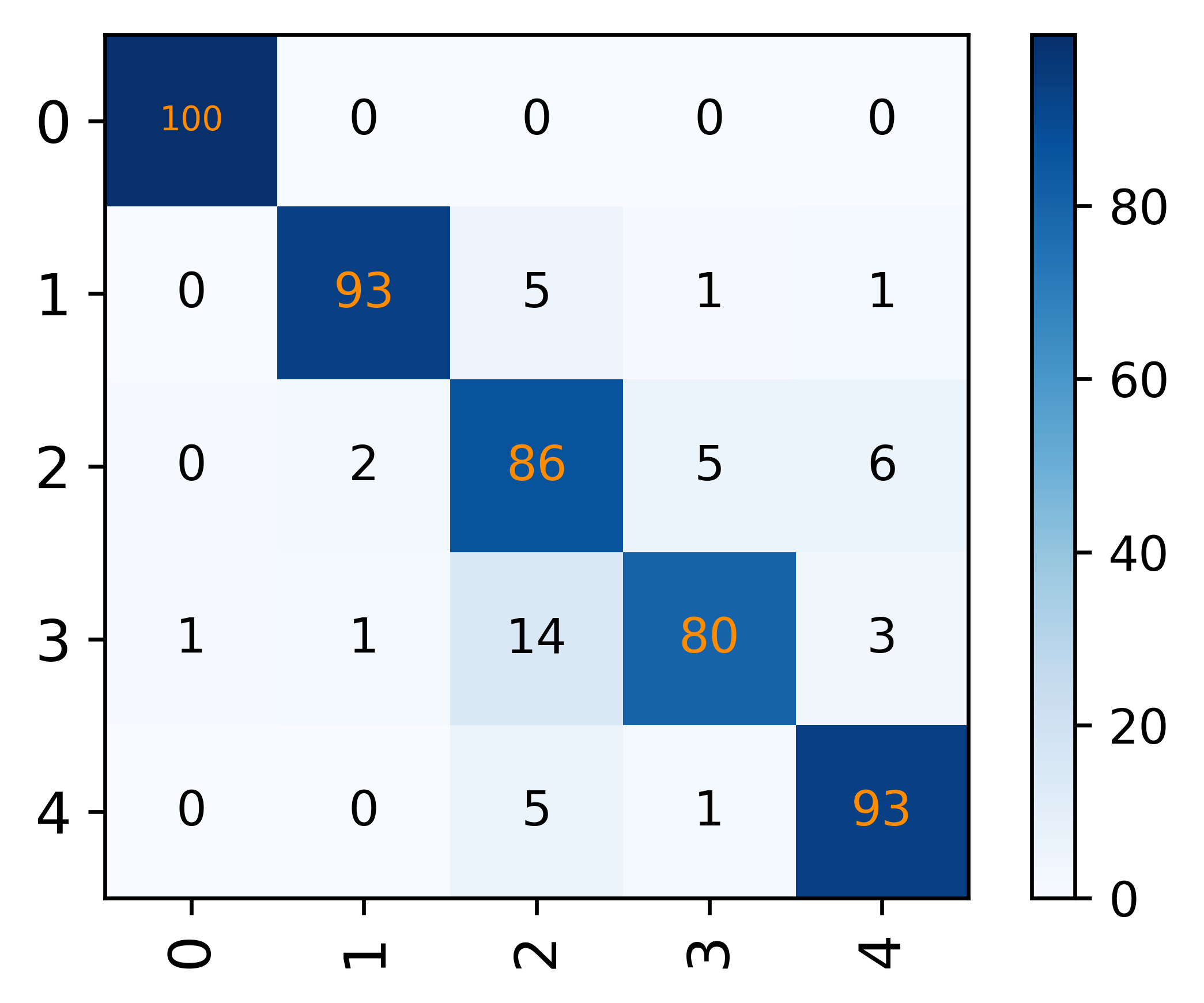}}
	\label{fig14a}
	\subfloat[]{\includegraphics[width=0.15\textwidth]{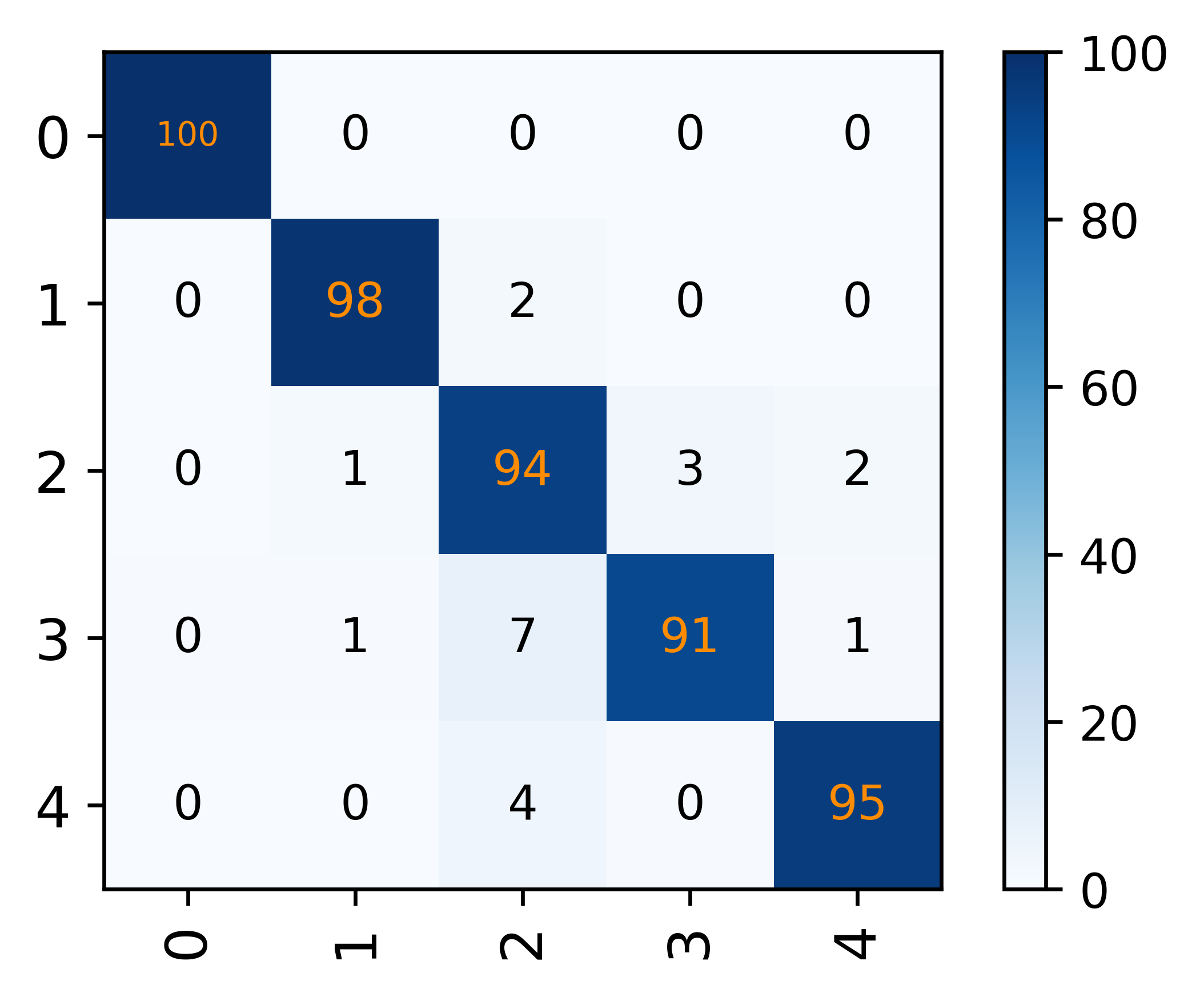}}
	\label{fig14b}
	\subfloat[]{\includegraphics[width=0.15\textwidth]{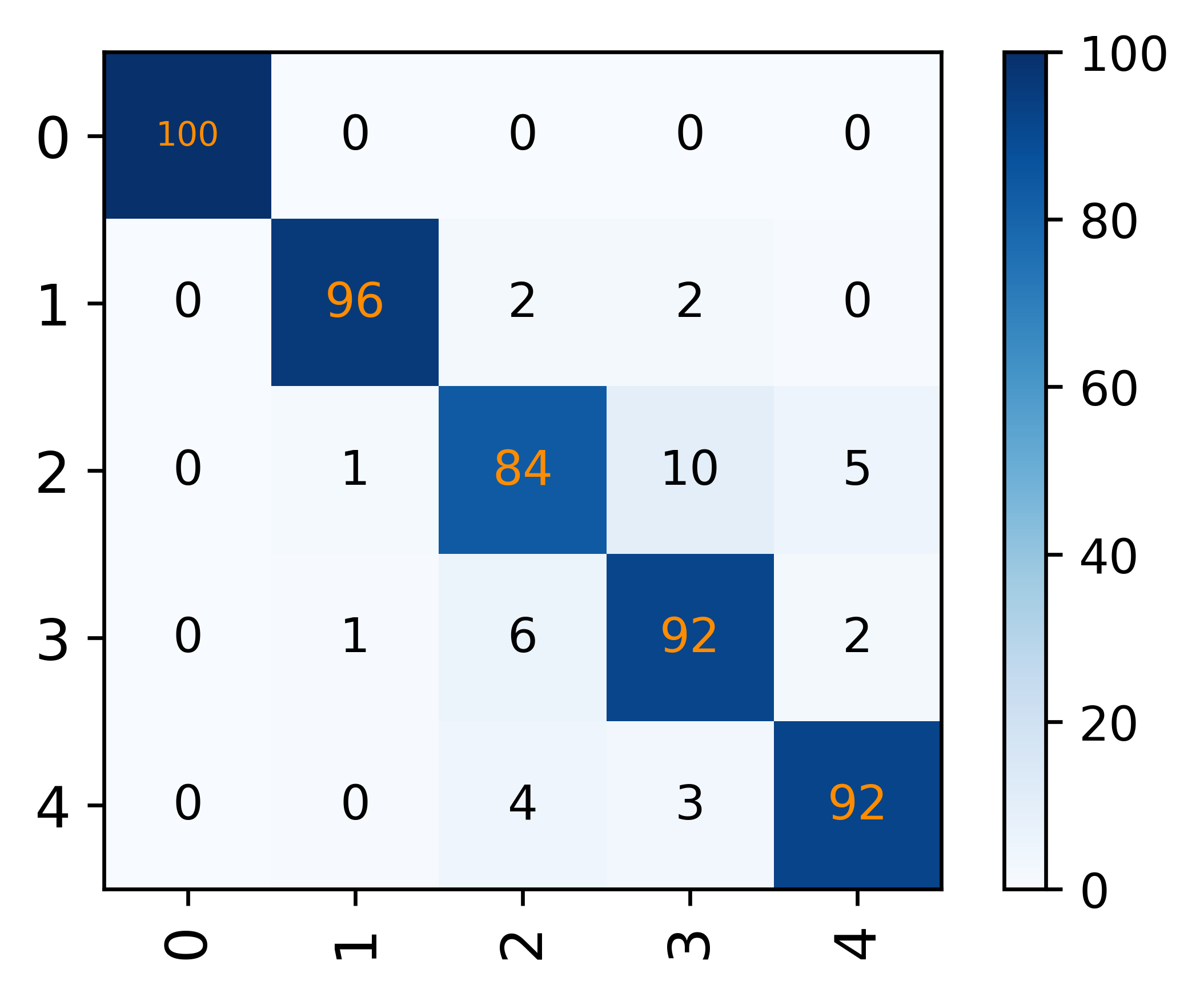}}
	\label{fig14c}
	\subfloat[]{\includegraphics[width=0.15\textwidth]{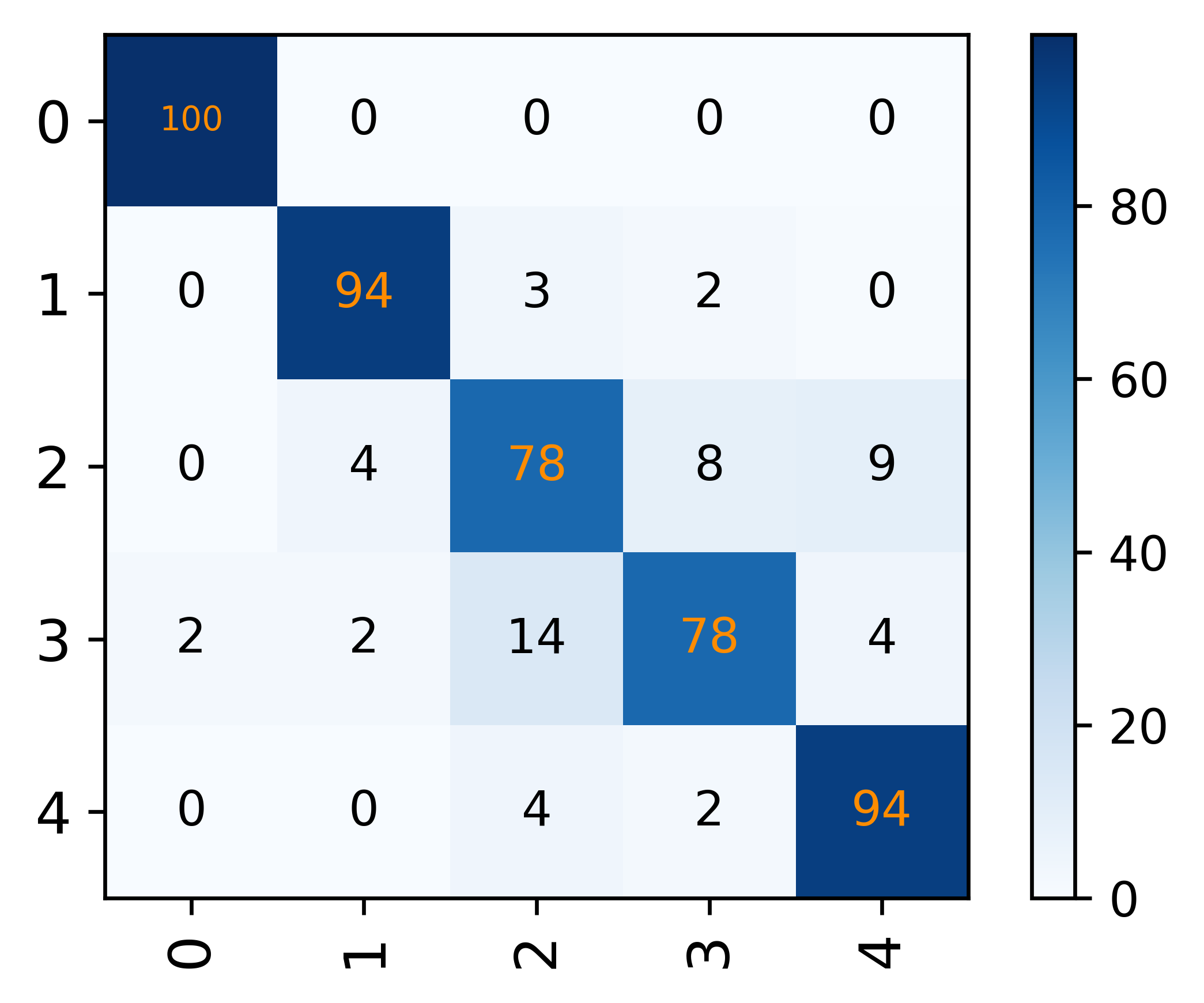}}
	\label{fig14d}\\
	\subfloat[]{\includegraphics[width=0.15\textwidth]{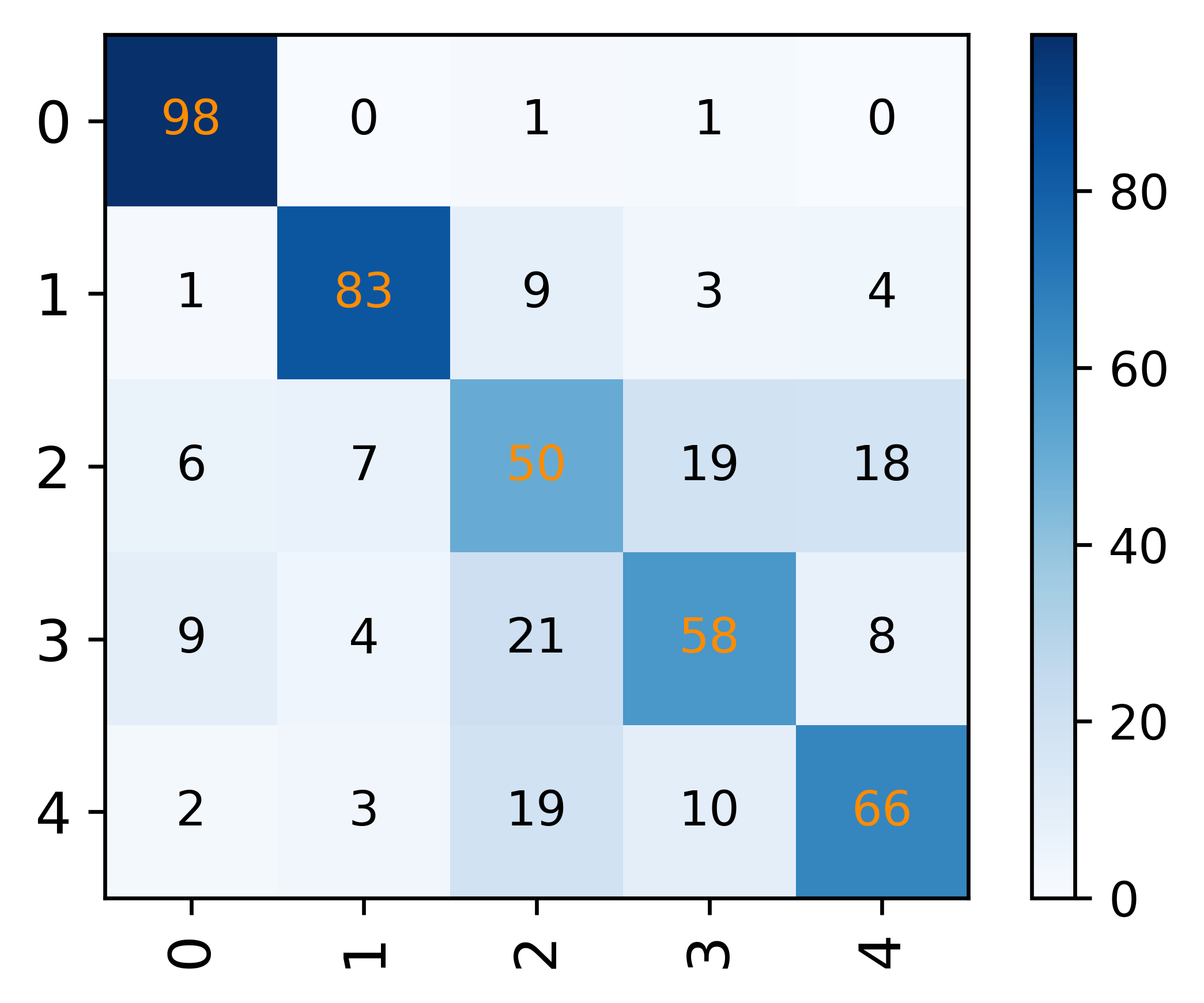}}
	\label{fig14e}
	\subfloat[]{\includegraphics[width=0.15\textwidth]{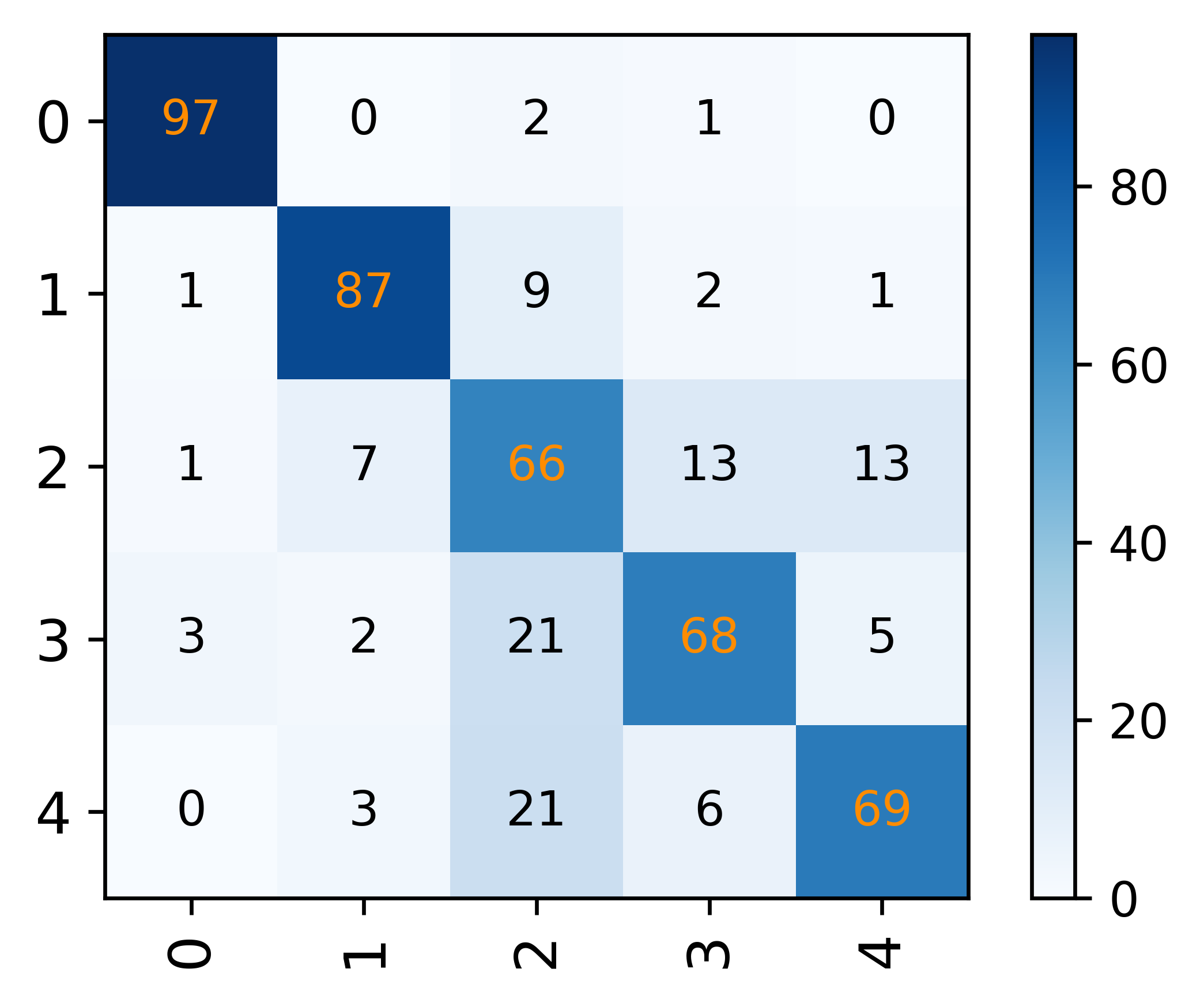}}
	\label{fig14f}
	\subfloat[]{\includegraphics[width=0.15\textwidth]{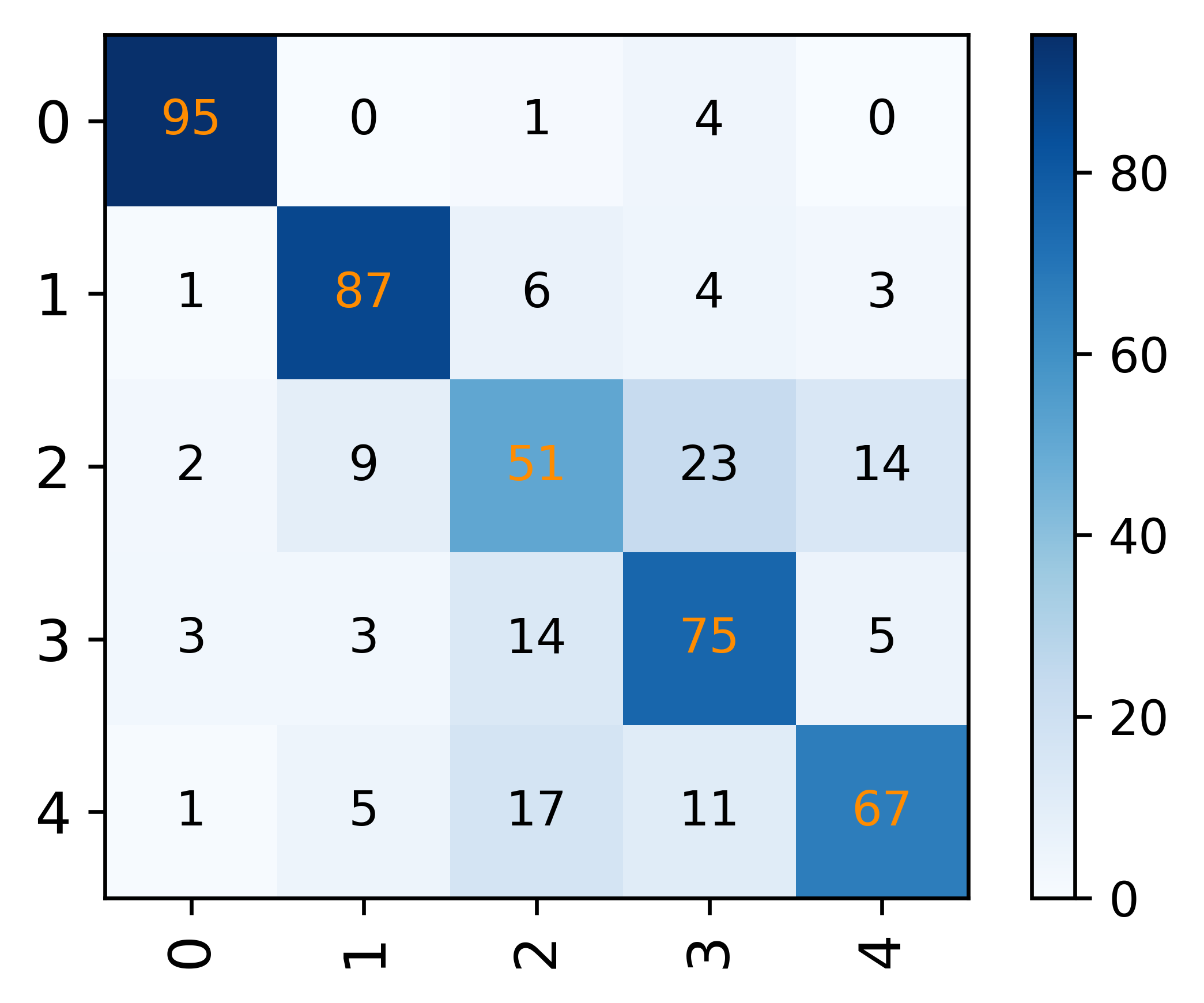}}
	\label{fig14g}
	\subfloat[]{\includegraphics[width=0.15\textwidth]{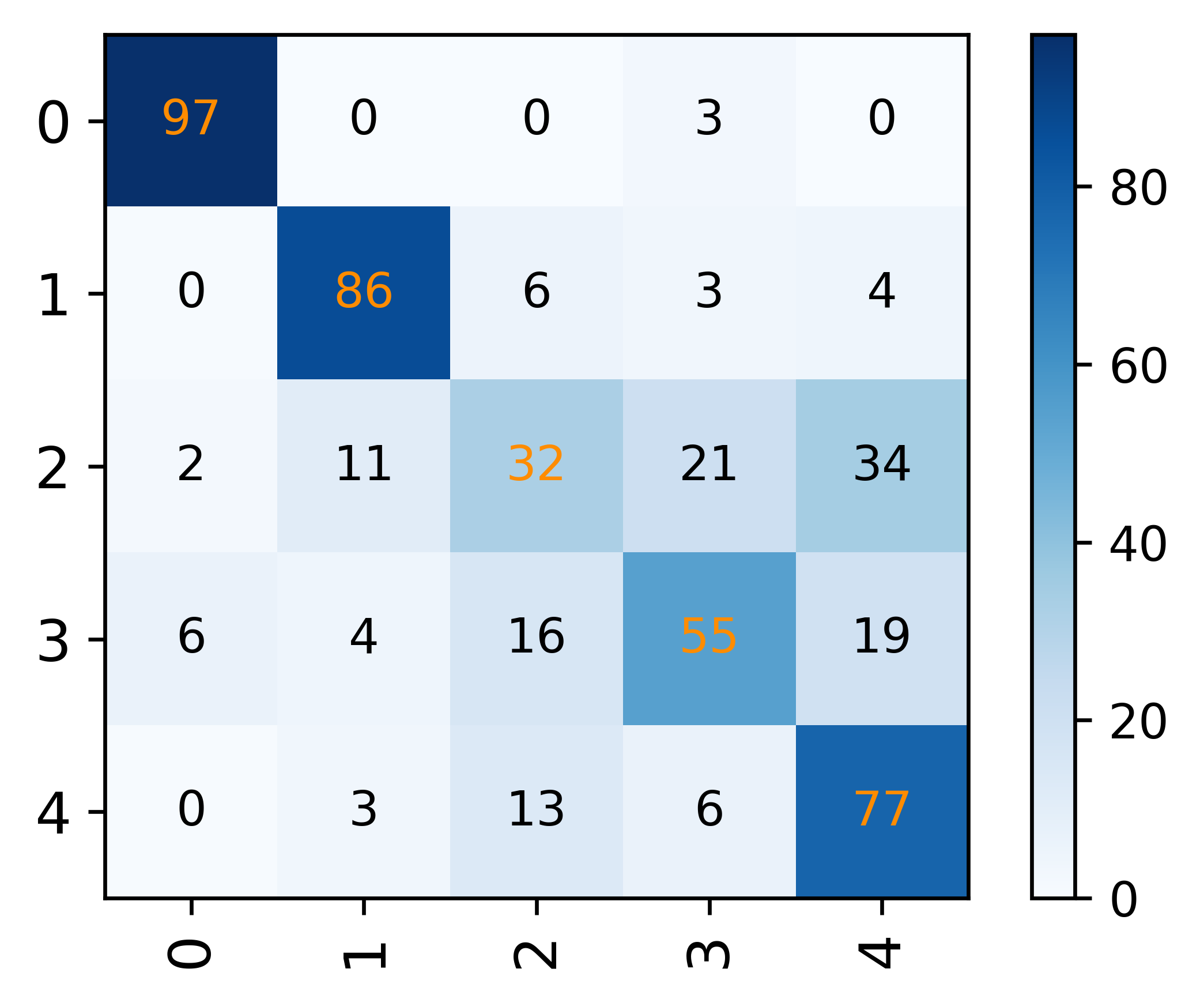}}
	\label{fig14h}\\
	\caption{Confusion matrix for radar modulation signals, where 0 to 4 are coherent pulse signal, barker code, polyphase barker code, frank code, and linear frequency modulation signal, respectively. (a) CLDNN -8 dB, (b) CNN -8 dB, (c) MCNET -8 dB, (d) RESNET -8 dB, (e) CLDNN -12 dB, (f) CNN -12 dB, (g) MCNET -12 dB, (h) RESNET -12 dB.}
	\label{fig14}
\end{figure*}

Fig.~\ref{fig14} shows the confusion matrix results of four methods, CLDNN, CNN, MCNET, and RESNET, for identifying five types of radar modulation signals at SNRs of -12 dB and -8 dB. With an SNR of -12 dB, the non diagonal confusion values of all methods are of a high level, and the correct recognition rate is usually less than 90\%, indicating that low SNR environments will significantly increase the difficulty of identifying radar modulation signals. The observed trends are specific to the evaluated radar waveform classes and should not be interpreted as universally representative of all radar AMR scenarios.

\subsection{Complexity Performance}
In this section, we tested the parameter count and floating-point operation count of several models to evaluate their computational complexity performance, which is shown in Table~\ref{tab12}. PETCGDNN achieves the lowest parameter count at 72.071K and MCNET achieves the lowest FLOPs at 5.654M, both demonstrating superior efficiency. In contrast, ResNet has the highest parameter count (3.099M), while ResNet also exhibits the largest computational overhead with 496.769M FLOPs. The MCLDNN achieves a good balance between model size and computational cost, with only 108.426K parameters and 53.079M FLOPs, outperforming traditional models like CNN, LSTM, and CLDNN in terms of efficiency.

\begin{table}[!h]
	\caption{Parameters and Flops comparison results. \label{tab12}}
	\renewcommand{\arraystretch}{1}
	\centering
	\resizebox{1.0\linewidth}{!}{
		\begin{tabular}{|c|c|c|c|c|c|}
			\hline
			Model & Parameters & Flops & Model & Parameters & Flops \\
			\hline
			CNN & 858.123K & 315.103M & GRU & 151.179K & 77.600M  \\
			\hline
			LSTM & 201.099K & 103.290M & PETCGDNN & 72.071K & 33.442M \\
			\hline
			MCNET & 121.611K & 5.654M & CLDNN & 519.243K & 255.080M \\
			\hline
			RESNET & 3.099M & 496.769M & MCLDNN & 108.426K & 53.079M  \\
			\hline
	\end{tabular}}
\end{table}

\subsection{Analysis of Performance Differences}
The performance gap among different architectures becomes more pronounced under low-SNR conditions. When the SNR is high, most discriminative modulation characteristics remain observable, and therefore different models tend to achieve similar recognition performance. However, as the SNR decreases, noise increasingly obscures local signal structures and reduces class separability.

CNN-based models primarily rely on local spatial patterns in constellation diagrams or time-frequency representations. Consequently, their performance may degrade significantly when local structures are corrupted by severe noise. In contrast, Transformer-based architectures can exploit long-range dependencies through self-attention mechanisms, allowing them to aggregate information from multiple signal segments and maintain stronger robustness under low-SNR conditions.

For raw I/Q sequence inputs, recurrent architectures such as RNNs, LSTMs, and GRUs may outperform image-based CNN models because modulation information is often encoded in temporal symbol transitions and phase evolution. Their recurrent structure explicitly models sequential dependencies, which are not fully preserved when the signal is transformed into image representations. Although Transformers generally achieve superior performance on large-scale datasets, their advantages may diminish when training data are limited. Unlike CNNs and RNNs, Transformers impose weaker structural priors and therefore require substantially larger datasets to learn robust representations. As a result, CNN-based architectures may remain competitive in small-data scenarios despite having lower representational capacity.

Performance differences across datasets are often associated with variations in modulation diversity, channel conditions, SNR distributions, and signal representations. Datasets containing a large number of modulation classes, severe channel impairments, or heterogeneous waveform families generally favor models with stronger representation learning capabilities. Conversely, relatively clean datasets with limited modulation diversity may not fully reveal the advantages of more sophisticated architectures.

The complexity performance comparison reveals that higher model complexity does not necessarily guarantee better recognition performance. While larger models typically possess stronger representation capacity, they may suffer from overfitting, increased training difficulty, and higher deployment costs. It should be noted that improved recognition performance does not necessarily imply lower computational complexity. Many deep learning architectures, particularly Transformers and foundation models, require significantly greater computational and memory resources than traditional feature-based methods. Therefore, the optimal AMR architecture should be selected according to the target operating environment, available computational resources, and recognition requirements rather than solely based on classification accuracy.

\subsection{Deployment Feasibility and Complexity Analysis}
Parameter count and FLOPs are widely used for evaluating model complexity, but they do not fully characterize deployment feasibility in practical AMR systems. Real-world deployment is additionally constrained by inference latency, memory footprint, energy consumption, hardware architecture, and communication overhead. Consequently, models with similar FLOPs may exhibit significantly different deployment performance on edge devices.

Inference latency is a critical factor for real-time AMR applications such as cognitive radio, spectrum monitoring, UAV sensing, and electronic warfare systems. CNN-based architectures generally achieve low inference latency because convolution operations can be efficiently parallelized. In contrast, recurrent architectures often exhibit longer execution times due to their sequential processing nature. Transformer and Mamba models provide stronger representation capabilities but may incur additional computational overhead for long observation sequences. Memory consumption is another important deployment consideration. Large Transformer-based models often require substantially more memory for storing parameters and intermediate attention maps than CNN-based models. This limitation can become critical for embedded platforms with restricted on-chip memory resources.

Energy efficiency is particularly important for battery-powered devices such as IoT sensors, underwater platforms, unmanned vehicles, and portable spectrum monitoring systems. Models with high computational complexity may achieve superior recognition accuracy but can significantly increase power consumption, thereby reducing operational lifetime. The suitability of an AMR model also depends on the target hardware platform. CNN-based models are generally well supported by GPUs, NPUs, and FPGA accelerators due to the regular structure of convolution operations. Transformer-based models may benefit from modern AI accelerators but often require greater memory bandwidth. Lightweight architectures are therefore attractive for deployment on resource-constrained edge devices. Table~\ref{tab:deployment} summarizes the deployment-oriented comparison of representative AMR models.

\begin{table*}[t]
	\caption{Deployment-Oriented Comparison of Representative AMR Models}
	\label{tab:deployment}
	\centering
	\renewcommand{\arraystretch}{1.2}
	\begin{tabular}{p{2.8cm}|p{2cm}|p{2cm}|p{2cm}|p{2.5cm}|p{4cm}}
		\hline
		\textbf{Model}
		&
		\textbf{Latency}
		&
		\textbf{Memory}
		&
		\textbf{Energy Efficiency}
		&
		\textbf{Hardware Friendliness}
		&
		\textbf{Typical Deployment Scenario}
		\\
		\hline
		
		SVM/RF
		&
		Low
		&
		Low
		&
		High
		&
		CPU-friendly
		&
		Small-scale embedded devices
		\\
		\hline
		
		CNN
		&
		Low
		&
		Moderate
		&
		High
		&
		GPU/FPGA-friendly
		&
		IoT, edge intelligence
		\\
		\hline
		
		RNN/LSTM/GRU
		&
		Moderate
		&
		Moderate
		&
		Moderate
		&
		Sequential execution
		&
		Temporal signal analysis
		\\
		\hline
		
		Transformer
		&
		High
		&
		High
		&
		Low
		&
		GPU/NPU-oriented
		&
		Cloud and high-performance platforms
		\\
		\hline
		
		Mamba
		&
		Moderate
		&
		Moderate
		&
		Moderate
		&
		Edge AI accelerators
		&
		Long-sequence AMR
		\\
		\hline
		
		Foundation Models
		&
		Very High
		&
		Very High
		&
		Low
		&
		Large-scale AI infrastructure
		&
		Cloud-based wireless intelligence
		\\
		\hline
		
	\end{tabular}
\end{table*}

\section{Open Issues and Future Directions}
Although AI technology has achieved significant success in AMR, there are still many issues that need to be addressed. Open issues and research directions will be discussed in this section. The open issues discussed in this section are not independent challenges, but rather arise from the theoretical limitations of existing AMR frameworks. Specifically, the unified formulation introduced in Section II highlights several fundamental difficulties associated with observation uncertainty, nuisance parameters, feature robustness, representation learning, and distribution mismatch. These theoretical gaps motivate the following research directions.

\subsection{Accuracy Enhancement in Complex Channel Environment}
As can be observed in Section \uppercase\expandafter{\romannumeral3}, Statistical ML and DL models have achieved great succeeds under high SNR ($>0$) conditions. Most work focuses on Gaussian white noise channel models, but actual communication environments are subject to pulse noise and strong multi-path signal interference, making high-precision modulation recognition in low SNR and complex channel environments a huge challenge. Existing algorithms often separate channel estimation and modulation recognition into independent steps, resulting in delayed compensation for channel distortion. In the future, an end-to-end joint model can be constructed to integrate channel state information, Doppler frequency shift, noise type, and other information into the DL model. Moreover, the traditional assumption of Gaussian noise is difficult to cover pulse noise, impulse noise, and even interference attack noise in scenarios such as radar and underwater acoustic communication. In the future, a combination of fractional low order statistics and deep learning can be used to preprocess non-Gaussian noise signals first, and then input them into deep learning models to extract robust features.

\subsection{Breakthrough in Few-shot AMR}
The few-shot AMR problem originates from insufficient sampling of the underlying modulation distributions. Limited observations prevent the learning model from accurately estimating class boundaries, especially in complex channel environments. Consequently, meta-learning, generative modeling, and foundation-model-based approaches may provide promising solutions for improving generalization under data scarcity. For example, a modulation signal training model based on meta-learning can be designed to construct meta tasks using local features of IQ sequences and time-frequency maps, and optimize the ability to quickly fine tune through model-agnostic meta-learning (MAML) \cite{Finn2017MAML}.

\subsection{Hybrid and Cross Domain Feature Fusion}
Hybrid-feature-based AMR mainly uses CNN to process image modalities such as constellation maps, time-frequency maps, and cyclic spectrograms, while RNN processes sequence modalities or simple combinations. However, there are problems such as fixed fusion modes, non-dynamically adjusted modal weights, and cross-modal feature misalignment. The fusion gain is limited at low SNRs. The potential future research direction is to design a dynamic attention fusion module for SNR perception, which calculates the classification contribution of each modality in real time, such as enhancing the modal weights of cyclic spectrograms at low SNRs and focusing on constellation modalities at high SNRs.

\subsection{Lightweight DL Models for AMR}
Lightweight AMR is fundamentally a resource-constrained optimization problem. Reducing model complexity often decreases representation capacity and feature resolution, thereby affecting classification accuracy. Future research should investigate the tradeoff among recognition performance, computational complexity, memory consumption, and inference latency.

\subsection{AMR Combining Different Modulation Methods and Waveforms}
Currently, most research on AMR predominantly operates under the assumption that the employed waveform is either OFDM-based or a straightforward single-carrier type. However, the increasing coexistence of heterogeneous communication and sensing waveforms introduces a mixed-waveform recognition problem. For instance, OTFS technology, which maps information onto the delay-Doppler domain, exhibits exceptional performance in high-speed mobile scenarios and time-varying channel environments. Other advanced waveforms, such as AFDM and orthogonal chirp division multiplexing (OCDM), also showcase unique merits in specific application scenarios like high-speed mobile communications. Against this backdrop, AMR is confronted with novel challenges: it now needs to simultaneously identify both waveform types (e.g., OFDM/OTFS) and subcarrier modulation schemes (e.g., QPSK/16QAM). Future methods should focus on scalable recognition frameworks capable of handling heterogeneous and evolving waveform spaces.

\subsection{AMR for ISAC Signals}
ISAC AMR introduces a fundamental identifiability problem because communication and sensing components are superimposed in the received observation. The receiver must determine whether the mixed signal can be uniquely decomposed into its constituent communication and sensing waveforms under realistic channel conditions. Therefore, future research should investigate signal separation, joint inference, and identifiability theory for ISAC signal recognition.

\subsection{Foundation Models for AMR}
Most existing AMR systems are trained for specific modulation sets, channel conditions, and signal types. As discussed in the preceding sections, these models often suffer from limited generalization capability under distribution shifts, unseen waveform classes, and data-scarce scenarios. Such limitations motivate the development of foundation-model-based AMR frameworks capable of learning more universal signal representations. Inspired by the success of foundation models in natural language processing and computer vision, recent wireless intelligence research has begun exploring wireless foundation models (WFMs) that are pre-trained on large-scale heterogeneous signal datasets and subsequently adapted to downstream tasks through fine-tuning or prompting. Unlike conventional task-specific AMR models, foundation models aim to learn transferable representations that generalize across modulation types, channel conditions, hardware platforms, and sensing environments. For AMR, foundation models have the potential to unify several traditionally separate learning paradigms, including modulation recognition, waveform classification, emitter identification, interference recognition, and ISAC signal analysis. By leveraging large-scale pre-training, they may significantly reduce the dependence on task-specific feature engineering and large labeled datasets.

\subsection{Real-World Datasets and Deployment Challenges}
Although substantial progress has been achieved in AMR through deep learning, the majority of reported results are still obtained using simulated datasets. Compared with computer vision and natural language processing, publicly available large-scale real-world signal datasets remain extremely limited. This shortage is particularly evident for radar sensing and ISAC applications, where data collection is constrained by hardware availability, spectrum regulations, operational security, and environmental complexity.

Beyond dataset availability, practical deployment introduces additional challenges, including computational constraints, energy consumption, hardware heterogeneity, synchronization errors, and online adaptation requirements. Future AMR systems should therefore be evaluated not only in terms of recognition accuracy but also with respect to robustness, reliability, latency, and deployment cost.

\section{Conclusion}
AI is a promising technology for AMR, benefiting from its strong capabilities of complex feature extraction. In this paper, we first investigate the modulation types commonly used in current communication and radar systems. Next, we provide a structured survey of AMR technologies from the perspectives of signal models, feature representations, learning paradigms, and application scenarios. Finally, we highlight open issues and propose future research directions for AMR.

\section*{Acknowledgments}
This work was supported in part by the National Natural Science Foundation of China under Grant 42404001, in part by the Shandong Provincial Natural Science Foundation under Grant ZR2023QF128, and in part by the Frontier Exploration Project of Hanjiang National Laboratory under Grant TS2024026. There is no conflict of interests.

\bibliographystyle{IEEEtran}
\bibliography{IEEEabrv,draft_hw}
\end{document}